\documentclass[a4paper,fleqn]{cas-dc}

\usepackage[numbers]{natbib}

\usepackage{amsmath}
\usepackage{amssymb}
\usepackage{algorithm}
\usepackage{algpseudocode}
\usepackage{booktabs}
\usepackage{bm}
\usepackage{etoolbox}
\usepackage{graphicx}
\usepackage{hyperref}
\usepackage[utf8]{inputenc}
\usepackage{siunitx}
\usepackage{svg}
\usepackage{xfrac}
\usepackage{tabularx}
\usepackage[dvips]{epsfig}
\usepackage{psfrag}
\usepackage{enumitem}

\usepackage{comment}
\usepackage[caption=false,font=normalsize,labelfont=sf,textfont=sf]{subfig} 

\newcommand{\sk}[1]{m^{\textrm{scan}}_{ij}(#1)}
\newcommand{\kflc}{k^{\textrm{ctrl}}}
\newcommand{\Tflc}{T^{\textrm{ctrl}}}
\newcommand{\krob}{k^{\textrm{rob}}}
\newcommand{\Trob}{T^{\textrm{rob}}}

\newcommand{\nxe}{n^{\textrm{h}}}
\newcommand{\nye}{n^{\textrm{v}}}

\newcommand{\Efeas}{E^{\textrm{feasible}}}

\newcommand{\nr}{n^{\textrm{r}}}
\newcommand{\nv}{n^{\mathrm{victim}}}

\newcommand{\Mgrid}{M^{\textrm{grid}}}
\newcommand{\Mdebris}{M^{\textrm{debris}}(k)}
\newcommand{\mdebris}{m^{\textrm{debris}}}
\newcommand{\op}{o_{ij}(k)}

\newcommand{\Mvictim}{M^{\textrm{victim}}(k)}
\newcommand{\mvictim}{m^{\textrm{victim}}}
\newcommand{\nvp}{n^{\mathrm{victim,p}}_{ij}(k)}
\newcommand{\Mscan}{M^{\textrm{scan}}(k)}
\newcommand{\mscan}{m^{\textrm{scan}}}
\newcommand{\Mstr}{M^\textrm{structure}}
\newcommand{\mstr}{m^\textrm{structure}}
\newcommand{\Mvel}{M^{\textrm{velocity}}(k)}
\newcommand{\mvel}{m^{\textrm{velocity}}}
\newcommand{\Mdir}{M^{\textrm{direction}}(k)}
\newcommand{\mdir}{m^{\textrm{direction}}}
\newcommand{\Mf}{M^{\textrm{fire}}(k)}
\newcommand{\mf}{m^{\textrm{fire}}}
\newcommand{\pf}{\pi^{\textrm{fire}}_{ij,lq}}
\newcommand{\kfipjp}{k^{\textrm{fire}}_{ij}}
\newcommand{\kfij}{k^{\textrm{fire}}_{lq}}
\newcommand{\fij}{f_{lq}}
\newcommand{\vij}{v_{ij,lq}}
\newcommand{\psiij}{\psi_{ij,lq}}

\newcommand{\va}{v^{\max}_r(\krob)}
\newcommand{\vg}{\bar{v}^{\max}_r}
\newcommand{\Np}{N^{\textrm{p}}}

\newcommand{\kone}{k^{2\textrm{min}}_{lq}}
\newcommand{\ktwo}{k^{10\textrm{min}}_{lq}}
\newcommand{\target}{\left( \tau_x , \tau_y \right)_{r,\krob}}

\newcommand{\thetaG}{\theta^{\textrm{ground}}_r(\krob)}
\newcommand{\thetaCorr}{\theta^{\textrm{correction}}_r(\krob)}

\newcommand{\dps}{\displaystyle}

\def\tsc#1{\csdef{#1}{\textsc{\lowercase{#1}}\xspace}}
\tsc{WGM}
\tsc{QE}
\tsc{EP}
\tsc{PMS}
\tsc{BEC}
\tsc{DE}

\begin{document}
\let\WriteBookmarks\relax
\def\floatpagepagefraction{1}
\def\textpagefraction{.001}
\shorttitle{Model Predictive Fuzzy Control}
\shortauthors{C. Maxwell et~al.}

\title [mode = title]{Model Predictive Fuzzy Control: A Hierarchical Multi-Agent Control Architecture for Outdoor Search-and-Rescue Robots}                      

\author[1]{Craig Maxwell}
\fnmark[1]

\affiliation[1]{organization={Department of Control and Operations, Delft University of Technology},
                city={Delft},
                postcode={2629 HS}, 
                country={The Netherland}}

\author[1]{Mirko Baglioni}
\cormark[1]
\fnmark[1]
\ead{M.Baglioni@tudelft.nl}

\author[1]{Anahita Jamshidnejad}
\cortext[cor1]{Corresponding author}
\fntext[fn1]{C.\ Maxwell and M.\ Baglioni are first co-authors.}

\begin{abstract}
Autonomous robots deployed in unknown search-and-rescue (SaR) environments can significantly improve the efficiency of the mission by assisting in fast localisation and rescue of 
the trapped victims. 
Control systems that coordinate and enable these robots to map the environment without direct supervision of humans 
face serious computational challenges. 
We propose a novel integrated hierarchical control architecture, called model predictive fuzzy control (MPFC),  
for autonomous mission planning of multi-robot SaR systems that should efficiently map an unknown environment:  
We combine model predictive control (MPC) and fuzzy logic control (FLC), where the robots are locally controlled by computationally efficient FLC controllers, and the parameters of these local controllers are tuned via a centralised MPC controller, in a regular or event-triggered manner.  
Although
MPC is computationally demanding, this does not result in a bottleneck for MPFC, due to the following reasons: First, the frequency of re-tuning the FLC controllers via the MPC layer is (much) smaller than the control frequency. 
Second, the system runs via the decisions of the FLC controllers, and is thus not dependent (directly) on the solutions of the MPC layer.
The proposed architecture provides three main advantages: 
(1) The control decisions are made by the FLC controllers, 
thus the real-time computation time is affordable. 
(2) The centralised MPC controller optimises the performance criteria with a global and predictive vision of the system dynamics, and updates the parameters of the FLC controllers accordingly. 
(3) FLC controllers are heuristic by nature and thus do not take into account optimality in their decisions, while 
the tuned parameters via the MPC controller can indirectly incorporate some level of optimality in local decisions 
of the robots.
A simulation environment for victim detection in a disaster environment was designed in MATLAB using discrete, 2-D grid-based models.
While being comparable from the point of computational efficiency, the integrated MPFC architecture improves the performance of the multi-robot SaR system compared to decentralised FLC controllers. 
Moreover, the performance of MPFC is comparable to the performance of centralised MPC for path planning of SaR robots, whereas MPFC requires significantly less computational resources, since the number of the optimisation variables in the control problem are reduced.%
\end{abstract}

\begin{keywords}
model predictive control \sep fuzzy logic control \sep hierarchical control \sep multi-agent systems \sep search-and-rescue robots
\end{keywords}

\maketitle

\section{Introduction} \label{sec:introduction}

In recent years, robots are increasingly being used for improving the outcome of search-and-rescue (SaR) missions, via contributing 
to various tasks, including mapping the unknown environments, and by locating the victims \cite{liu_nejat_2013, grogan_pellerin_gamache_2018, tanzi_et_al_2016}. 
The main benefits of using SaR robots are reduced cost, improved time efficiency, and less risks for the SaR crew.
While humans use their heuristics to deal with unknown or uncertain SaR situations, robots can optimise the mission plans 
and safely explore the environment through systematic mathematical approaches.

In this paper, we introduce a mission planning approach for multi-agent systems, particularly for a multi-robot SaR team 
that should autonomously map an unknown SaR environment.
Our proposed approach includes a novel two-layer control architecture, 
where fuzzy logic control (FLC) is used locally for the path planning of each 
individual agent and model predictive control (MPC) is used globally at the top layer for online tuning of the fuzzy logic 
controllers, in order to improve the predicted combined performance of the agents. 
The resulting integrated architecture is called model predictive fuzzy control (MPFC).
Moreover, we use an agent-based simulation environment in order to model
various SaR scenarios, and to assess 
desired performance of a multi-agent team of drones that autonomously maps the environment using MPFC.

\subsection{Motivations}

Disasters may cause rapid damage across a widespread area, leaving many victims trapped in the environment.
SaR is the process of mapping the disaster environments and rescuing the trapped victims, 
with the objective of saving as many victims as possible. 
Therefore, reducing the SaR mission time, while increasing the coverage of the unknown area, are very crucial \cite{rom_kelman_2020}.
Some of the main challenges regarding SaR in the aftermath of disasters include the large size of the disaster environments, 
which may span an entire city, the unknown location of the SaR targets (e.g., the trapped victims), 
the life-threatening dangers for the SaR crew, especially due to uncertainties about, e.g., fire propagation, explosions, and unstable structures, as well as areas that are inaccessible or difficult to access for humans.
These factors cause SaR operations to be extremely resource intense and slow.

Robotic systems that are designed to assist SaR missions have the potential to save many lives by automating 
the mission planning and by allowing for a more efficient allocation of the SaR resources \cite{sampedro_et_al_2019, verykokou_et_al_2016}.
Therefore, in recent years, the use of robots for assisting the SaR crew in dangerous or complicated stages of SaR missions has been growing \cite{shah_choset_2004}. 
In addition to their contribution to the time-efficiency of SaR missions, deploying robots allows human rescuers to contribute to other complex operations, including providing logistics or medical support \cite{murphy_et_al_2008}.

On the one hand, FLC is a heuristic control technique that can integrate human knowledge into the control system, including highly nonlinear systems. 
In fact, by including rules given by human experts, no data or physical knowledge of the system is necessary, while being a white box approach. 
This results in efficient real-time decision making with low computational costs \cite{dubois_et_al_2014}. 
On the other hand, MPC systematically incorporates various state and input constraints into 
the decision making process, while it optimises one or several (possibly competing) objectives \cite{rawlings_mayne_2017}. 
All these characteristics of both MPC and FLC are highly relevant for autonomous controllers 
that are designed for SaR robots. 
Thus, we introduce MPFC, which exploits the benefits of both control approaches. The purpose is 
to develop a novel intelligent mission planning architecture for SaR robots that makes them more autonomous, and that improves their performance. MPFC is particularly suited for systematic mapping of unknown or partially known SaR environments.%

Multi-agent systems are mainly controlled via one of the following architectures: 
centralised, decentralised (when there is no interaction among the agents that allows them to directly influence the decisions of other agents), and distributed (when agents can exchange information and can thus influence the decisions of each other) \cite{parker_et_al_2016}. 
Each of these architectures faces advantages and drawbacks: While centralised control can provide global optimality for the performance, it is computationally very 
expensive (especially for large-scale systems) and the control system collapses, whenever the centralised controller fails during the mission \cite{matoui_et_al_2020}. 
Besides, decentralised control architectures are usually computationally less expensive than the other two architectures, 
but if non-negligible influences and inter-dynamics exist among the agents, the performance of this architecture is worse 
due to lack of coordination between the agents \cite{raibail_et_al_2022}. 
Finally, distributed control may be used to break the centralised control 
problem into smaller local control problems with links and inter-dynamics. 
Distributed control, however, requires exchange of information among the agents 
(which may be very costly or even impossible for SaR robots) and computationally complicated algorithms 
for coordination of the agents. 
Our proposed architecture, MPFC, integrates the advantages of all these architectures through its bi-layer structure.

\section{Main contributions and structure of the paper} \label{sec:contributions}

The main contributions of this paper include:
\begin{itemize}
    \item 
    A novel control architecture, called MPFC, which integrates 
    and combines the advantages of MPC and FLC 
    to include global optimality and predictive decision making of MPC and time-efficient, human-inspired decision making of FLC in one control system
    \item 
    An event-triggered tuning and control architecture that performs via a decentralised architecture in real time, 
    but still incorporates desired levels of coordination and performance improvement 
    at the global level within the multi-agent decision making system
    \item 
    Application of MPFC for autonomous multi-robot path planning in SaR missions, proposing a grid-based 
    2D model of a dynamic SaR environment that includes fire spread dynamics, 
    as well as performance analysis and assessment of different configurations of the MPFC system compared to 
    different control systems for multi-agent SaR robotics
\end{itemize}

The rest of the paper is organised as follows.
Section \ref{sec:background} details the existing research related to the topic of this paper.
Section \ref{sec:methodologies} introduces the statement of the problem identified in this paper, and details the modelling of the disaster environment, agents, and controller.
Section \ref{sec:case_study} presents the case study that we have analyzed and implemented,
where in Section \ref{sec:results} we present the results of the simulations outcomes of the case study and in Section \ref{sec:discussion} we discuss them.
Section \ref{sec:conclusion} concludes the findings of this paper and gives the main recommendations for future research to expand upon the work in this paper.
Moreover, Table~\ref{tab:notation} shows the mathematical notation frequently used in the paper.%

\section{Background} \label{sec:background}

The focus of this paper is on the mission planning for a team of SaR flying robots (also called unmanned aerial vehicles or UAVs). 
This involves path planning with the aim of 
mapping the SaR environment, i.e. to determine the path of the robots and to detecting the victims' location and other elements (e.g., obstacles) in the SaR environment.
For multi-robot systems, the problem of appropriate coordination and interaction of the robots should be addressed. 
In particular, the mission planning approaches used for multi-agent (or multi-robot) systems can be divided into \emph{centralised} \cite{bento_et_al_2013} 
(when a single supervisory system controls all the robots), \emph{decentralised} \cite{balch_arkin_1998} 
(when each robot has a local controller that operates without communicating with the controller of other robots), 
and \emph{distributed} \cite{ferranti_et_al_2022} (when there are local controllers for different robots, but these 
controllers communicate relevant information with each other to provide coordination and cooperation among the robots). 
Finally, \emph{hierarchical} \cite{ocampo_et_al_2013} control systems involve more than one control layer, where 
different categories of multi-agent control methods can be combined (e.g., distributed local controllers at the bottom layer of control and a supervisory global controller at the top layer of control).
Alternatively, approaches can be distinguished into \emph{cooperative} approaches, where there is sharing of information between the robots, and in \emph{non-cooperative} approaches, where there is not.

Firstly, in those cases in which the entire robot team is supervised from a single point of control 
(e.g., in SaR missions where it is possible to have all the
available information to solve the coordination problem, 
all the measurements of the complete system can be received by a central location, 
a model that describes the dynamics of the complete system is available, 
there is a precise point from which the robots can be observed, 
there is the need for transferring expensive onboard operations to the more powerful ground station, 
or the fused map has to be created on the central station and presented to the first responders), \emph{centralised} approaches are used.
In these methods (e.g., task allocation \cite{geng_et_al_2018} and path planning \cite{matoui_et_al_2020}),
the central station communicates with all the robots from a higher point of view. 
In state-of-the-art approaches where centralised control architectures are presented and compared to other types of architecture,
for example using the centralised controller to determine a sequence of reference way-points per UAV,
it is shown that centralised approaches can outperform the others in area coverage, while requiring more time for the coordination than decentralised approaches.
Similarly, other centralised approaches achieve an optimal assignment in task allocation, but they too computationally expensive, and therefore they become unsuitable for dynamic SaR environments. 
Heuristic control methods, e.g. fuzzy-logic-based, for discrete-space path planning, are also used in a centralised way, 
in which for example a map of the search area is generated where places with a higher risk and with a higher probability 
for finding the trapped victims are specified. 
These methods achieve medium results: 
it is shown that distributed approaches have better performance in smaller maps, because of how they are designed, and they are faster in areas with high priority (in terms of risk and probability of victim presence),
while the centralised approach performs better in speed in areas with low priority.
Moreover, hierarchical cooperative, reinforcement-learning-based control methods, can be used so that robots learn to cooperate and explore and identify the victims within the disaster environment. 
The control system determines which regions and victims should be explored and identified by each robot.
Comparing cooperative and non-cooperative exploration for multi-robot teams, the results show that the robots that learn a cooperative approach perform better in the sense of covering an area and finding victims, while not increasing the mission time in terms of exploration steps.
In some cases, e.g. when a SaR multi-UAV algorithm for collaborative exploration and mapping of a forest environment is considered, 
a centralised architecture is implemented in order to perform heavy computations at the SaR ground station. 
Hence, a limited communication bandwidth is faced, 
and therefore a map compression scheme that compresses onboard data is used. 
Increasing the interactions among the robots is not necessary when the search regions do not overlap with each other, 
but in order to, e.g., assign robots to different search areas at different times, explicit multi-robot coordination is needed \cite{alotaibi_et_al_2019, sanJuan_et_al_2018, liu_nejat_2016, tian_et_al_2020}.

Secondly, when due to practical reasons (e.g., in SaR missions this may be due to Wi-Fi interruptions or sensor limitations, 
the need for reducing energy consumption, limited communication ranges, or the risks associated with security reasons) communication is not possible among various robots, or when this communication 
is too costly, \emph{decentralised} approaches are used. In such methods (e.g., flocking \cite{gervasi_prencipe_2004} and rendezvous \cite{flocchini_et_al_2013}), coordination of the robots should occur 
without direct communication among the robots.  
With regard to state-of-the-art approaches based on decentralised control architectures, 
e.g. in methods for exploration where the robots have to visit waypoints, 
the coordination is required to make agents visit the waypoint at the same time.
Here, an initial time cost for pairing the robots is required.
In the end, the suggestion is that more complex robots that can visit the waypoints independently is a better solution, when the cost of individual robots is less than twice the cost of robots that visit waypoints together as a couple.
In addition, when the sensory range is low, issues are faced, i.e. not perceiving waypoints.
Decentralised control strategies are also available for connectivity maintenance of the communication graph (that describes the communication links among robots), using only local information, 
where maintenance of the local connectivity between the robots is not required.
It is shown that the connectivity maintenance control action allows maintaining connectivity even if one robot is forced to stop.
A relevant issue that needs to be investigated is the complexity of the decentralised estimation and computation.
When a multi-robot collaborative manipulation
task is addressed, decentralised coordination and load distribution among the robots is achieved without communication,
making the coordination in accordance to the force capability of each robot (i.e. the maximum force that the robot can execute), and achieving accurate trajectory tracking by the multi-robot system.
For multi-robot collision avoidance purposes, 
a decentralised message selector for scenarios with low communication bandwidth is presented. 
Selective communication is useful for a large number of robots with limited communication capabilities, and can be fruitfully exploited when the information inside the message is required, while avoiding communication when this information is not required.
The limitations are the assumptions that other robots have perfect sensing and the computational complexity of the method.
A hierarchical controller architecture with a FLC decentralised layer and a centralised MPC for multiple SaR robots can improve the performance compared to a purely centralised controller, while being robust to failures.
In addition, the area coverage is similar to coverage-oriented approaches, and the computation time is comparable to non-cooperative approaches \cite{schermerhorn_scheutz_2006, sabattini_chopra_secchi_2013, yan_et_al_2021, zhai_et_al_2021, deKoning_jamshidnejad_2022}.

Lastly, if it is not possible to rely on a central location for coordinating the robots
(e.g., in missions: where the central SaR station does not have sufficient computational power, or is not fast enough;
where the multi-robot system is too complex, or there are several uncertainties involved) 
and communication among/between robots is possible 
(e.g., in SaR missions where a wireless network is available in the environment, or when an exchange of information between the robots is required due to the extent of the dynamic coupling between them), 
then \emph{distributed} approaches are used. 
These coordination methods are used for performing several tasks related to SaR mission planning 
(e.g., coverage-oriented path planning \cite{carr_wang_2022}), dynamic coverage with cooperative exploration \cite{franchi_et_al_2009} and persistent monitoring \cite{boldrer_et_al_2022a}, target detection \cite{robin_lacroix_2016}, navigation, connectivity maintenance or rendezvous \cite{boldrer_et_al_2022b}).
As a result, distributed methods receive more attention in research than centralised or decentralised methods. 
The most commonly distributed approaches for path planning of SaR robots include heuristics techniques, e.g., bug algorithms \cite{husain_et_al_2021}, potential fields methods \cite{cooper_2020}, FLC \cite{din_et_al_2018}, particle swarm optimisation \cite{kumar_et_al_2022}.
All these approaches involve some degree of communication among the robots, at least in a neighbourhood of each agent, and eventually with a central station of the SaR responders.
Among state-of-the-art distributed approaches, optimisation-based methods using genetic algorithms for path planning can be found, 
and they achieve minimisation of the mission completion time and can be tuned to increase area coverage or connectivity as a trade-off. 
A similar hybrid approach, that is both centralised and distributed, achieves an improvement in mission completion time. 
It is centralised for performing SaR search task (i.e. designing coverage paths), inform task (i.e. carrying the target location information to the base station), and monitor task (i.e. connecting the SaR target with the base); 
it is distributed for the replanning task that ensures connectivity, that is the goal.
Distributed MPC can be used for path planning of a cooperative multi-drone system, where the UAVs explore and search a given outdoor dynamic environment, and where the wind flow is modelled.
Here, MPC is used to generate the optimal commanded airspeeds and roll angles (control inputs) that maximise the reward obtained by visiting the cells of the environment, and to minimize the fluctuation of the control inputs.
In multi-robot coordination approaches for clearing a road blocked by obstacles after a disaster, that are robust because in case of failure a robot can be replaced by another one in its task,
it is shown that in scenarios with heavy obstacles it is required to form a coalition (i.e. a group of agents that cooperate to execute a task), and therefore the distributed approach is well suited.
Finally, in a distributed control system for multiple UAVs in a SaR mission where, depending on the phase of the mission, some tasks are implemented in a centralised or distributed way (planning, collision avoidance and video streaming) while others do not require coordination (navigation and detection), the benefit is that the approach can adapt to various scenarios or to different operator demands, and it is proved to be robust in case of failure of a robot, because the joining or leaving of the network by a robot does not affect the mission goal.
\cite{hayat_et_al_2017, hayat_et_al_2020, deAlcantaraAndrade_et_al_2019, nath_arun_niyogi_2019, scherer_et_al_2015}.

\section{Proposed methodologies} \label{sec:methodologies}

In this section, the methodology of MPFC is explained: We first present the problem definition, including 
the assumptions and the mathematical models of the environment and the robots. 
Next, we develop the mathematical formulations of MPFC.

\begin{table*}
    \caption{Table of notation.}
    \label{tab:notation}
    \centering
    \begin{tabular}{l|l}
        \toprule
        Variable            & Description                   \\
        \midrule
        $a_{ij}(\kflc)$     & attraction value for cell $(i, j)$ at $\kflc$    \\
        $\Efeas_r$          & feasible part of the environment for robot $r$    \\
        $k$               & simulation time step counter of the dynamic model of the SaR environment (i.e., global time step) \\
        $\kflc$             & control time step counter for the controller(s) \\
        $\krob$             & simulation time step counter of the dynamic model per robot (i.e., local time step)  \\
        $\mathbb{K}_k$      & set of local simulation time steps that occur between global simulation time steps $k$ and $k+1$ \\
        $\ell^{\textrm{x}}_{(i,j)}$ & length of side $x$ of the cell $(i, j)$   \\
        $\ell^{\textrm{y}}_{(i,j)}$ & length of side $y$ of the cell $(i, j)$   \\
        $\Mdebris$          & debris occupancy matrix at global time step $k$         \\
        $\Mdir$             & wind direction matrix at global time step $k$          \\
        $\Mf$               & fire state matrix at global time step $k$                    \\
        $\Mgrid$            & grid matrix                               \\
        $\Mscan$            & scan certainty matrix at global time step $k$                \\
        $\Mstr$             & structure matrix          \\
        $\Mvel$             & wind velocity matrix at global time step $k$        \\
        $\Mvictim$          & victim probability matrix at global time step $k$ \\
        $\mdebris_{ij}(k)$  & $(i,j)^{\mathrm{th}}$ element of $\Mdebris$ \\
        $\mdir_{ij}(k)$     & $(i,j)^{\mathrm{th}}$ element of $\Mdir$ \\
        $\mf_{ij}(k)$       & $(i,j)^{\mathrm{th}}$ element of $\Mf$ \\
        $\mscan_{ij}(k)$    & $(i,j)^{\mathrm{th}}$ element of $\Mscan$ \\
        $\mstr_{ij}$        & $(i,j)^{\mathrm{th}}$ element of $\Mstr$ \\
        $\mvel_{ij}(k)$     & $(i,j)^{\mathrm{th}}$ element of $\Mvel$ \\
        $\mvictim_{ij}(k)$  & $(i,j)^{\mathrm{th}}$ element of $\Mvictim$ \\
        $\nr$               & number of SaR robots          \\
        $\nv$               & maximum number of victims in each cell      \\
        $\nvp$              & perceived number of victims in cell $(i, j)$ at global time step $k$ \\
        $\nxe$              & 2D grid $x$ dimension (horizontal)          \\
        $\nye$              & 2D grid $y$ dimension (vertical)          \\
        $\sk{k}$            & scan certainty state of cell $(i, j)$ at global time step $k$ \\
        $T$                 & time step for the simulation of the SaR environment dynamics \\
        $\Tflc$             & control sampling time for the controller(s)     \\
        $\Trob$             & simulation sampling time of the dynamic model per robot(s)   \\
        $\va$               & maximum possible air velocity of robot $r$ at time step $\krob$ \\  
        $x_i$               & $x$ coordinate of the cell $(i, j)$ \\
        $y_j$               & $y$ coordinate of the cell $(i, j)$ \\
        $\bm{\theta}_{r} (\kflc)$   & vector of all tuning parameters for the FLC controller corresponding to robot $r$, and optimization variables \\
        & \; of the MPC problem \\
        $\tau_x$            & $x$ coordinate of the target cell \\
        $\tau_y$            & $y$ coordinate of the target cell \\
        \bottomrule
    \end{tabular}
\end{table*}

\subsection{Problem statement} \label{sec:problem_statement}

\begin{figure}
    \centering
    \includegraphics[width = \linewidth]{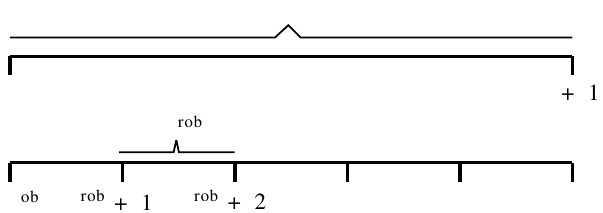}
    \caption{Different time scales used for modelling the SaR environment: 
    All local simulation time steps starting from $\krob$ that fall within the global simulation 
    time steps $k$ and $k+1$ are shown by set $\mathbb{K}_k$.}
    \label{fig:time_scale_model}
\end{figure}

\begin{figure*}
    \centering
    \begin{tabular}{@{}c@{}c@{}}
        \subfloat[]{\includegraphics[width=.3\textwidth]{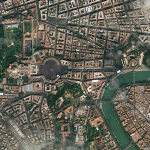}
        \label{fig:satellite_image}}
        \hfil
    \end{tabular}
    \begin{tabular}{@{}c@{}c@{}}
        \subfloat[]{\includegraphics[width=.3\textwidth]{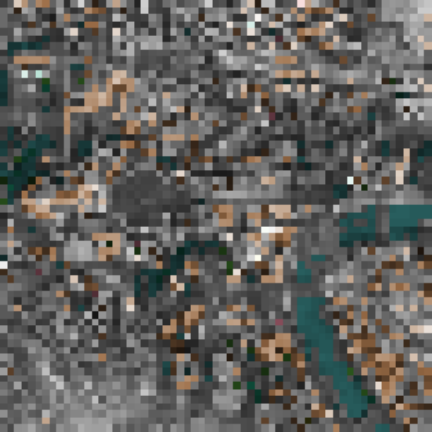}
        \label{fig:pixels_image}}
    \end{tabular}
    \begin{tabular}{@{}c@{}c@{}}
        \subfloat[]{\includegraphics[width = .3\linewidth]{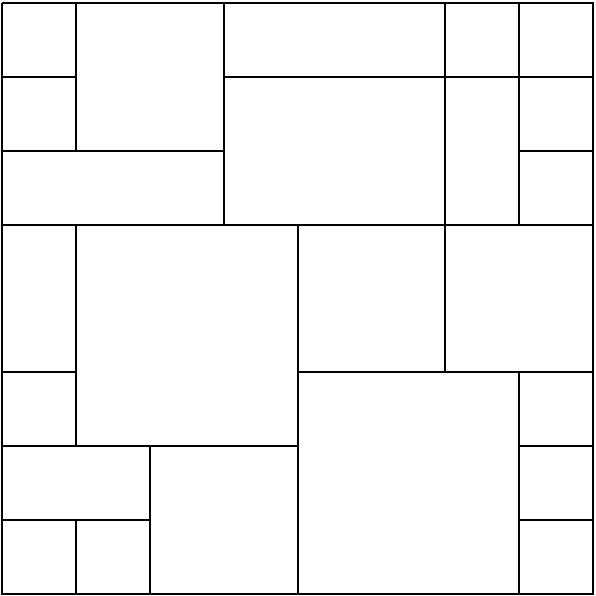}
        \label{fig:grid_cells}}
    \end{tabular}
    \caption{Illustration of a SaR environment given via a 2D satellite image (a) and modelled by 2D grids of cells (b), and 2D representation of an environment that has been discretised via cells of different sizes (c).}
    \label{fig:environment_images}
\end{figure*}

The modelling and control paradigms of this paper are formulated for a discrete space and discrete time framework. 
We consider $\nr$ agents (e.g., flying robots) that explore and map an outdoor SaR environment, which after discretisation is denoted by  $\mathcal{E}$, 
in which a given number of victims, propagating fire and debris exist. 
We use $\nv$ to represent the maximum number of victims per cell of the environment.

The time discretisation is performed at 2 levels in the simulation of the SaR mission: 
(i) a time step $\Trob$ and the corresponding step counter $\krob$ (called the local time step) for the 
simulation of the dynamics of each SaR robot, and  
(ii) a time step $T$ and the corresponding step counter $k$ (called the global time step) for the 
simulation of the dynamics of the SaR environment. 
For the sake of simplicity, we assume that $T$ is a multiple of $\Trob$ (see Figure~\ref{fig:time_scale_model}), and 
that $k=0$ and $\krob = 0$ coincide. Moreover, all local simulation time steps that occur between the global 
simulation time steps $k$ and $k+1$ are given by the set $\mathbb{K}_k$. 

Each flying robot is equipped with a proper sensor that allows it to scan the cells. The robot performs either of the tasks 
``traverse'' or ``scan'' per local time step with respect to the cell(s) that the robot flies over within the given time. 
In case the associated task for local time step $\krob$ is ``scan'', 
a map of the sub-area of $\mathcal{E}$ that is perceived by the robot is output after a given scan time, 
which depends on the robot and the size of the scanned area. 
We assume that each robot has a perfect knowledge of its measured states per local time step.

\subsection{Modelling the SaR environment} \label{sec:environment_model}

The SaR environment is discretised into a 2D grid that is represented by its cells. 
Figure~\ref{fig:environment_images} shows a satellite image of the environment that has been discretised into $\nxe \times \nye$ cells. 
As it is ahown in Figure~\ref{fig:grid_cells}, the cells may have different sizes after discretisation. 
In that case, the size $\nxe \times \nye$ of a matrix called the grid matrix $\Mgrid$ is selected based on the maximum number of the cells in the horizontal and vertical directions, and the values corresponding to those elements of $\Mgrid$ that do not correspond to a cell in the discretised environment will be set to zero.
In order to model the discretised environment, the $\nxe \times \nye$ grid matrix $\Mgrid$ will be used, where each element $(i,j)$ of the matrix stores the coordinates $(x_i, y_j)$ of the centre, the horizontal $\ell^{\textrm{x}}_{(i,j)}$ and vertical $\ell^{\textrm{y}}_{(i,j)}$ lengths of its corresponding cell in the environment.

The SaR robots build up and regularly update a map of the environment, that is represented via matrices of size $\nxe \times \nye$. 
These matrices correspond the information relevant per simulation time step to every cell (where cells are identified via the grid matrix $\Mgrid$). 
The following static and dynamic matrices define the environment map:
\paragraph{Structure matrix} \label{sec:structure_matrix}
The matrix $\Mstr$ is time-invariant and its $(i,j)^{\mathrm{th}}$ element $\mstr_{ij}$ corresponds to a parameter that determines the combustibility of the corresponding cell of the environment, based on the materials in the structure of a cell and their ratios.  
The combustibility parameter adopts a value within the range $[0,1]$, with $1$ corresponding to a combustible structure, $0.6$ corresponding to a fire-preventive structure, and $0$ corresponding to a fire-proof structure.
The combustibility of a cell that is composed of different materials is estimated by multiplication of the percentage of the material that builds the largest proportion of the cell and the combustibility of the corresponding material. 
For example, if a cell is composed of $80\%$ combustible and $20\%$ fire-proof material, the combustibility of the cell is estimated to be $0.8$. 
\paragraph{Scan certainty matrix} \label{sec:scan_certainty_matrix} 
In the matrix $\Mscan$, the $(i,j)^{\mathrm{th}}$ element $\mscan_{ij}(k)$ at 
global simulation time step $k$ represents the scan certainty state $\sk{k} \in [0,1]$ of cell $(i, j)$ for this time step, where for $i = 1, \ldots, \nxe$ and $j = 1, \ldots, \nye$ we have:
\begin{align} \label{eq:agentScanState}
    & \sk{k} = \\
    &\begin{cases} 
        \max\{\sk{k-1}-\sigma, 0\} \\ 
        \qquad \text{if cell $(i,j)$ is not scanned at global simulation} \\
        \qquad \text{time step $k$}\\
        \max\{\sk{k-1} - \sigma , \eta \} \\ 
        \qquad \text{if cell $(i,j)$ is scanned at global simulation} \\
        \qquad \text{time step $k$ by a SaR robot, for which the} \\
        \qquad \text{sensor accuracy is $\eta$}\\
    \end{cases}
    \nonumber
\end{align}
Note that in \eqref{eq:agentScanState}, we assume that due to the dynamics of the environment the scan certainty state reduces 
by a fixed ratio $\sigma$ per global simulation time step, unless the cell is scanned at the current simulation time step. 
Moreover, $\eta$ is the accuracy of the sensor of the SaR robots and belongs to $[0,1]$.
\paragraph{Victim probability matrix} \label{sec:victim _probability_matrix}
The elements of the matrix $\Mvictim$ represent the perceived probability of existence of the $\nvp$ victims per cell of the environment (with $\nvp \in \{1, ..., \nv\}$).
The victim probability matrix $\Mvictim$ is initialised with zeros for the SaR robots.
Per global simulation time step, the elements of matrix $\Mvictim$ will be updated 
for $i  = 1, ..., \nxe$ and $j = 1, ..., \nye$, according to: 
\begin{align} \label{eq:update_Mvictim}
    &\mvictim_{ij}(k) = \\
    &\begin{cases}
    \frac{\nvp \sk{k}}{\nv} \\
    \qquad \text{if cell $(i,j)$ is scanned at global simulation} \\
    \qquad \text{time step $k$ and $\nvp > 0$ victims are} \\ 
    \qquad \text{perceived} \\
    1 - \sk{k}  \\
    \qquad \text{if cell $(i,j)$ is scanned at global simulation} \\
    \qquad \text{time step $k$ and no victim is perceived} \\
    \qquad \text{(i.e., $\nvp = 0$)} \\
    \mvictim_{ij} (k-1) \\
    \qquad \text{if cell $(i,j)$ is not scanned at global }\\
    \qquad \text{simulation  time step $k$} 
    \end{cases}
    \nonumber 
\end{align}
\paragraph{Debris occupancy matrix} \label{sec:debris_occupancy_matrix}
The element  $\mdebris_{ij}(k)$  of the matrix $\Mdebris$ includes the perceived percentage $\op \in [0,1]$ of that cell $(i , j)$ of the environment that is occupied by debris (e.g., rubble, wall, building, etc.). 
While we assume that the position of the debris is fixed during the simulation, 
the elements of the debris occupancy matrix $\Mdebris$ may be updated per 
global  simulation time step for all $i = 1, ..., \nxe$ and $j = 1, ..., \nye$, according to the following equation:
\begin{align} \label{eq:update_Mdebris}
    &\mdebris_{ij}(k) = \\
    &\begin{cases}
    \op \sk{k} \\
    \qquad \text{if cell $(i,j)$ is scanned at global simulation} \\
    \qquad \text{time step $k$ and $\op > 0$ percentage is} \\
    \qquad \text{perceived to be filled with debris} \\
    1 - \sk{k}  \\
    \qquad \text{if cell $(i,j)$ is scanned at global simulation} \\
    \qquad \text{time step $k$ and no debris is perceived} \\
    \qquad \text{(i.e., percentage $\op = 0$)} \\
    \mdebris_{ij}(k-1) \\
    \qquad \text{if cell $(i,j)$ is not scanned at global }\\
    \qquad \text{simulation time step $k$}
    \end{cases}
    \nonumber
\end{align}
\paragraph{Wind velocity matrix} \label{sec:wind_velocity_matrix}
The element $\mvel_{ij}(k)$  of the matrix $\Mvel$ includes the magnitude of the velocity of the wind in cell $(i, j)$ 
for global simulation time step $k$. 
We assume that the magnitude of the velocity is measured/given per global simulation time step. 
\paragraph{Wind direction matrix} \label{sec:wind_direction_matrix}
The element $\mdir_{ij}(k)$  of the matrix $\Mdir$ includes the 
angle (in counterclockwise direction, with reference to the east) of the wind in cell $(i, j)$ for global simulation time step $k$. 
We assume that the direction of the wind is measured/given per global simulation time step. 
\paragraph{Fire state matrix} \label{sec:fire_state_matrix}

\begin{table}
    \centering
    \caption{Possible values for the fire state per cell}
    \label{tab:m_f_states}
    \begin{tabularx}{.5\textwidth}{l|l|X}
        \toprule
        Value   & State             & Description           \\
        \midrule
        0       & Non flammable     & The cell is unburnable    \\
        1       & Flammable         & The cell is not burning yet, but it has the possibility of burning        \\
        2       & Catching fire     & The cell is catching fire, but has no ability to spread the fire    \\
        3       & Burning           & The cell is on fire and has the ability to spread the fire        \\
        4       & Extinguished      & The cell is extinguished and cannot spread the fire (anymore) \\
        \bottomrule
    \end{tabularx}
\end{table}

The element $\mf_{ij}(k)$ of the matrix $\Mf$ includes the state of the fire per cell $(i, j)$ 
for global simulation time step $k$. 
The state of the fire is represented via one of the five values that are given in table \ref{tab:m_f_states}.

\subsection{Modelling the fire dynamics} 
\label{sec:fire_model}

Before SaR starts, the fire state for all the cells is initialised by values $0$ and $1$, 
depending on whether or not a cell is flammable. In addition, a small amount of cells is initialised with state $2$, 
with these cells representing the ignition points where the area starts to burn. 
These state values are regularly updated using the dynamic model that is explained next.

\begin{figure}
    \begin{center}
        \includegraphics[width = 0.9\linewidth]{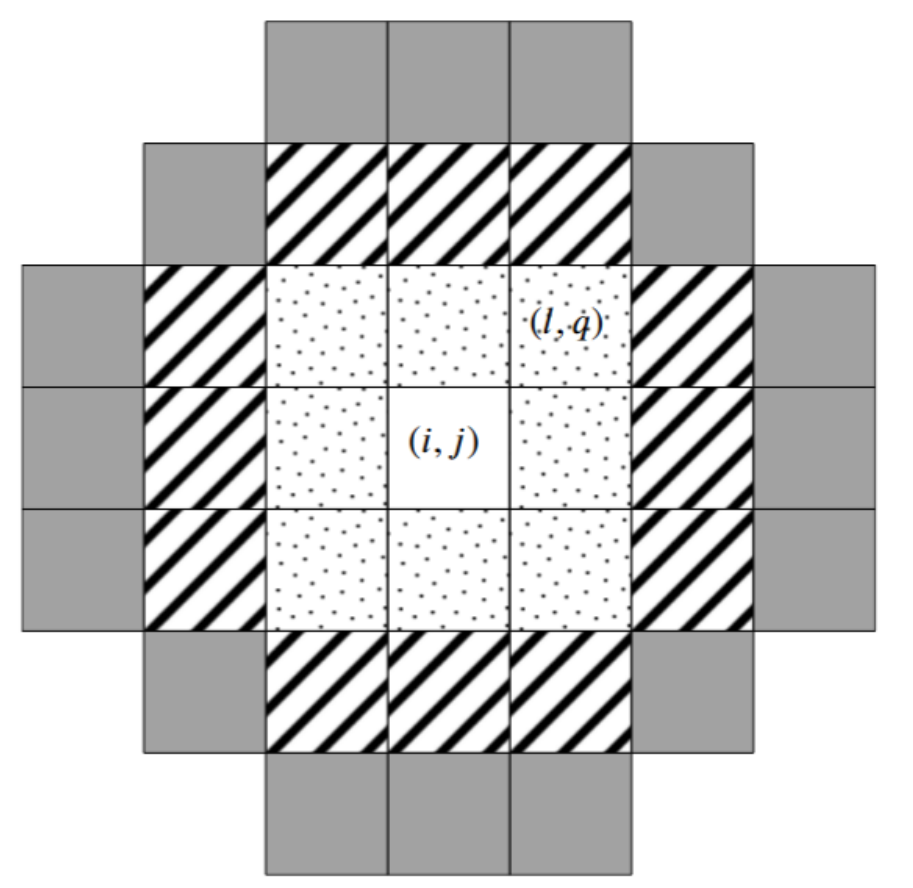}
        \caption{All the $(l, q)$ neighbouring cells of cell $(i, j)$ from which the fire can spread to $(i, j)$, based on the wind speed: dotted cells for $\mvel_{lq}(k) < 1~\frac{\textrm{m}}{\textrm{s}}$; dotted and lined cells for $1~\frac{\textrm{m}}{\textrm{s}} \leq \mvel_{lq}(k) \leq 5~\frac{\textrm{m}}{\textrm{s}} $; dotted, lined and solid cells for $\mvel_{lq}(k) > 5~\frac{\textrm{m}}{\textrm{s}}$.}
        \label{fig:ohgai_cells}
    \end{center}
\end{figure}

The dynamics of the fire state is modelled using a discrete cellular automata approach based on Ohgai et.\ al.\ \cite{ohgai_gohnai_watanabe_2007} and Freire and DaCamara \cite{freire2019using}. 
The probability that at global simulation time step $k$ the fire spreads to cell $(i, j)$ from cell a neighbouring cell $(l, q)$, 
where, based on the wind speed:

\begin{align} 
    (l, q) \in
    \begin{cases} 
        \text{dotted cells in Figure~\ref{fig:ohgai_cells},} \\
        \qquad \text{for $\mvel_{lq}(k) < 1~\frac{\textrm{m}}{\textrm{s}}$} \\
        \text{dotted and lined cells in Figure~\ref{fig:ohgai_cells},} \\
        \qquad \text{for $1~\frac{\textrm{m}}{\textrm{s}} \leq \mvel_{lq}(k) \leq 5~\frac{\textrm{m}}{\textrm{s}} $} \\
        \text{dotted, lined and solid cells in Figure~\ref{fig:ohgai_cells},} \\
        \qquad \text{for $\mvel_{lq}(k) > 5~\frac{\textrm{m}}{\textrm{s}}$} \\
    \end{cases}
\end{align}
is given by: 
\begin{align} \label{eq:fireSpreadProbability}
    \pf(k) = & 
    \ \alpha_1 \mstr_{ij} \op \fij \left(k \right) \cdot  \\
    & \exp\Big(\alpha_2 \delta \big (l, q) , (i,j) \big) + \alpha_3 \vij(k) + \nonumber \\
    & \alpha_4 \vij(k) \cos\psiij(k) \Big) \nonumber
\end{align}
where $\alpha_1$, $\alpha_2$, $\alpha_3$, $\alpha_4$ are tuning parameters, 
$\delta: \mathcal{E}^2 \rightarrow \mathbb{R}$ is a function (e.g., a norm) that represents the distance of the input cells, 
$\vij(k)$ and $\psiij(k)$ are, respectively, the average magnitude and the angle between the average direction 
of the wind velocity and the direction of the fire propagation for global simulation time step $k$, given as the mean of $\mvel_{lq}(k)$ and $\mvel_{ij}(k)$, and mean of 
$\mdir_{lq}(k)$ and $\mdir_{ij}(k)$, respectively, and 
$\fij\left(k \right)$ the ability of cell $(l,q)$ to cause the fire to spread to its neighbouring cells, 
including  cell $(i, j)$, for this time step.
This function is computed via:

\begin{equation} \label{eq:p_t_bt}
\begin{split}
    &\fij \left(k \right) = \\
    &\begin{cases} 
        \frac{\dps 4 (k - \kfij) + 0.2 \ktwo - 4.2 \kone}{\dps \ktwo - \kone} 
        \\
        \qquad \kone + \kfij \leq k \leq 0.2 \ktwo + 0.8 \kone + \kfij\\
        1.25\frac{\dps \ktwo - k+ \kfij}{\dps \ktwo - \kone }  
        \\ 
        \qquad 0.2 \ktwo + 0.8 \kone + \kfij \leq k\leq \ktwo + \kfij
    \end{cases}
\end{split}
\end{equation}
where $\kone$ and $\ktwo$ are the global simulation time steps corresponding to, respectively, 
$2$ minutes (i.e., when the cell becomes able to spread the fire) and $10$ minutes 
(i.e., when the cell is already burnt out) after ignition of fire in cell $(l,q)$, and $\kfij$ 
the global simulation time step when cell $(l, q)$ catches fire.%

Finally, the fire state of cell $(i, j)$ is updated via the following equation (due to any neighbouring cell $(l, q)$ as in Figure~\ref{fig:ohgai_cells}):
\begin{align}
    \label{eq:fire_state_update}
    &\mf_{ij}(k) = \\
    &\left\{
    \begin{array}{ll} 
        2 
        & \text{If $\mf_{ij}(k-1)=1$ \& $\mf_{lq}(k)=3$}\\
        & 
        \text{ \& $\pf\left(k \right)>\zeta$} \\
        3 
        & \text{If $\mf_{ij}(k-1)=2$ \& $k \geq \kfipjp + k^{2\textrm{min}} $} \\
        4 
        & \text{If $\mf_{ij}(k-1)=3$ \& $k \geq \kfipjp + k^{10\textrm{min}} $} \\
        \mf_{ij}(k-1) & \text{otherwise}
    \end{array} \nonumber
    \right.
\end{align}
where $\zeta$ is a given threshold for the cells to catch fire and $k^{2\textrm{min}}$ and $k^{10\textrm{min}}$ 
are the integer values that show the (ceiling) of the number of global simulation time steps within a period of, 
respectively, $2$ minutes and $10$ minutes.

\subsection{Modelling the dynamics of the SaR robots} \label{sec:agents_model}

\begin{figure}
    \centering
    \includegraphics[width = \linewidth]{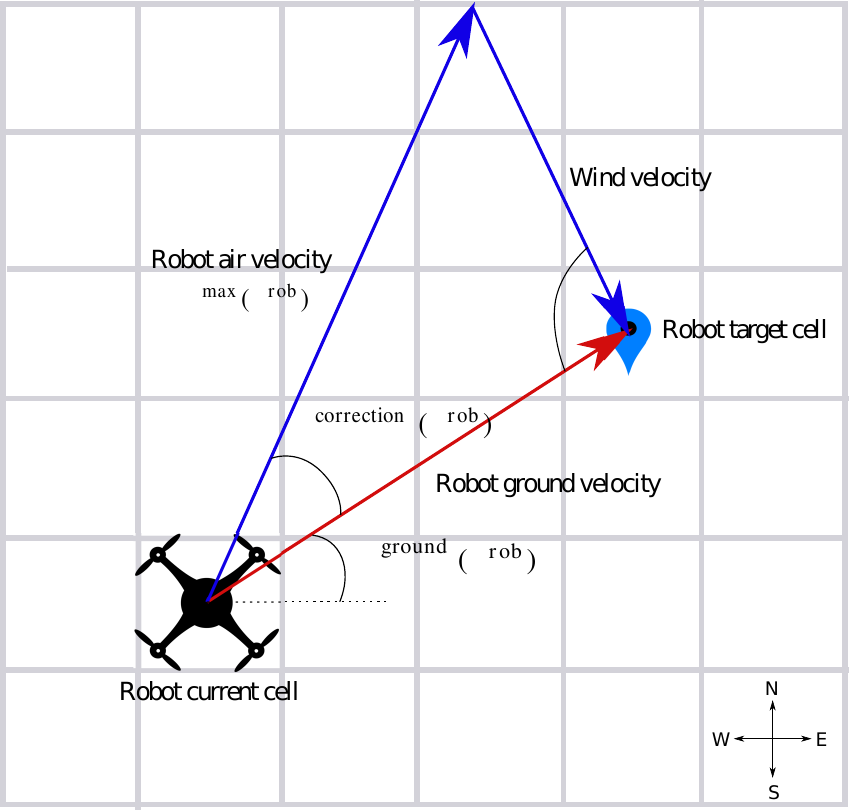}
    \caption{Corrected heading angle for achieving an air velocity for the flying robot that, despite the wind flow, allows the robot to fly in the desired direction (aligned with the ground velocity).}
    \label{fig:movement_in_wind}
\end{figure}

Each SaR robot $r$ (with $r = 1, \ldots, \nr$) is assigned a target cell $\target$, decided by the FLC controller, per local simulation time step $\krob$. 
The robot should reach this cell as quickly as possible and scan the cell. 
Thus, the robot moves from its current position to the centre of the target cell along a straight line, with its maximum possible velocity. An action $a_{r,\krob}$ is executed by robot $r$ per local simulation time step. The robot may need more than one local simulation sampling time $\Trob$ to reach the target cell. For the cells across the trajectory of the robot that are not the target of the robot, this action is ``traverse'' and otherwise is ``scan''.

The set of all cells that robot $r$ can reach with respect to its current position $(i,j)_{r,\krob}$ (for $r = 1, \ldots, \nr$) is given by: 
\begin{align}
    &\Efeas_r(\krob) = \\
    &\left\{(l, q) | (l, q)\in \mathcal{E}, \delta\left( (l, q) , (i,j)_{r,\krob} \right) \leq v^{\max}_r(\krob) \Tflc_r\right\} \nonumber
\end{align}
with $v^{\max}_r(\krob)$ the maximum possible velocity for the robot for local simulation time step $\krob$ (for instance, the speed of the wind may affect this velocity), 
$\Tflc_r$ the control sampling time of the robot, which implies how frequently the target of the robot is updated and  
includes a multiple of the local simulation sampling time $\Trob$ (the control time discretisation is further explained in Section~\ref{sec:controller_design}). 
For the action ``scan'' the robot spends the following time within the target cell $\target$ to scan the cell:
\begin{equation}
    t^{\textrm{scan}}_r(\krob) = \bar{t}_r \ell^{\textrm{x}}_{\target} \ell^{\textrm{y}}_{\target}
\end{equation} 
where $\bar{t}_r$ is a constant value for robot $r$ that gives the scan time of the robot per square metre. 
Each robot $r$ scans the target cell using a sensor with accuracy $\eta_r \in [0, 1]$, which is used to update the matrix $M^{\textrm{scan}}_r(k)$ (that is updated separately per robot $r$, i.e., each robot has a different matrix that tracks its own scan certainty).

A 2D motion dynamics, based on \cite{daidzic2016general}, is considered for each flying robot, where  
the position of the robot per local simulation time step is determined based on the coordinates of the cell over which the robot flies. 
We assume that after assigning a new target for robot $r$, it moves with its maximum possible air velocity $\va$,  
which is determined based on the maximum ground (i.e., nominal) velocity $\vg$ of the flying robot and the velocity of the wind for a given time step (see Figure~\ref{fig:movement_in_wind}). 
For $\krob \in \mathbb{K}_k$ (see Figure~\ref{fig:time_scale_model}), suppose that robot $r$ is in cell $(i, j)_{r,\krob}$ and should reach and scan the target cell $(\tau_x, \tau_y)_{r,\krob}$.  
These cells belong to the sub-area $\Efeas_r(\krob )$, where the average wind velocity and direction are, respectively, 
$\bar{m}^{\textrm{velocity}}(k)$ and $\bar{m}^{\textrm{direction}}(k)$ given by:
\begin{align*}
    &\bar{m}^{\textrm{velocity}}(k) = \textrm{mean} \{ \mvel_{ij}(k) \}
    \\
    &\bar{m}^{\textrm{direction}}(k) = \textrm{mean} \{ \mdir_{ij}(k) \}
\end{align*}
Then the actual heading of the robot (measured with reference to the east, in counterclockwise direction) within the local time frame is given by:
\begin{equation}
    \theta^{\textrm{heading}}_{r}(\krob) = \thetaG + \thetaCorr
\end{equation}
where we have:
\begin{align}
    \label{eq:ground_angle}
    &\thetaG = \arctan \left( \frac{\dps \tau_y - j}{\dps \tau_x - i} \right)_{r,\krob} 
    \\
    \label{eq:correction_angle}
    &\thetaCorr = \\ 
    &\arcsin \left( \frac{\dps \bar{m}^{\textrm{velocity}}(k)}{\dps \va}  \sin\left( \thetaG - \bar{m}^{\textrm{direction}}(k) \right) \right)  \nonumber
\end{align}

Moreover, the time that the robot needs to reach the target cell is computed via:
\begin{equation}
\label{eq:travel_time}
    t^{\textrm{travel}}_r(\krob) = \frac{\dps \sqrt{ \delta\left( (i,j)_{r,\krob}  , (\tau_x,\tau_y)_{r,\krob} \right) }}
    {\dps v^{\textrm{ground}}_r(\krob)}
\end{equation}
with:

\begin{align}
    & \left( v^{\textrm{ground}}_r(\krob) \right)^2 = \left( \va \right)^2 + \left( \bar{m}^{\textrm{velocity}}(k) \right)^2 + \\ 
    & \ 2 \va \bar{m}^{\textrm{velocity}}(k) \cos\left( \bar{m}^{\textrm{direction}}(k) - \theta^{\textrm{heading}}_{r}(\krob) \right) \nonumber
\end{align}

\subsection{Model predictive fuzzy control (MPFC) for multi-agent systems} 
\label{sec:controller_design}

The proposed control architecture, MPFC, has two control layers (see Figure~\ref{fig:MPFC}): At the lower layer, FLC systems are used to   
locally control each SaR robot via a decentralised way, while at the higher layer, MPC is used in order to 
tune certain parameters of the local FLC controllers, whenever certain performance criteria is triggered,  
based on a globally optimal point-of-view within a finite prediction horizon.
Next, we explain the details of the MPFC architecture and its two different layers.%

\subsubsection{Control architecture} \label{sec:control_architecture}

\begin{figure}
    \centering
    \includegraphics[width = 0.9\linewidth]{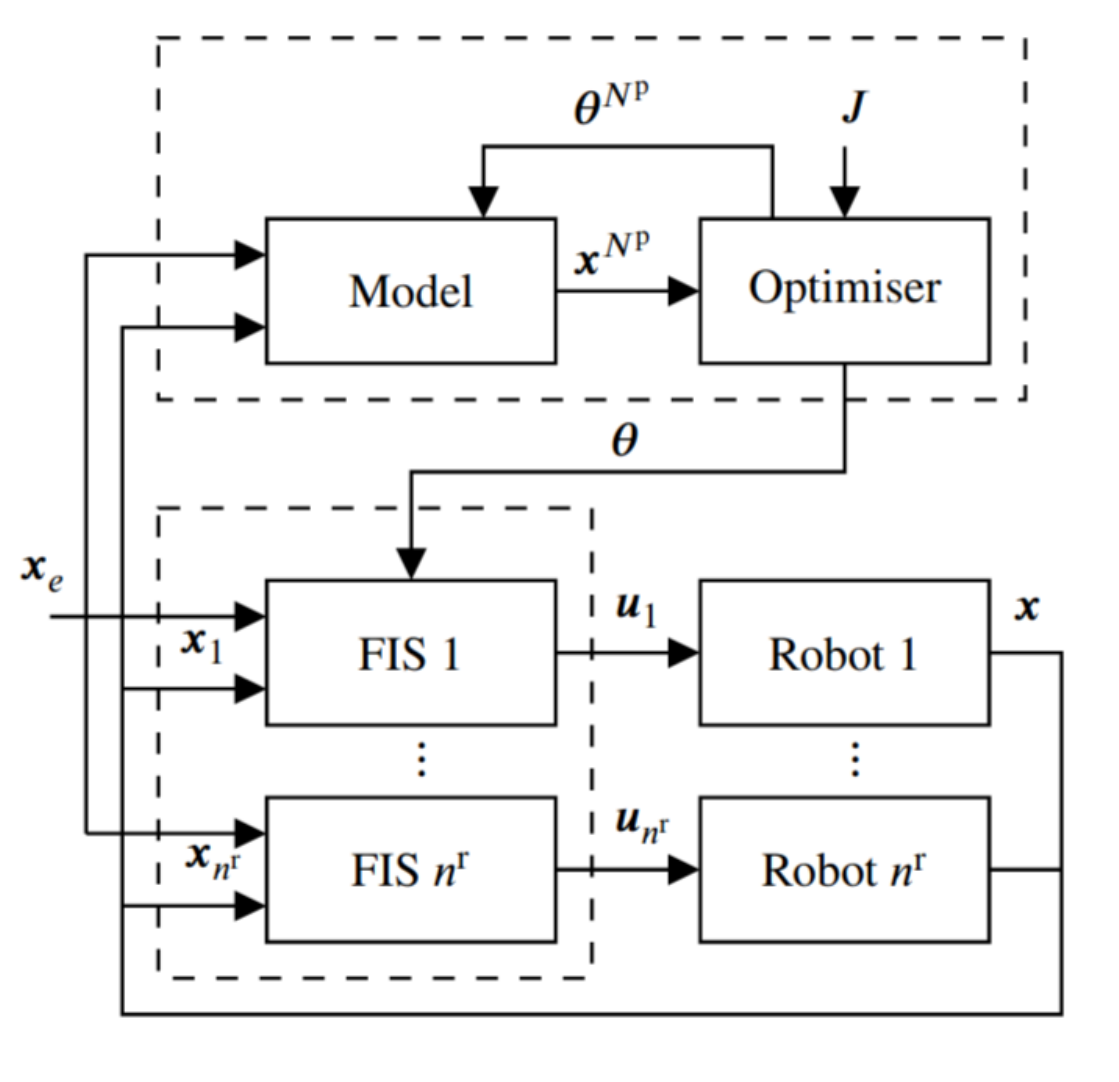}
    \caption{MPFC architecture at time step $\kflc$, with two layers specified by dashed rectangles: 
    The fuzzy inference systems control the multi-agent  system  directly, in a decentralised way. 
    The model predictive control layers tunes, with a frequency that is generally lower than the control frequency, the parameters of the local fuzzy inference systems in order to improve the global performance of the multi-agent system.}
    \label{fig:MPFC}
\end{figure}

\begin{figure}
    \begin{center}
        \includegraphics[width = \linewidth]{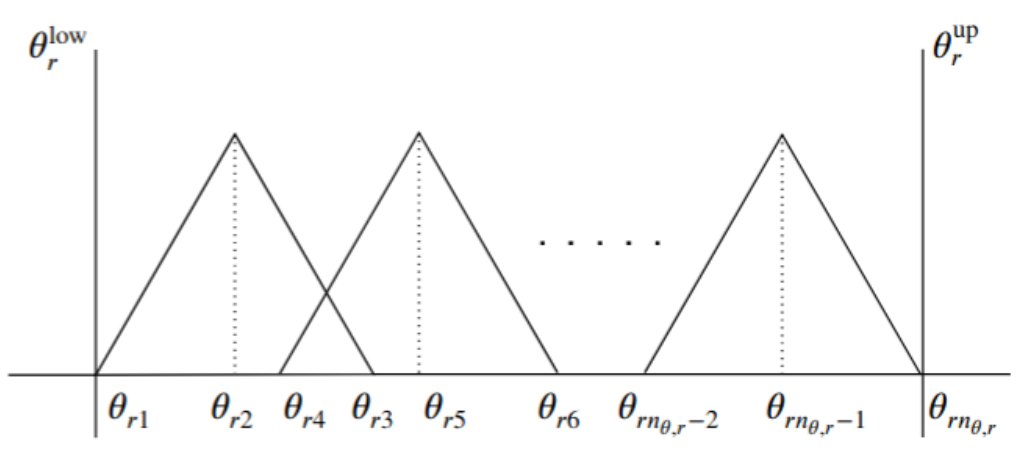}
        \caption{The candidate parameters of the FLC output surfaces, i.e. the optimisation variables of the MPC problem.}
        \label{fig:mpc_opt_variables}
    \end{center}
\end{figure}

MPFC (see Figure~\ref{fig:MPFC}) operates based on a discrete time framework, with a sampling time $\Tflc$ 
and the corresponding control time step counter $\kflc$. 
For the sake of simplicity for the following equations, we assume that $\Tflc$ is a multiple of $\Trob$. 
The computations for generating a path per robot are performed per control time step 
via the FLC controllers, whereas tuning of the parameters of the FLC controllers (see Figure~\ref{fig:mpc_opt_variables})
is performed via the MPC layer in a regular or 
triggered way, thus not necessarily at every control time step.

\subsubsection{Decentralised FLC layer} \label{sec:flc_layer}

Each robot has its local FLC-based path planning controller, which performs in a decentralised way 
with respect to the FLC controllers of other robots, i.e., there is no interaction or exchange of information among the FLC controllers.

The rule-based policies of the FLC controllers are formulated such that each controller includes a number of tuning parameters. 
More specifically, these rules for the FLC controller of robot $r$, when the input corresponding to a candidate target cell $(\tau_x,\tau_y)\in\Efeas_r(\kflc)$ is $I_r((\tau_x,\tau_y),\kflc)$, 
are given by:
\begin{itemize}
    \item 
    Rule 1: If $I_r((\tau_x,\tau_y), \kflc)$ is $A^{\textrm{fuzzy}}_1$, 
    then output is $f_1 \left( I_r( (\tau_x,\tau_y) , \kflc) , \bm{\theta}_{r} (\kflc) \right)$. 
    \item [] \vdots 
    \item 
    Rule $n^{\textrm{rule}}$: If $I_r( (\tau_x,\tau_y) , \kflc)$ is $A^{\textrm{fuzzy}}_{n^{\textrm{rule}}}$, 
    then output is $f_{n^{\textrm{rule}}} \left( I_r( (\tau_x,\tau_y) , \kflc) , \bm{\theta}_{r} (\kflc) \right)$. 
\end{itemize}
with ${\bm{\theta}_{r}} (\kflc) , \ldots , \bm{\theta}_{r}(\kflc)$ 
the most updated values of the tuning parameters for the FLC controller. 
Then the vector of all tuning parameters for the FLC controller corresponding to robot $r$ is given by 
$ \bm{\theta}_r (\kflc) = [ {\bm{\theta}^{\top}_{r}} (\kflc) , \ldots , \bm{\theta}^{\top}_{r}(\kflc) ]^{\top}$. 

A target cell per robot $r$ is determined every control time step $\kflc$ via its FLC controller, based on the attraction of the cells within $\Efeas_r(\kflc)$.
The attraction value $a_{\tau_x\tau_y}(\kflc)$ for cell $(\tau_x,\tau_y) \in \Efeas_r(\kflc)$ is determined via:
\begin{equation} \label{eq:calc_M_att}
    a_{\tau_x\tau_y}(\kflc) = 
        \textrm{FLC}_r \left(I_r( (\tau_x,\tau_y) , \kflc), \bm{\theta}_r(\kflc)\right) 
\end{equation}
A type-$1$ fuzzy inference system is used to calculate the attraction of the cells.
In other words, the inputs are mapped into a linear output surface, where the 
inputs $I_r( (\tau_x,\tau_y) , \kflc)$ to the FLC controller of robot $r$ for control time step $\kflc$ include:
\begin{enumerate}
    \item Cell time efficiency:  
    Time needed for robot $r$ at control time step $\kflc$ to reach cell $(\tau_x,\tau_y)$, determined via \eqref{eq:travel_time}
    \item Cell priority: 
    Probability of presence of a victim in cell $(\tau_x,\tau_y)$, given by $\mvictim_{ij}(k)$ (see Section~\ref{sec:victim _probability_matrix})
    \item Cell fire risk time: 
    Time left  until cell $(\tau_x,\tau_y)$ catches fire, determined via Algorithm~\ref{alg:fire_risk_time}
    \item Cell scan certainty: 
    For cell $ (\tau_x,\tau_y)$ this certainty is determined via \eqref{eq:agentScanState} 
\end{enumerate}
All these input values are normalised within the range $[0,1]$ before being injected into the FLC controller.

\subsubsection{Centralised MPC layer} \label{sec:mpc_layer}

MPC is composed of two main elements: a prediction model and an optimiser (see Figure~\ref{fig:MPFC}).
The inputs to the prediction model of the MPC layer in MPFC at control time step $\kflc$ include 
the measured states  of all the robots at this control time step, 
i.e., $[\bm{x}_1^{\top}(\kflc) , \ldots , \bm{x}_{\nr}^{\top}(\kflc)]^{\top}$, 
the environment state vector at control time step $\kflc$ (see Section~\ref{sec:environment_model} for details), 
and the vector of the candidate parameters of the FLC output surfaces (determined via the optimiser) 
based on the  most recently updated values for these parameters (i.e., $[\bm{\theta}^{\top}_1(\kflc), \ldots, 
\bm{\theta}^{\top}_{\nr}(\kflc)]^{\top}$).
The output of the prediction model includes the predicted value for the future states of the environment over the prediction horizon.
We assume that MPC has a perfect model of the environment and the robots, and 
has complete information about the structure of the path planning FLC controllers. 
If these are not the case, then different variants of MPC, e.g., robust MPC \cite{bemporad_morari_1999, langson_chryssochoos_rakovic_mayne_2004}, may be used. 

The predicted values of the states over the prediction horizon are input to the optimiser, which also receives 
a cost function and the relevant constraints of the optimisation problem.
The optimisation variables of the MPC problem are the parameters $[\bm{\theta}^{\top}_1(\kflc), \ldots, 
\bm{\theta}^{\top}_{\nr}(\kflc)]^{\top}$ of the FLC output surface for all robots $r = 1, \ldots, \nr$ (see Figure~\ref{fig:MPFC}).

The constrained optimisation problem that is solved by the MPC layer of MPFC at $\kflc$, 
when tuning has been triggered, is given by: 
\begin{align} 
    \label{eq:MPC}
    &\ \max_{[\bm{\theta}^{\top}_1(\kflc), \ldots, 
\bm{\theta}^{\top}_{\nr}(\kflc)]^{\top}}\sum_{\kappa=\kflc}^{\kflc+\Np} 
\sum_{r = 1} ^ {\nr} a_{(\tau_x,\tau_y)_{r,\kappa}}(\kappa)  \\
    & \textrm{such that for } \kappa = \kflc, \ldots, \kflc + \Np \nonumber \\
        &\begin{cases}
            (\tau_x,\tau_y)_{1,\kappa} \in \Efeas_1 (\kappa) \\
            a_{(\tau_x,\tau_y)_{1,\kappa}}(\kappa) = \textrm{FLC}_1 (I_1( (\tau_x,\tau_y)_{1,\kappa} , \kappa), \bm{\theta}_1(\kflc)) \\
            \theta_{11}, \ldots, \theta_{1 n_{\theta,1}} \in [\theta^{\textrm{low}}_1, \theta^{\textrm{up}}_1] \\
            \theta_{1, i-1} < \theta_{1, i} < \theta_{1, i+1}, i \in \{5, 8, 11, \ldots, n_{\theta,1} - 4\} \\
        \end{cases} \nonumber \\
        & \phantom{=====} \vdots \nonumber \\
        &\begin{cases}
            (\tau_x,\tau_y)_{\nr,\kappa} \in \Efeas_{\nr} (\kappa) \\
            a_{(\tau_x,\tau_y)_{\nr,\kappa}}(\kappa) = \textrm{FLC}_{\nr} (I_{n_r}( (\tau_x,\tau_y)_{\nr,\kappa} , \kappa), \bm{\theta}_{\nr}(\kflc )) \\
            \theta_{\nr 1}, \ldots, \theta_{\nr n_{\theta,\nr}} \in [\theta^{\textrm{low}}_{\nr}, \theta^{\textrm{up}}_{\nr}] \\
            \theta_{\nr, i-1} < \theta_{\nr, i} < \theta_{\nr, i+1}, i \in \{5, 8, 11, \ldots, n_{\theta,\nr} - 4\} \\
        \end{cases} \nonumber
\end{align}
where $\Np$ is the prediction horizon, $a_{(\tau_x,\tau_y)_{r,\kappa}}(\kappa)$ is the attraction value 
of the candidate target cell $(\tau_x,\tau_y)_{r,\kappa}$ for robot $r$ at time step $\kappa$, 
and  $n_{\theta,r}$, $\theta^{\textrm{low}}_r$, and $\theta^{\textrm{up}}_r$  
are, respectively, the number of the tuning parameters within the FLC controller of robot $r$ and the lower and upper bounds for these parameters with $r=1,\ldots,\nr$.
In fact, \eqref{eq:MPC} determines the tuning parameters of the decentralised FLC controllers, 
such that the global cumulative value of the 
attraction of the scanned cells for a given prediction horizon is maximised. 

\section{Case study} \label{sec:case_study}

\subsection{Assumptions}

In the case study, the following assumptions are made:

\setlist[enumerate,1]{leftmargin=1.3cm}
\begin{enumerate} [label=\textbf{A\arabic*}]
    \item The 2D grid of the environment that we consider in the case study is all composed by identical cells.
    The dimension of the cell is such that the radius of the sensor of each robot can precisely scan one cell at the same time (the cell where the robot is located), during the \textrm{``scan''} action.
    \item The wind velocity and the wind direction are constant through all the environment, and therefore so are the elements $(i, j)$ of the matrices $\Mvel$ and $\Mdir$.
    The wind velocity $\mvel_{ij}(k)$ is less than $1 \frac{\textrm{m}}{\textrm{s}}$ for any cell $(i, j)$ in the environment and for any time step $k$, therefore the neighbourhood to which the fire can spread from a cell with an active fire only consists of dotted cells in Figure~\ref{fig:ohgai_cells}.
    \item The matrices of the robot model are coarsened, i.e., the environment map is divided into cells of larger dimensions than the cells of the environment matrices, for computational efficiency. 
    For this reason, each robot has two additional parameters: \textit{robot location}, i.e., the location of the robot within a coarsened cell, and  \textit{robot target}, i.e., the target coarsened cell.
    \item All the simulation time steps are synchronised and identical.
    \item A type-$1$ TSK fuzzy inference system \cite{lilly_2011_chap6} is used in the FLC controllers.
    \item The MPC is called regularly after a certain amount of time has passed, to tune the parameters $\bm{\theta}_r$ of the FLCs in order to improve their performance.
\end{enumerate}

\subsection{Comparison approaches}
\label{sec:comparison_approaches}

\paragraph{MPC comparison method}

The MPC approach is like the MPFC, but each robot has a parameter defining a queue of target cells that the robot can schedule in advance, to travel to and scan them.

\paragraph{Centralised and decentralised architectures}

For the comparison in the case study, a centralised architecture (for both MPFC and MPC) and a decentralised architecture (for both MPFC and MPC) are defined.

\paragraph{Prediction modes of the supervisory controllers}

For the supervisory MPC controllers, two prediction modes (``probability threshold'' and ``exact'') are used to estimate the future states of the MPCs, and they are used in the comparisons of the case study.

\subsection{Estimation of the environment matrices}

\paragraph{Estimation of the victim probability matrix}

In the case study, the elements of the victim probability matrix are estimated as follows:
\begin{align}
    &\hat{m}^{\text{victim}}_{ij}(k) = \\
    &\begin{cases}
    \mvictim_{ij}(k) & \text{if cell $(i, j)$ is scanned} \\
    c^{\text{population}} \ell^{\textrm{x}}_{(i,j)} \ell^{\textrm{y}}_{(i,j)} \op & \text{if cell $(i, j)$ is not scanned}
    \end{cases}
    \nonumber
\end{align}
where the matrices $\hat{M}^{\text{victim}}(k)$ and $\Mvictim$ (that respectively contain the elements $\hat{m}^{\text{victim}}_{ij}(k)$ and $\mvictim_{ij}(k)$) are coarsened, and $c^{\text{population}}$ is the average population density of the disaster environment.

\paragraph{Estimation of the fire matrix}

The elements of the fire matrix are estimated as follows. Firstly, the estimation of the elements of the downwind map is given by:
\begin{equation}
    \hat{m}^{\text{downwind}}_{ij}(k) = f^{\text{downwind}}(\hat{m}^{\text{fire}}_{ij}(k))
\end{equation}
where $\hat{m}^{\text{fire}}_{ij}(k)$ are the elements of the estimated fire state matrix, $\hat{M}^{\text{fire}}_{ij}(k)$, that is coarsened, and $f^{\text{downwind}}$ is a function that executes Algorithm~\ref{alg:downwind_map} (see Appendix \ref{app:algorithms}).
Lastly, the elements of the estimated fire matrix are given by:
\begin{equation}
    \hat{m}^{\text{fire}}(k) = f^{\text{fire}}(m^{\text{fire}}(k-1))
\end{equation}
where $f^{\text{fire}}$ is a function that executes \eqref{eq:fire_state_update}.

\paragraph{Computation of the cell time efficiency}

In the case study the cell time efficiency (input 1 of the FLC) is computed via the following equation:
\begin{align}
    M^{\textrm{response}}(k) = \frac{f^{\textrm{response}}(k)}{t^{\textrm{response, max}}}
\end{align}
\begin{align}
    f^{\textrm{response}}&(k) = \\
    &\begin{cases}
    t^{\textrm{travel}}_r(k) + t^{\textrm{scan}}_r(k) + t^{\textrm{travel}} \big( (\tau_x,\tau_y)_r, (i,j)_r \big) \\
    \qquad \text{if the task is ``travel''} \\
    t^{\textrm{scan}}_r(k) + t^{\textrm{travel}} \big( (\tau_x,\tau_y)_r, (i,j)_r \big) \\
    \qquad \text{if the task is ``scan''} \\
    \end{cases}
    \nonumber 
\end{align}
where $t^{\textrm{travel}}_r(k)$ is the remaining travel time of robot $r$ at time step $k$, $t^{\textrm{travel}} \big( (l, q), (i, j) \big)$ is a function which returns the travel time for a robot between cells $(l, q)$ and $(i, j)$, while the maximum response time $t^{\textrm{response, max}}$ is defined as the longest time a robot would require to travel between any two cells in the environment and perform a scan.

\subsection{FLC membership functions}

\begin{table}
\centering
\caption{MF parameters for FLC inputs.}
\begin{tabular}{@{}clc@{}}
\toprule
\textbf{Name} & \textbf{Type} & \textbf{Parameters} \\
\midrule
low    & Triangular & [0, 0, 0.5] \\
medium & Triangular & [0, 0.5, 1] \\
high   & Triangular & [0.5, 1, 1] \\
\bottomrule
\end{tabular}
\label{tab:flc_input_mf_parameters}
\end{table}

\begin{figure*}
\centering
\includegraphics[width = 0.7\linewidth]{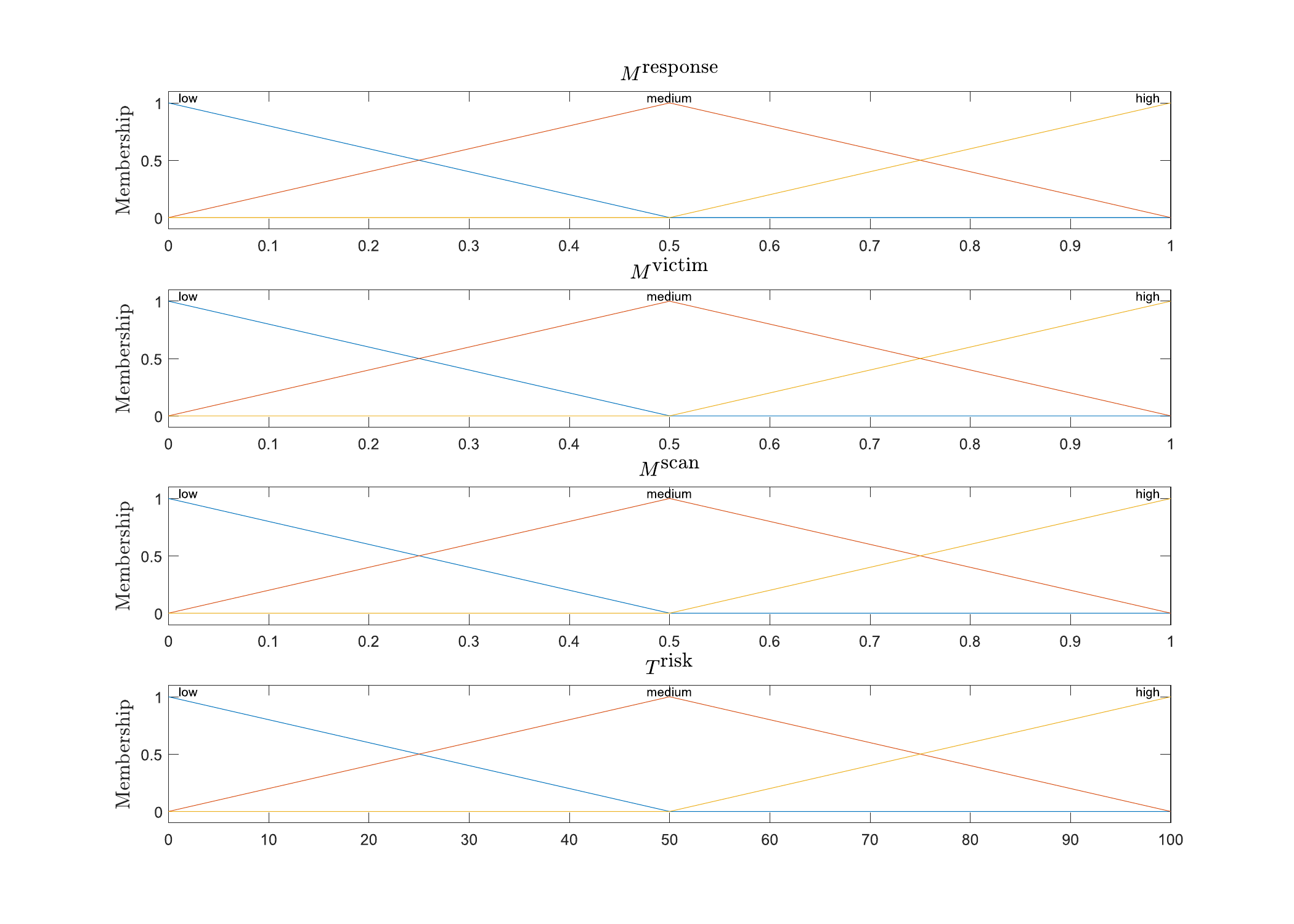}
\caption{FIS input MFs.}
\label{fig:fis_input_mf}
\end{figure*}

\begin{table}
\caption{Default output MF parameters.}
\centering
\begin{tabular}{@{}ccc@{}}
\toprule
\textbf{Name} & \textbf{Type} & \textbf{Parameters (coefficients)} \\
\midrule
low    & linear & \( -1, 1, -1, -1, 0 \) \\
medium & linear & \( -1, 1, -1, -1, 0.5 \) \\
high   & linear & \( -1, 1, -1, -1, 1 \) \\
\bottomrule
\end{tabular}
\label{tab:output_mf}
\end{table}

\begin{figure*}
\centering
\includegraphics[width=0.7\linewidth]{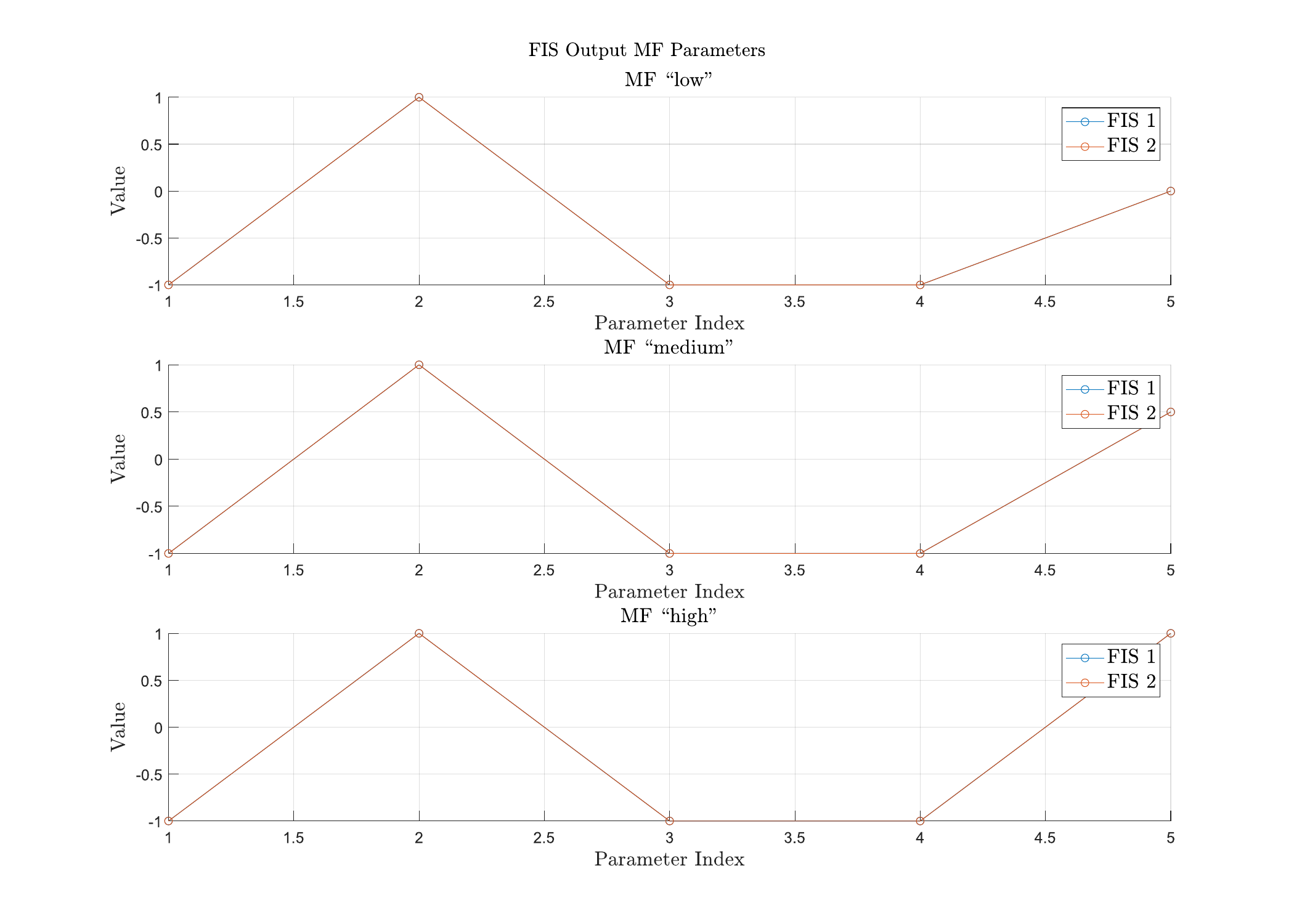}
\caption{Parallel coordinate plot of default FIS Output MFs.}
\label{fig:fis_output_mf_parallel_coordinate}
\end{figure*}

\begin{figure*}
\centering
\includegraphics[width=0.7\linewidth]{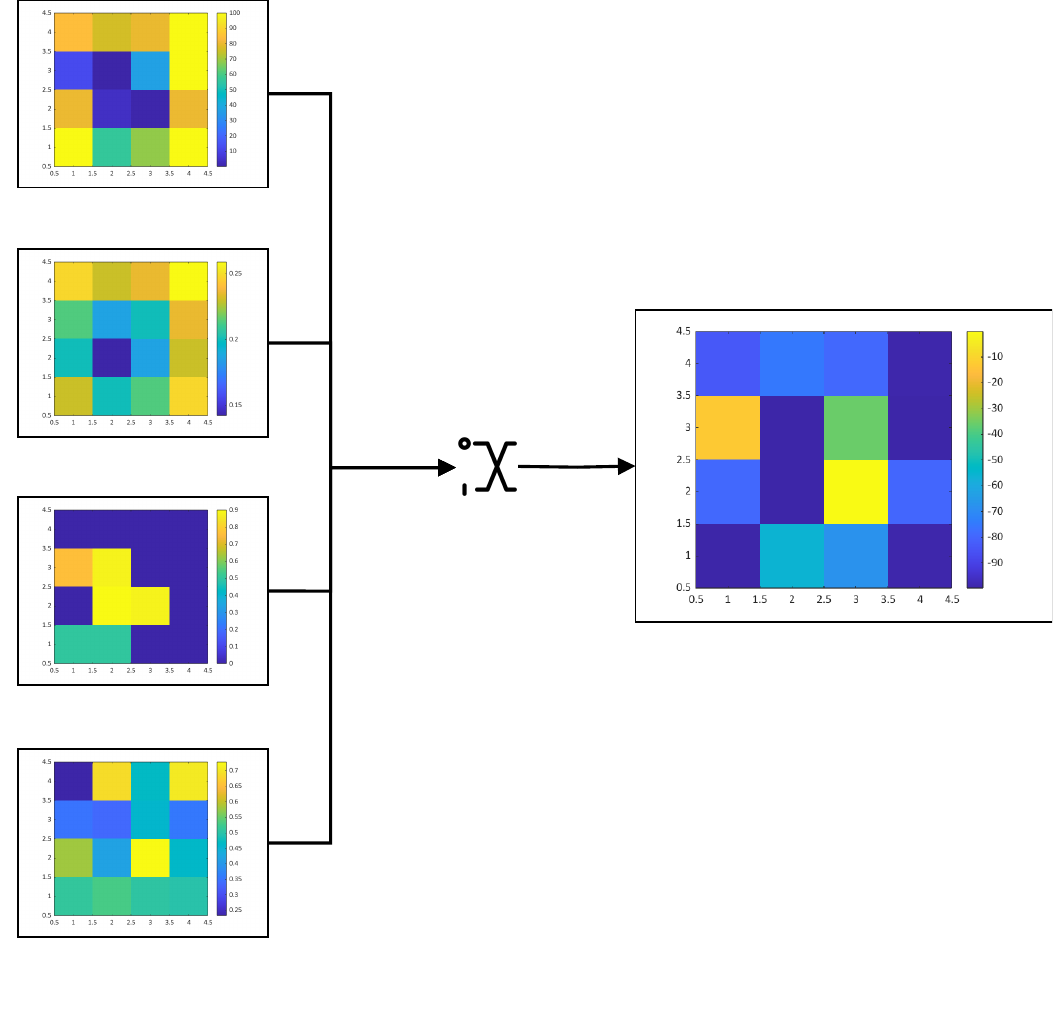}
\caption{Inputs and outputs of FLC.}
\label{fig:fis_inputs_outputs}
\end{figure*}

In the FLC, all input MFs are set as triangular MFs and initialised with linear spacing across the range of the input, as shown in Table \ref{tab:flc_input_mf_parameters}, where the parameters represent the vertices of each of the MFs.
This results in the full set of input MFs shown in Figure \ref{fig:fis_input_mf}.
Instead, the output MFs are chosen to be linear and are initialised as shown in Table \ref{tab:output_mf} and Figure \ref{fig:fis_output_mf_parallel_coordinate}.
Lastly, Figure \ref{fig:fis_inputs_outputs} displays an example of the inputs to and resulting output of the FLC for a robot $a$.
In the example, the robot spatial information of the disaster environment is represented in a $(4 \times 4)$ grid.
We can see from the \textit{fire-risk} input, $T^{\text{risk}}$, that there are active fires in the centre of the disaster environment.
The current robot position in cell $(2,2)$ is visible in $M^{\text{response}}_r(k)$.
In the output $M^{\text{a}}_r(k)$ it can be seen that the next cell the robot will prioritise will be cell $(3,2)$.
Inspecting the inputs, this does not appear to be the logical cell to prioritise, as the cell already has a high scan certainty.
Under these conditions, MPFC may conclude that the FLC should raise the weighting of scan certainty and reduce the influence of response time, potentially causing the robot to select cell $(3,2)$ as its next target instead.

\subsection{MPC objective function}

For the MPC controller in the MPFC, in the case study a different objective function is used:
\begin{align} \label{eq:mpc_obj_func}
    \max_{[\bm{\theta}^{\top}_1(\kflc), \ldots, \bm{\theta}^{\top}_{\nr}(\kflc)]^{\top}}
    & \sum_{\kappa=\kflc}^{\kflc+\Np} 
    \sum_{i = 1}^{\nxe}
    \sum_{j = 1}^{\nye}
    \mvictim_{ij}(k) \mscan_{ij}(k) \cdot \\
    & \big( c_{\textrm{o}_1} - c_{\textrm{o}_2} m^{\textrm{downwind}}_{ij}(k) \big) \nonumber
\end{align}
where $m^{\textrm{downwind}}_{ij}(k)$ is an element of the matrix $M^{\textrm{downwind}}(k)$, which is the \textit{downwind matrix}, that represents the relative proximity of each cell in the disaster environment to an active or catching fire, normalised in range [0, 1] by considering the map dimension.
It is generated according to Algorithm~\ref{alg:downwind_map} (see Appendix \ref{app:algorithms}) using the following components:
\begin{align} \label{eq:m_downwind_dir}
    m^\textrm{downwind direction}_{ij}&(k) = \exp \Bigg( \mvel_{ij}(k) \bigg( c_{\textrm{w}_1} + \\ 
    & c_{\textrm{w}_2} \Big( \cos \big( \mdir_{ij}(k) - \theta^{\textrm{fire}}_{ij} \big) - 1 \Big) \bigg) \Bigg) \nonumber
\end{align}
where $\theta^{\textrm{fire}}_{ij}$ is the angle between cell $(i, j)$ and the cell $s^{\textrm{fire}}$ where an active fire is located, given by:
\begin{align} 
    \theta^{\textrm{fire}}_{ij} = 
    \arctan \left( \frac{\dps j - s^{\textrm{fire,y}}}{\dps i - s^{\textrm{fire,x}}} \right)
\end{align}
For any $s^{\textrm{fire}}$, we calculate $\theta^{\textrm{fire}}_{ij}$ for all the cells $(i, j)$ in the environment.
\begin{align} \label{eq:m_downwind_dist}
    m^\textrm{downwind distance}_{ij}(k) = 1 - \frac{\sqrt{(i - s^{\textrm{fire,x}})^2 + (j - s^{\textrm{fire,y}})^2}}{\sqrt{{\nxe}^2 + {\nye}^2}}
\end{align}

\subsection{Setup}

See Appendix \ref{app:examples} for some example simulations.

\subsubsection{Simulation configuration}

The case study is implemented in MATLAB and all simulations are run on an AMD Ryzen 5 3500U CPU.
For each simulation, we rely on a dedicated set of configuration scripts to define all simulation parameters, that can be found in the 4TU repository \cite{maxwell_dataset_2024} or the GitHub repository \cite{maxwell_github_2024}.
Within the simulation environment, we reformat the objective function \eqref{eq:mpc_obj_func} so that a lower value is considered better, therefore all optimisations become minimisations.

\subsubsection{Global simulation parameters configuration}

A standard setup for the global simulation parameters is used across all simulations with the only exceptions being the simulation time, the MPC step size and prediction horizon.
The purpose of the various simulations performed in this case study is not to explore the influence of these parameters, even though they can significantly impact the overall performance of the system and the results of the simulation.
The simulation step size is set as $T = \SI{15}{\second}$, the coarsening factor is set to $5$, and the objective function constants are set as $c_{\textrm{o}_1} = 1$ and $c_{\textrm{o}_2} = 1$.

\subsubsection{Environment configuration}

The standard environment model configuration is defined as:
\begin{itemize}
    \item Environment cell length, $\ell^{\textrm{x}}_{(i,j)} = \ell^{\textrm{y}}_{(i,j)} = \SI{10}{\meter}$.
    \item The debris occupancy matrix $\Mdebris$ has the elements initialised to $0.5$.
    \item The structure matrix $\Mstr$ has the elements initialised to $1$.
    \item Wind velocity, $\mvel_{ij}(k) = \SI{0}{\meter \per \second}$.
    \item Wind direction, $\mdir_{ij}(k) = -\frac{\pi}{4}$
    \item Fire spread constants, $c^{\text{fs1}} = 0.2$ and $c^{\text{fs2}} = 0.2$.
    \item Ignition time, $k^{2\textrm{min}} = \SI{120}{\second}$.
    \item Burnout time, $k^{10\textrm{min}} = \SI{600}{\second}$.
    \item Wind spread radius, $r^{\text{wind}} = 3$.
    \item Wind model constants, $c^{\text{wm1}} = 0.1$, $c^{\text{wm2}} = 0.1$, and $c^{\text{wmd}} = 0.4$.
    \item Max number of victims per search map cell, $\nv = 5$.
\end{itemize}

Given the environment model dimensions ($\nxe, \nye$), a standard initialisation for environment map parameters is also defined.
The three scenarios used in the case study are described in detail in Appendix \ref{app:scenarios}.

\subsubsection{Robots configuration}

We define the following standard robot model configuration:
\begin{itemize}
    \item Robot maximum velocity, $\va = \SI{5}{\meter \per \second}$.
    \item Robot scan time per square meter, $t^{\textrm{scan}}_r = \SI{0.01}{\second \per \square \meter}$.
    \item Robot sensor accuracy, $\eta = 0.9$.
    \item Scan certainty loss per time step, $\sigma = 0.01$.
    \item The robot tasks is always initialised with a ``scan'' task.
\end{itemize}

With MPFC, the robot target cells are initialised as the current robot locations, $(\tau_x,\tau_y)_{r,k} = (i, j)_{r,k}$. 
In contrast, for the MPC controller, the first set of target cells are initialised as the current robot locations and the remaining cells are set as ones.

\subsubsection{Controller Configuration}

Three separate controller configurations are selected for the simulations, including an MPFC, MPC, and Pre-tuned FLC.
In real-time control problems, a key constraint is the optimisation computation time, which must be completed before determining the next set of control inputs.
Therefore, a set of optimisation constraints are imposed to reflect this and limit the time spent for each optimisation.

\paragraph{MPFC Configuration}

The MPFC controller optimisation is configured to use a classic \textit{pattern search} algorithm, which is selected due to its suitability for handling non-smooth, non-linear, and discontinuous optimisation problems without requiring gradient information. 
This makes it particularly well-suited for the complex and probabilistic nature of the environment model used in our simulations. 

The key constraints for the optimisation are defined by the maximum number of function evaluations to $100$ and the maximum number of iterations to $100$.
The optimisation bounds are set as lower bounds all to $-1$ and upper bounds all to $1$.

\paragraph{MPC Configuration}

The MPC controller optimisation is configured to use a \textit{genetic algorithm}, which is selected over \textit{pattern search} due to its ability to handle discrete optimisation problems effectively and is well-suited for optimisation problems where the search space is large, complex, and discontinuous. 
The key constraints for the optimisation are defined by the maximum number of function evaluations to $100$ and the population size to $100$.
The optimisation constraints are set as $[1, 1]$ for the set of lower bounds and as the coarsened environment dimention for the set of upper bounds.

\paragraph{Pre-tuned FLC Configuration}

The Pre-tuned FLC is configured as defined in Section \ref{sec:flc_layer}.
The Pre-tuned FLC parameters are fixed throughout the simulation and provide the baseline performance for a simple local robot controller.

\subsubsection{Performance parameters}

The parameters selected to analyse the performance of the controller are the global objective function \eqref{eq:mpc_obj_func}, $J(k)$, and the optimisation time, $t^{\text{opt}}$, which is the time required for the predictive controller optimisation to complete.

Due to the probabilistic nature of the environment model and optimisation algorithm, the performance of the multi-robot system is inherently variable over repeated simulations.
Therefore, we evaluate the mean values and confidence intervals on the performance metrics to quantify the expected performance and variability over multiple simulations.
By calling $n^{\text{sim}}$ the total number of simulations, the instantaneous mean objective function is:

\begin{equation}
\overline{J}(k) = \frac{1}{n^{\text{sim}}} \sum_{s=1}^{n^{\text{sim}}} J_{s}(k)
\end{equation}
and the instantaneous mean optimisation time is:

\begin{equation}
\overline{t}^{\text{opt}}(k) = \frac{1}{n^{\text{sim}}} \sum_{s=1}^{n^{\text{sim}}} t^{\text{opt}}_{s}(k)
\end{equation}

We calculate 95\% confidence intervals by taking the mean time series and adding/subtracting $1.96 \times \text{SEM}$, where SEM is the standard error of the mean. 
We denote these intervals as $(\overline{J}_{0.025}(k)$, $\overline{J}_{0.975}(k))$ and $(\overline{t}^{\text{opt}}_{0.025}, \overline{t}^{\text{opt}}_{0.975}(k))$, respectively.

To ensure a fair comparison between different controller architectures, we define a sequence of simulation seeds for each set of simulations. 
These seeds are used to initialise the random number generator for each controller architecture in each simulation, guaranteeing that the probabilistic environment variables remain consistent across equivalent simulations for each controller architecture.

\subsection{Results} \label{sec:results}

\subsubsection{Standard results format}

\begin{figure}
\centering
\includegraphics[width=0.7\linewidth]{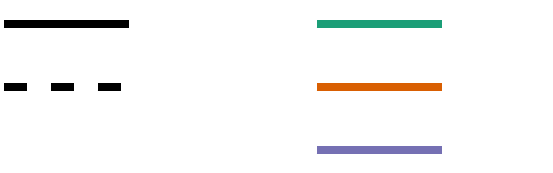}
\caption{Line styles for simulation results.}
\label{fig:standard_line_format}
\end{figure}

The simulation parameters are displayed using line plots for each of the simulation cases.
To establish consistency across the visualisation of simulation results, we define a standard results format for each controller type and controller architecture, as shown in Figure \ref{fig:standard_line_format}.
A solid line is used for centralised controllers, and a dashed line is used for decentralised controllers.
FLC results are coloured green, MPFC results are coloured orange, and MPC results are coloured purple.

\subsubsection{MPFC performance analysis} \label{subsec:performanceAnalysis}

\begin{figure}[p]
\centering
\includegraphics[width=.9\linewidth]{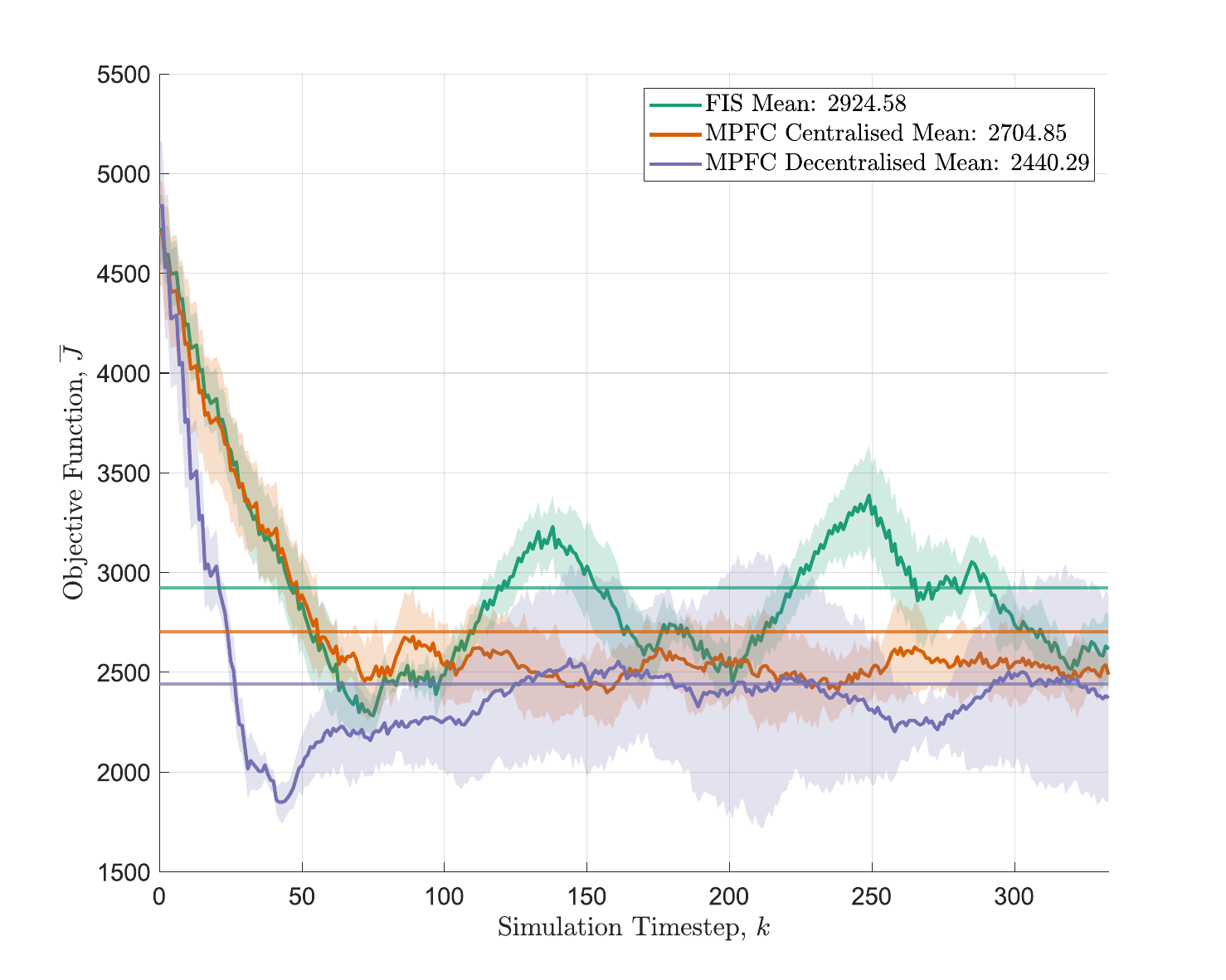}
\caption{$\overline{J}(k)$ for a two-robot system in small static disaster environment, 5 simulations.}
\label{fig:static_2_agents_obj}
\end{figure}

\begin{figure}[p]
\centering
\includegraphics[width=.9\linewidth]{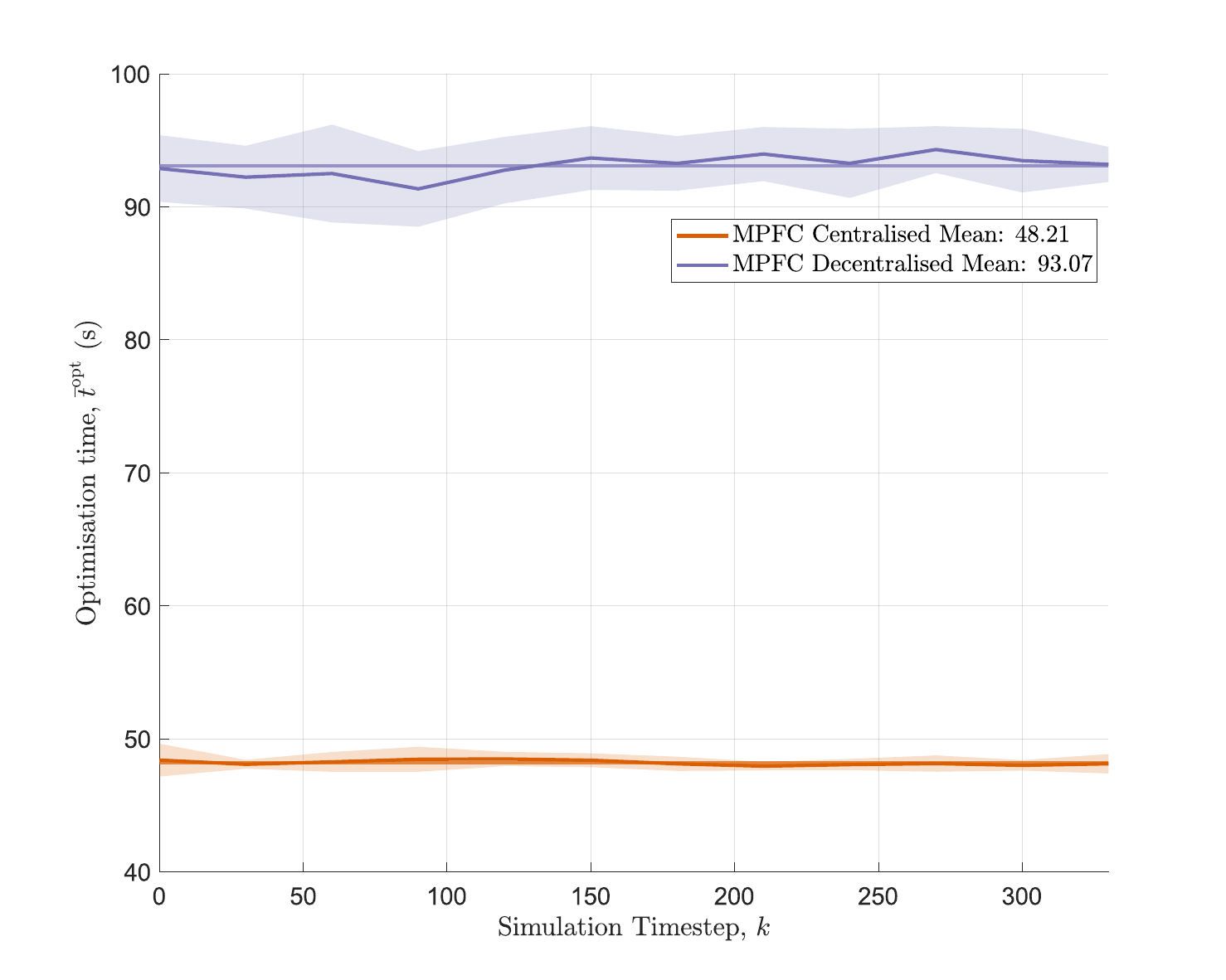}
\caption{$\overline{t}^{\text{opt}}(k)$ for a two-robot system in small static disaster environment, 5 simulations.}
\label{fig:static_2_agents_t_opt}
\end{figure}

\begin{figure}[p]
\centering
\includegraphics[width=.9\linewidth]{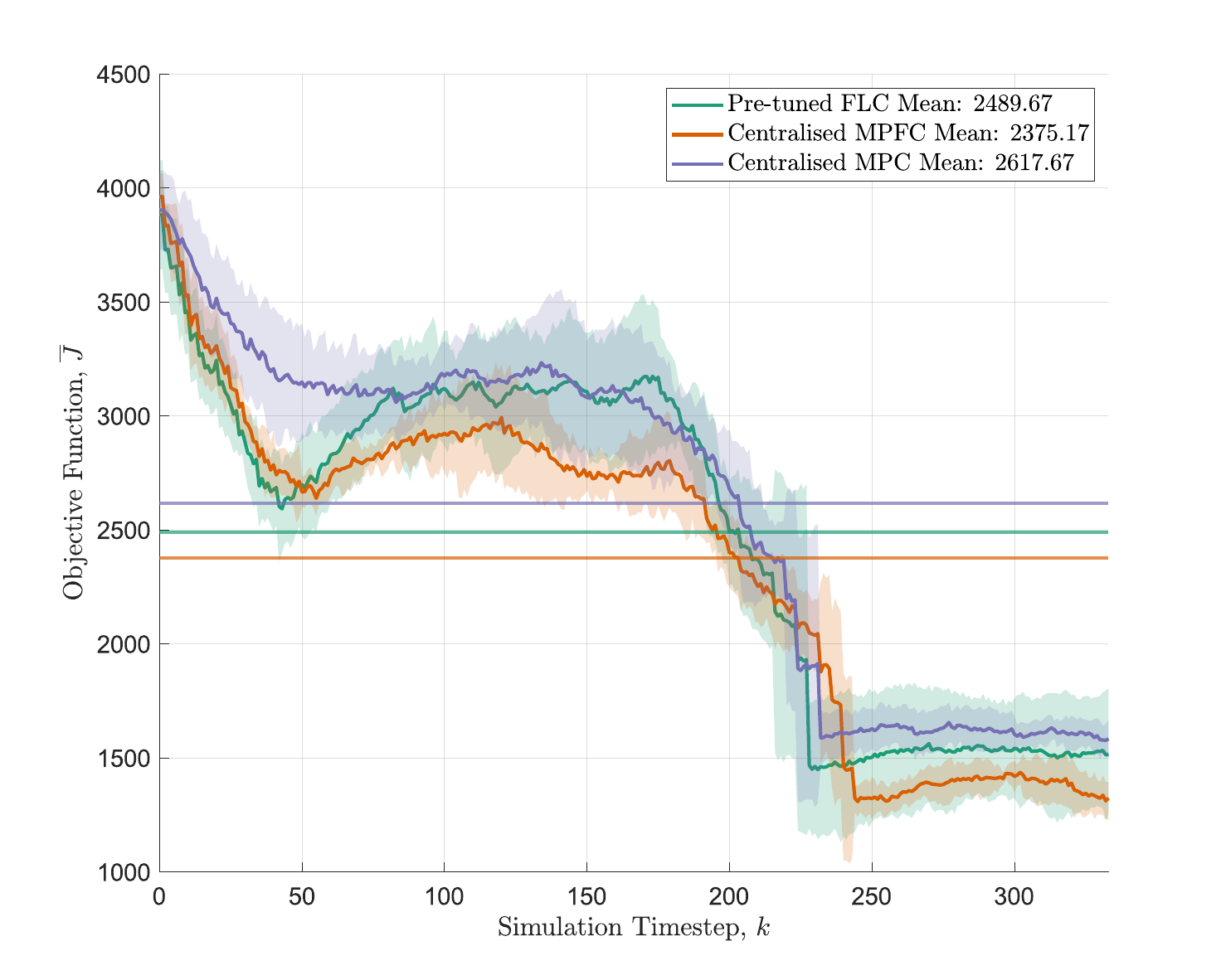}
\caption{$\overline{J}(k)$ for two-robot system in small dynamic disaster environment, 5 simulations.}
\label{fig:dynamic_2_agents_obj}
\end{figure}

\begin{figure}[p]
\centering
\includegraphics[width=.9\linewidth]{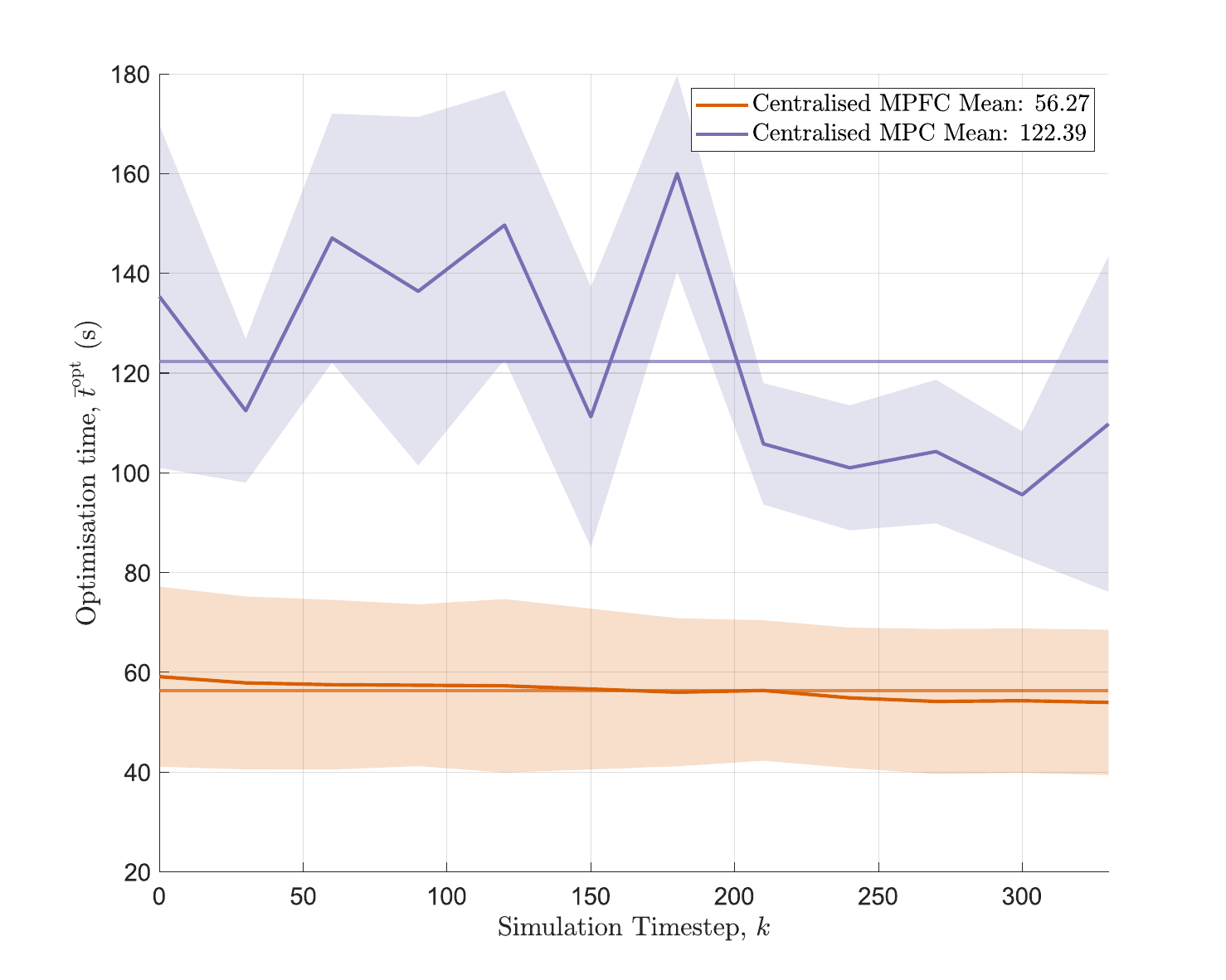}
\caption{$\overline{t}^{\text{opt}}(k)$ for a two-robot system in small dynamic disaster environment, 5 simulations.}
\label{fig:dynamic_2_agents_t_opt}
\end{figure}

\begin{figure}[p]
\centering
\includegraphics[width=.9\linewidth]{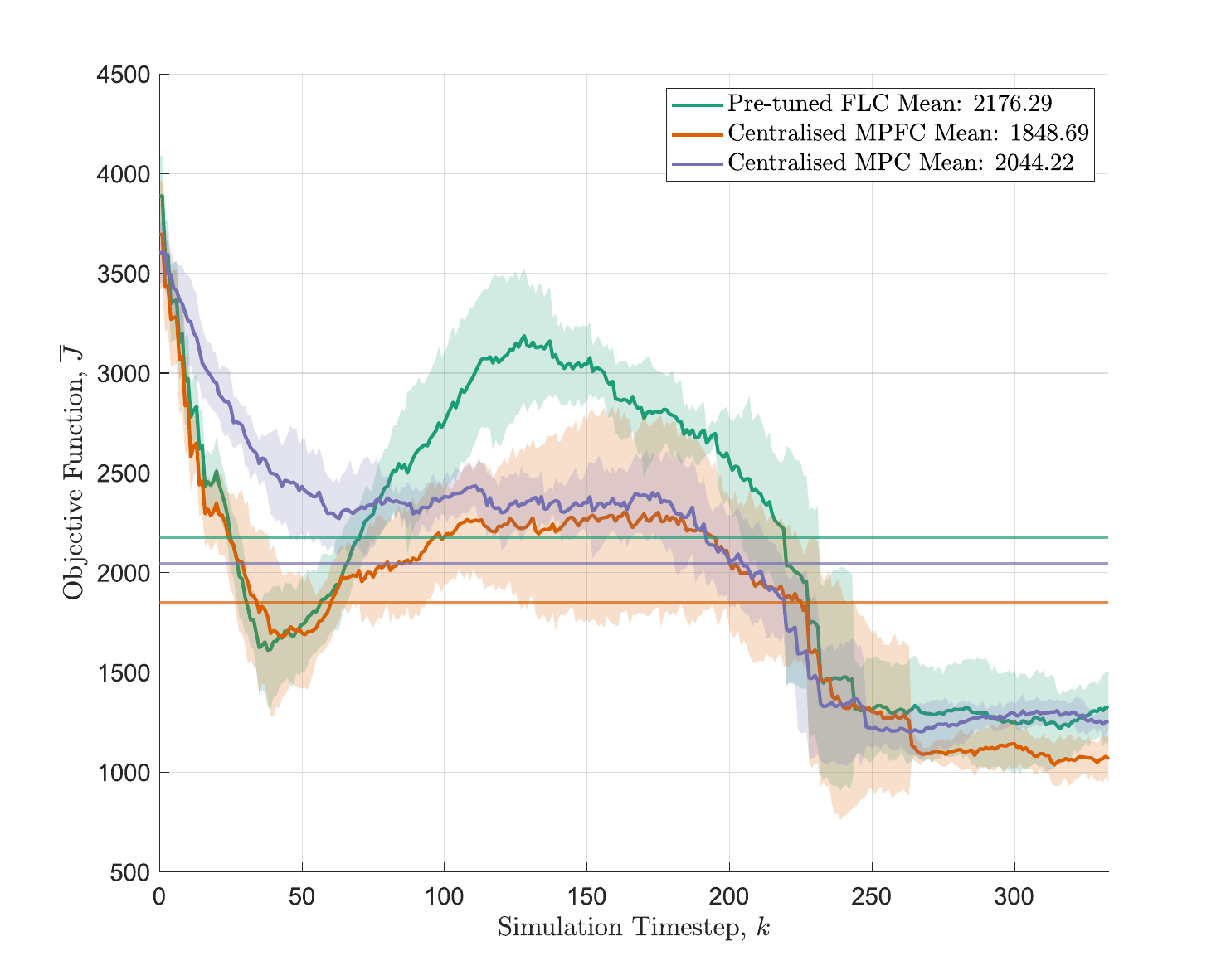}
\caption{$\overline{J}(k)$ for four-robot system in small dynamic disaster environment, 5 simulations.}
\label{fig:dynamic_4_agents_obj}
\end{figure}

\begin{figure}[p]
\centering
\includegraphics[width=.9\linewidth]{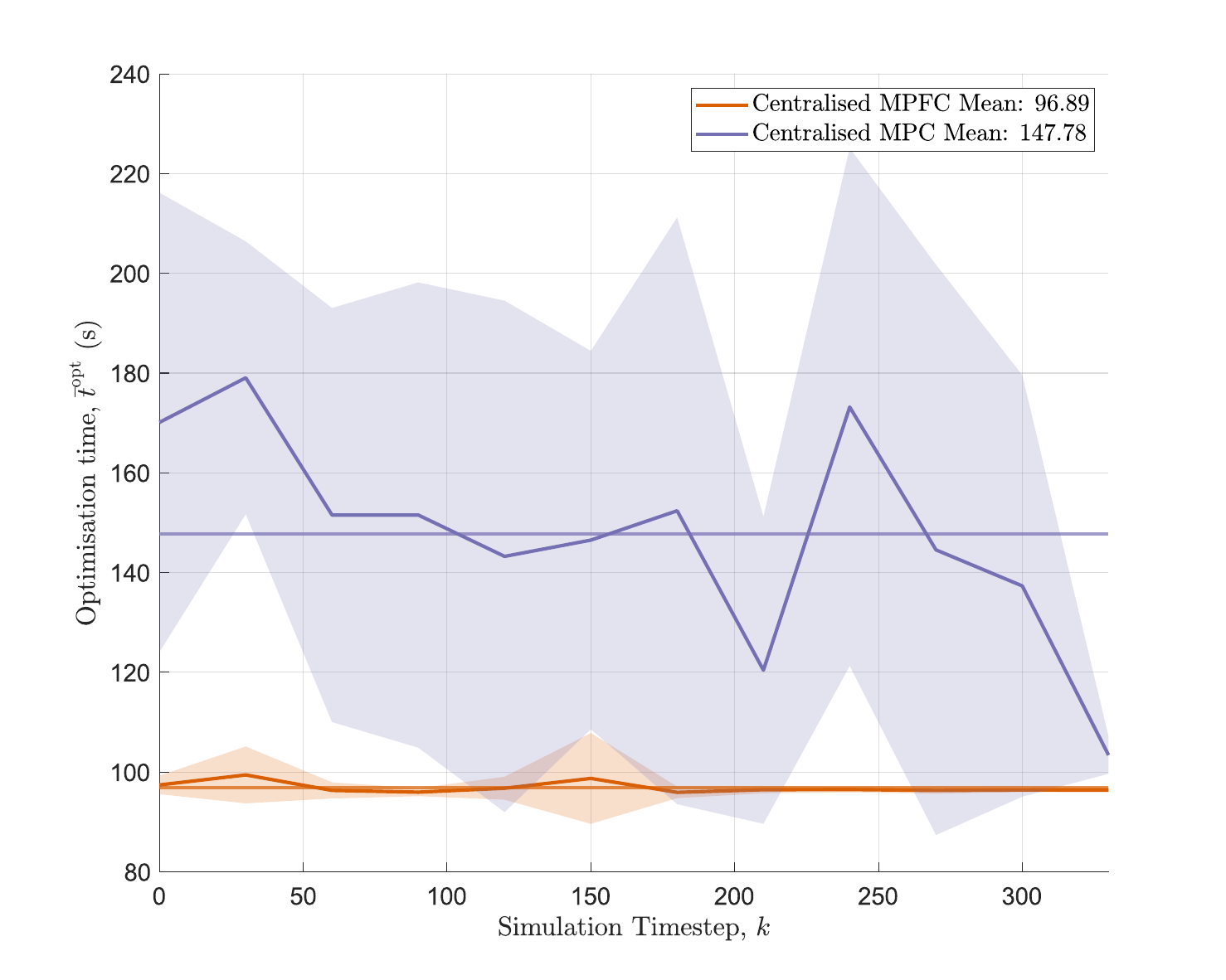}
\caption{$\overline{t}^{\text{opt}}(k)$ for a four-robot system in small dynamic disaster environment, 5 simulations.}
\label{fig:dynamic_4_agents_t_opt}
\end{figure}

In the MPFC performance analysis, MPFC is simulated and assessed against the MPC and Pre-tuned FLC.
This is done by beginning with simple simulation cases to isolate controller performance from the modelling complexities and gradually introducing more complex simulations.
Figure \ref{fig:static_2_agents_obj} presents the instantaneous objective function of the centralised MPFC, centralised MPC, and Pre-tuned FLC controllers for a two-robot system in a small static environment, simulated over $\SI{5000}{\second}$.
Figure \ref{fig:static_2_agents_t_opt} shows the computational optimisation time for each MPC step over the simulation.
Figure \ref{fig:dynamic_2_agents_obj} presents the instantaneous objective function of the centralised MPFC, centralised MPC, and Pre-tuned FLC controllers for a two-robot system in a small dynamic environment, simulated over \SI{5000}{\second}.
Figure \ref{fig:dynamic_2_agents_t_opt} shows the computational optimisation time for each MPC step over the simulation.
Figure \ref{fig:dynamic_4_agents_obj} presents the results for the same configuration defined above for Figure \ref{fig:dynamic_2_agents_obj} where the number of robots is increased to $\nr = 4$.
Figure \ref{fig:dynamic_4_agents_t_opt}, likewise, shows the computational optimisation time.

\paragraph{Decentralised vs centralised MPFC controller architectures}

\begin{figure}[p]
\centering
\includegraphics[width=.9\linewidth]{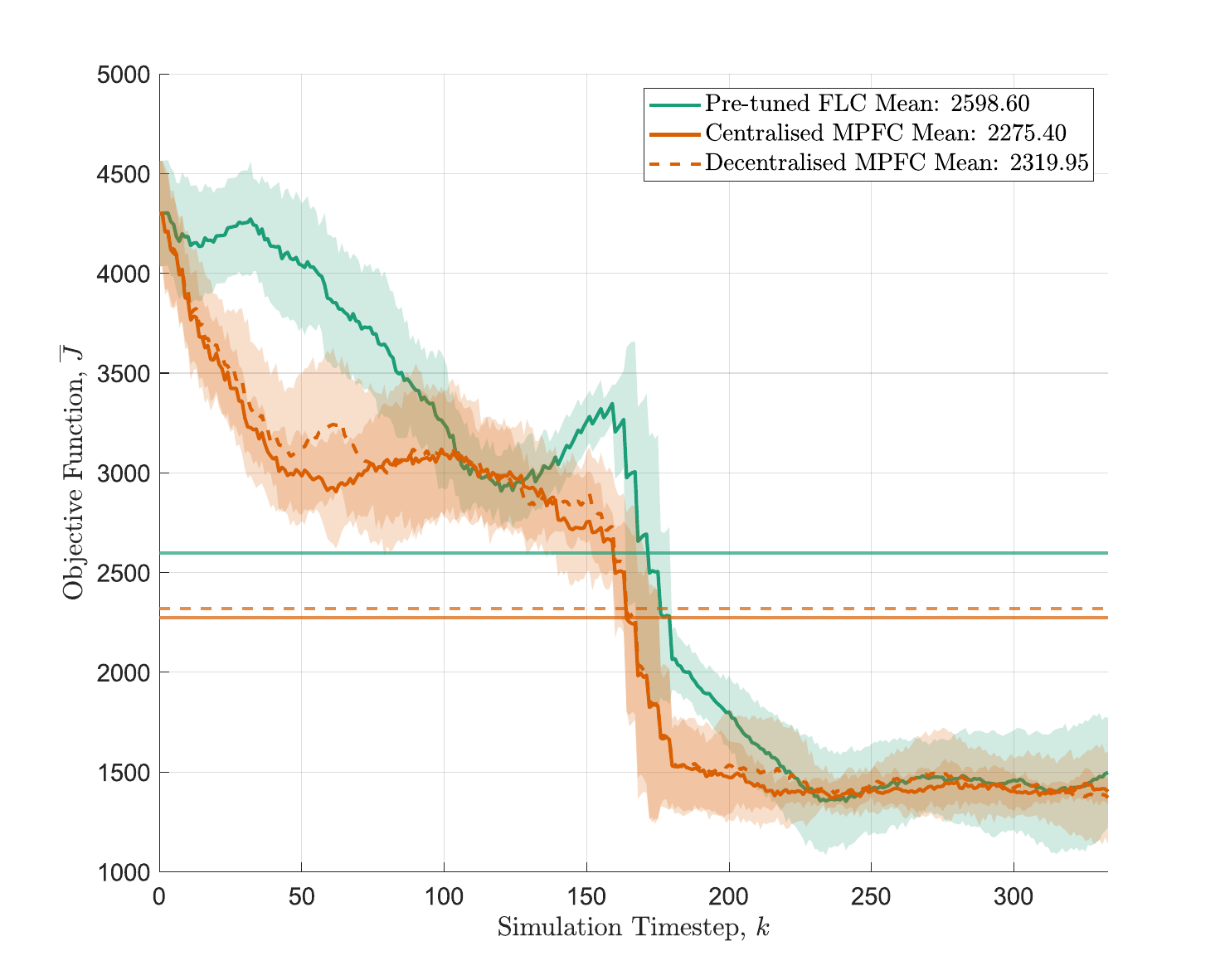}
\caption{$\overline{J}(k)$ for centralised vs decentralised MPFC with $\nr = 2$, 5 simulations.}
\label{fig:sim_perfo_architecture_n_a_2_obj}
\end{figure}

\begin{figure}[p]
\centering
\includegraphics[width=.9\linewidth]{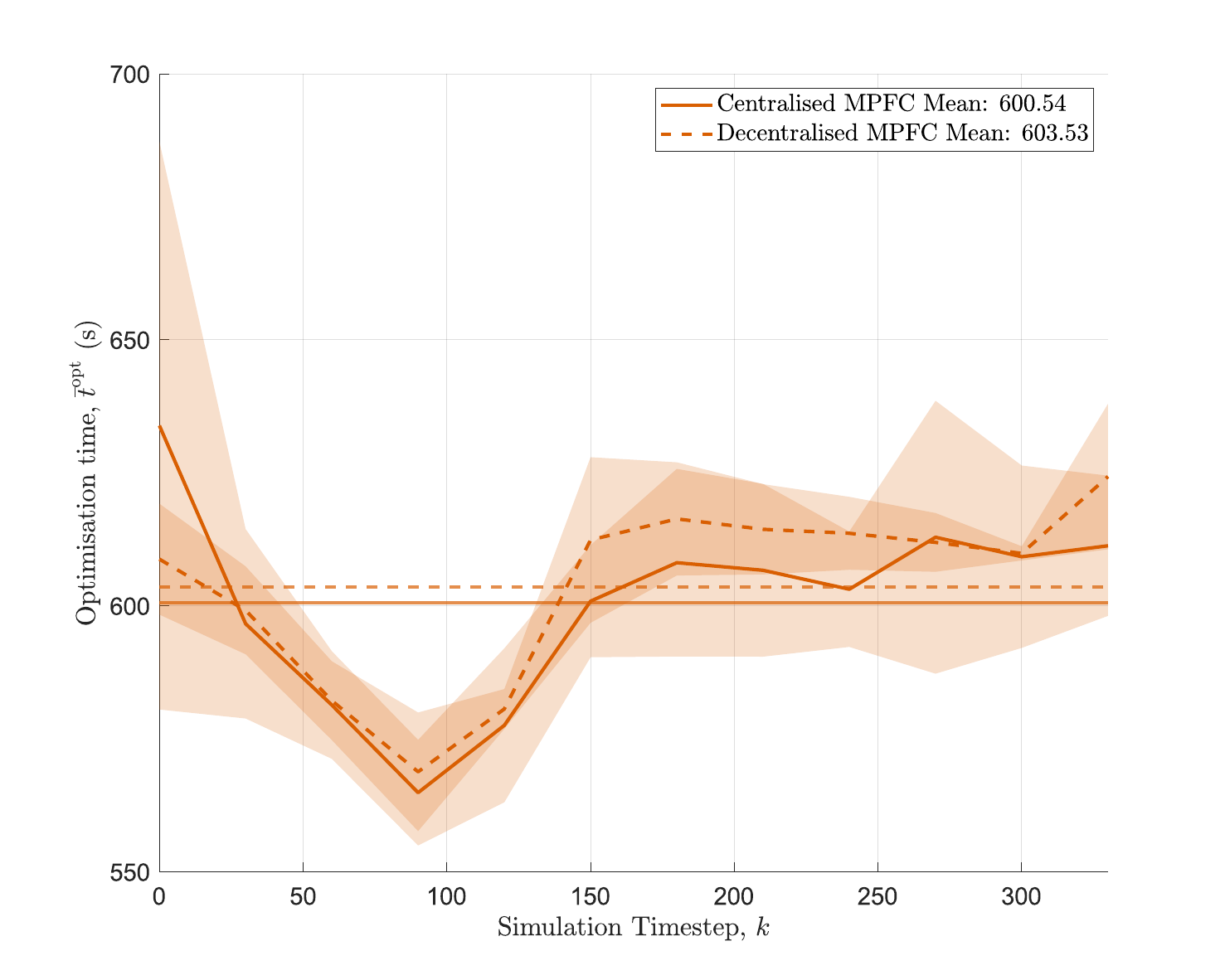}
\caption{$\overline{t}^{\text{opt}}(k)$ for centralised vs decentralised MPFC with $\nr = 2$, 5 simulations.}
\label{fig:sim_perfo_architecture_n_a_2_t_opt}
\end{figure}

\begin{figure}[p]
\centering
\includegraphics[width=.9\linewidth]{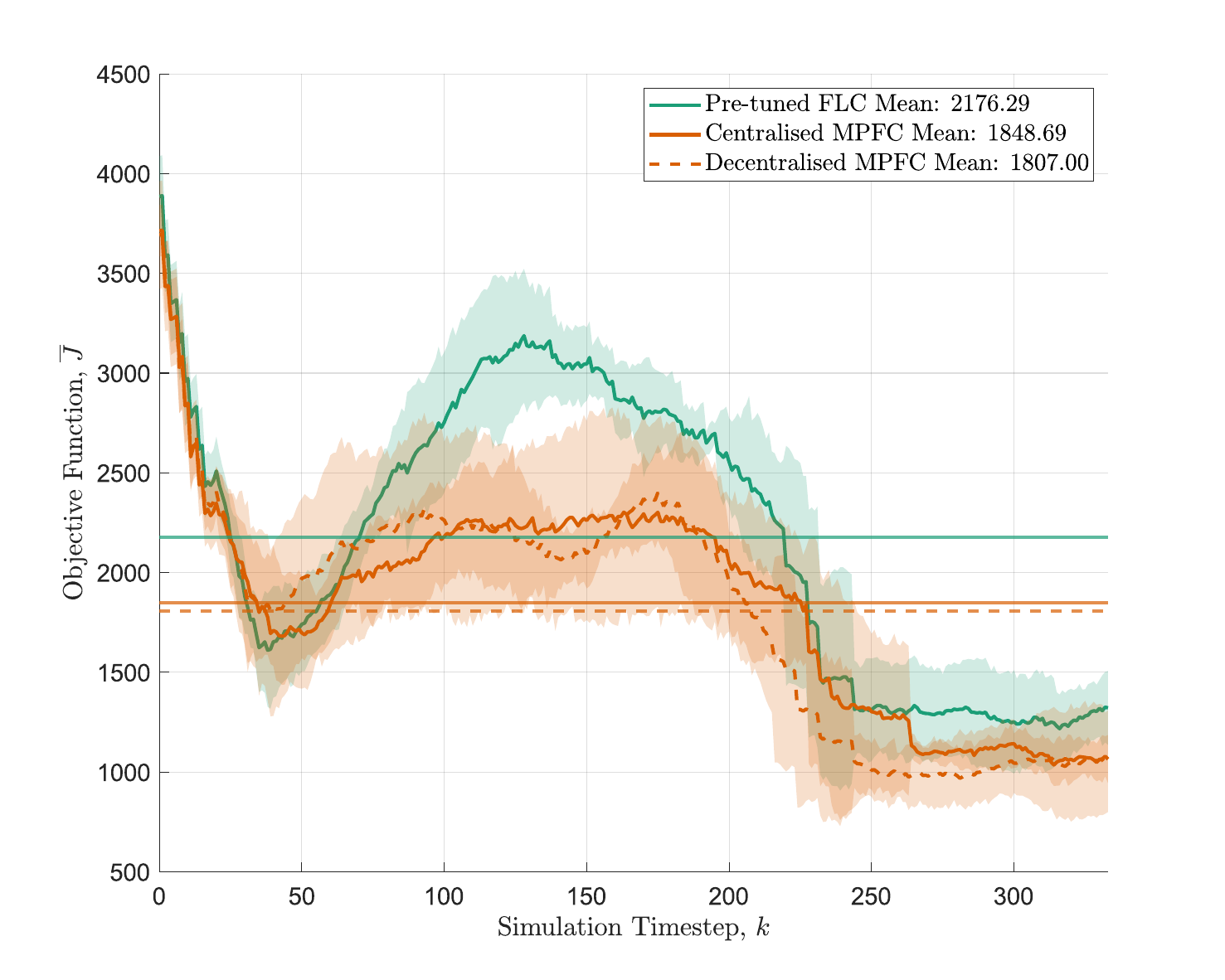}
\caption{$\overline{J}(k)$ for centralised vs decentralised MPFC with $\nr = 4$, 5 simulations.}
\label{fig:sim_perfo_architecture_n_a_4}
\end{figure}

\begin{figure}[p]
\centering
\includegraphics[width=.9\linewidth]{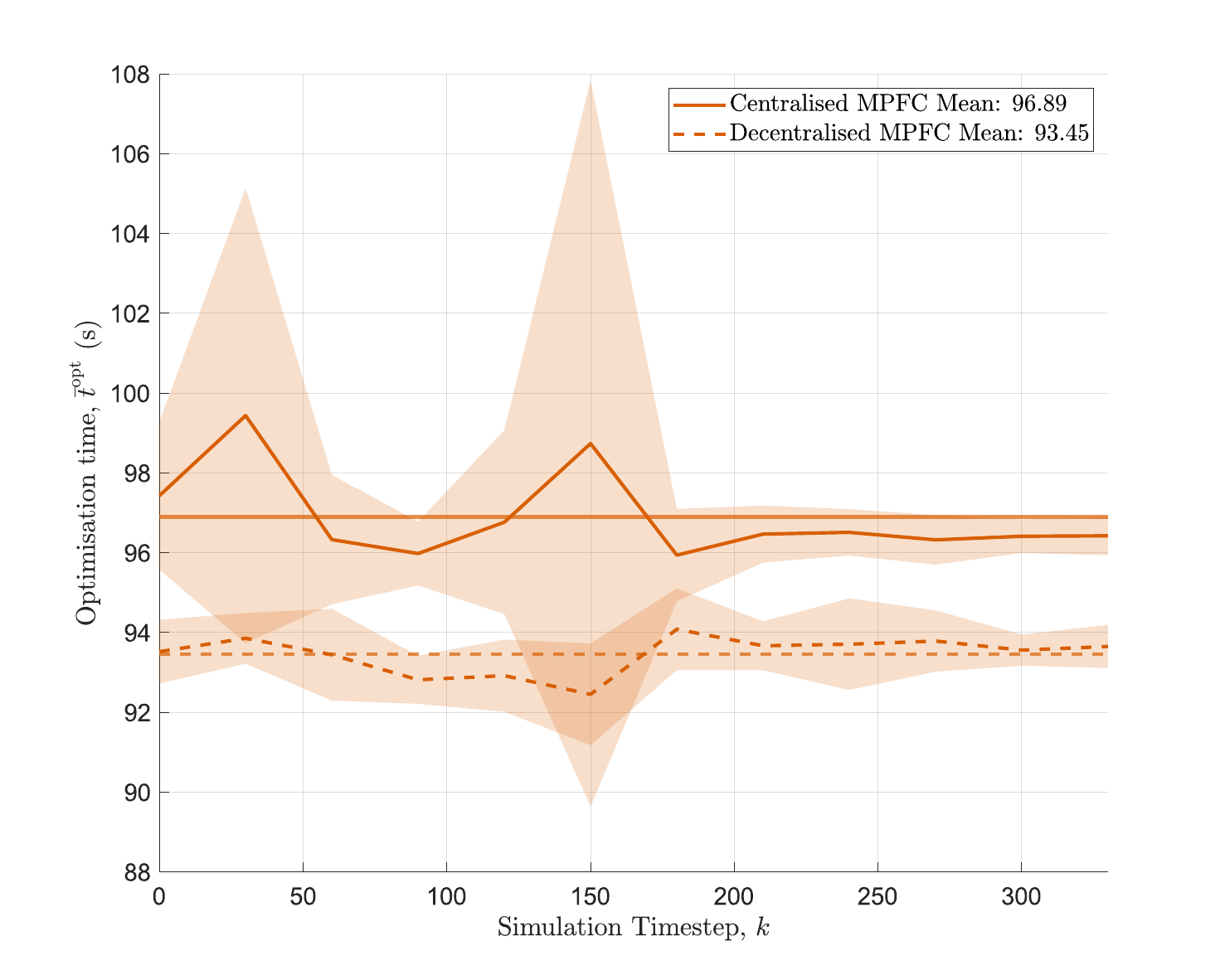}
\caption{$\overline{t}^{\text{opt}}(k)$ for centralised vs decentralised MPFC with $\nr = 4$, 5 simulations.}
\label{fig:sim_perfo_architecture_n_a_4_t_opt}
\end{figure}

Maintaining the same basic small dynamic disaster environment model, the centralised MPFC and decentralised MPFC architectures are simulated to compare performance.
Figure \ref{fig:sim_perfo_architecture_n_a_2_obj} presents the instantaneous objective function of the centralised MPFC and decentralised MPFC controllers for a two-robot system in a small dynamic environment, simulated over \SI{5000}{\second}.
Figure \ref{fig:sim_perfo_architecture_n_a_2_t_opt} shows the corresponding computational optimisation time.
Figure \ref{fig:sim_perfo_architecture_n_a_4} presents the results for the same configuration defined above for Figure \ref{fig:sim_perfo_architecture_n_a_2_obj} where the number of robots is increased to $\nr = 4$.
Figure \ref{fig:sim_perfo_architecture_n_a_4_t_opt}, likewise, shows the computational optimisation time.

\paragraph{Complex dynamic disaster environment}

\begin{figure}[p]
\centering
\includegraphics[width=.9\linewidth]{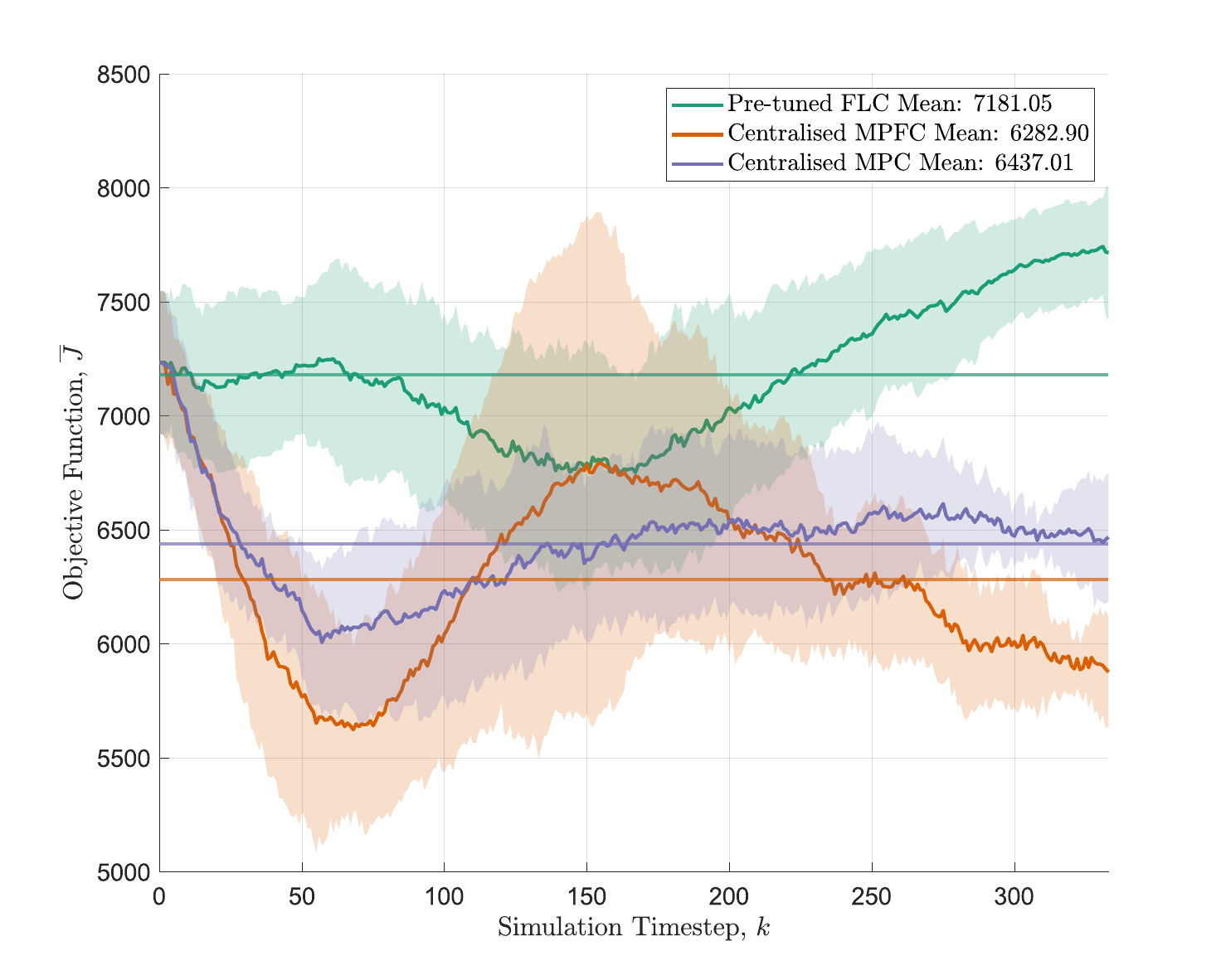}
\caption{$\overline{J}(k)$ for a two-robot system in complex disaster environment, 5 simulations.}
\label{fig:complex_2_agents_obj}
\end{figure}

\begin{figure}[p]
\centering
\includegraphics[width=.9\linewidth]{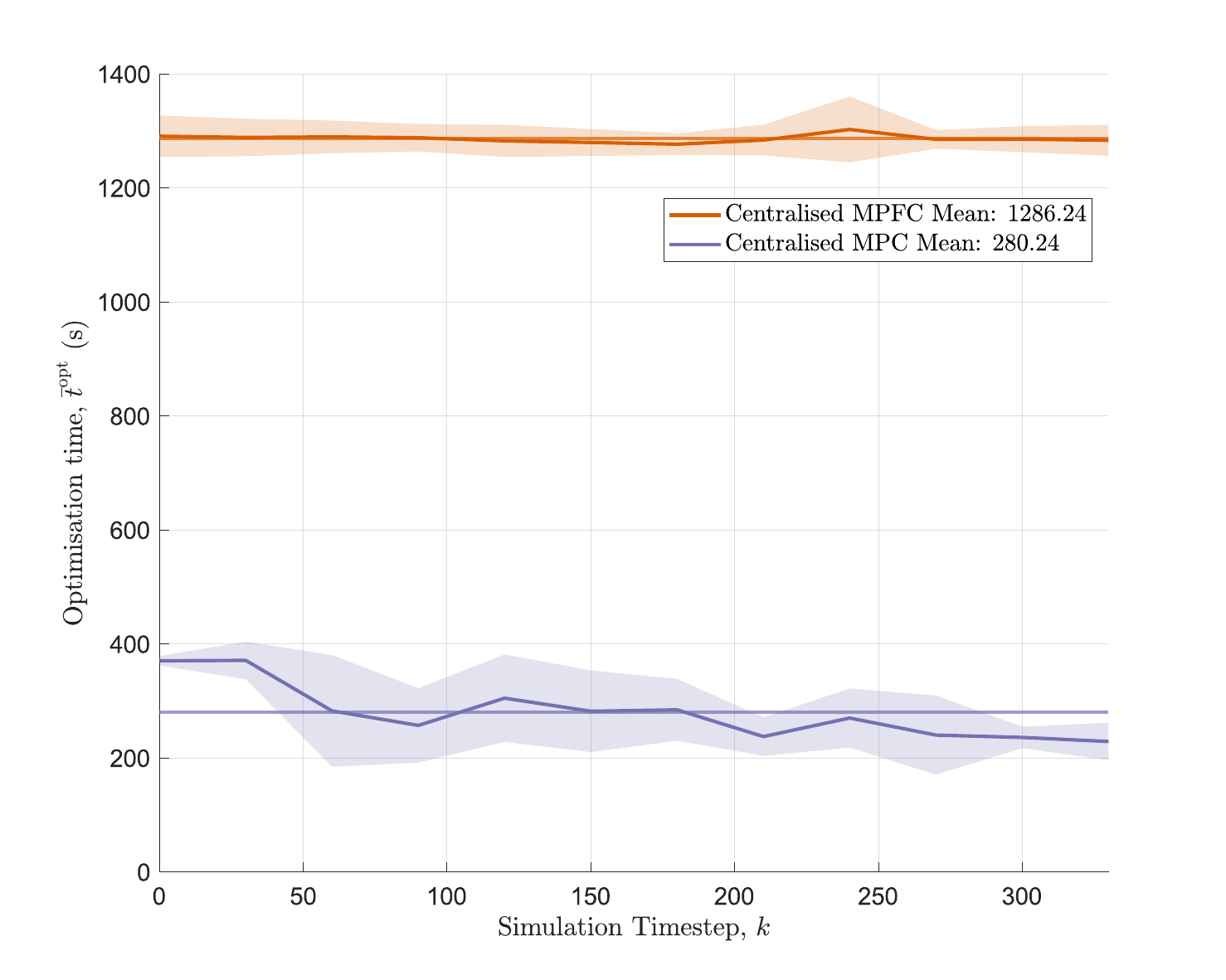}
\caption{$\overline{t}^{\text{opt}}(k)$ for a two-robot system in complex disaster environment, 5 simulations.}
\label{fig:complex_2_agents_t_opt}
\end{figure}

In the previous performance analysis simulations, the environment variables were simplified as much as possible to isolate controller performance in a controlled environment.
In this simulation, we analyse controller performance under more complex conditions by introducing more complex environment parameters.
Figure \ref{fig:complex_2_agents_obj} presents the instantaneous objective function while Figure \ref{fig:complex_2_agents_t_opt} shows the computational optimisation time.

\subsubsection{MPFC sensitivity analysis} \label{subsec:sensitivityAnalysis}

The objective of the sensitivity analysis is to investigate the performance of the supervisory controllers over a range of values for a given design parameter and identify the strengths and weaknesses of MPFC compared to the alternative controller architectures.
Four parameters are selected, each of which is a key design parameter in the sizing of the prediction step of the supervisory controllers, and an individual sensitivity analysis is performed for each parameter across a defined range of values.
All sensitivity analyses are performed for the basic dynamic environment initialisation, with the design parameters adjusted according to Table \ref{tab:sensitivity_analysis_parameters}.

\begin{table}
\centering
\caption{Sensitivity analysis parameters.}
\label{tab:sensitivity_analysis_parameters}
\renewcommand{\arraystretch}{1.5}
\begin{tabular}{|c|c|}
\hline
\textbf{Parameter} & \textbf{Range} \\
\hline
Number of robots ($\nr$) & 2, 3, 4 \\
\hline
Disaster environment size ($\nxe \cdot \nye$) & 20, 40, 60 \\
\hline
MPC step size ($\Tflc$) & 30, 75, 225, 450, 675, 900 \\
\hline
Prediction horizon ($\Np$) & 450, 675, 900, 1125 \\
\hline
\end{tabular}
\end{table}

\paragraph{Number of robots} \label{sim_sens_n_a}

\begin{figure}[t]
\centering
\includegraphics[width=.9\linewidth]{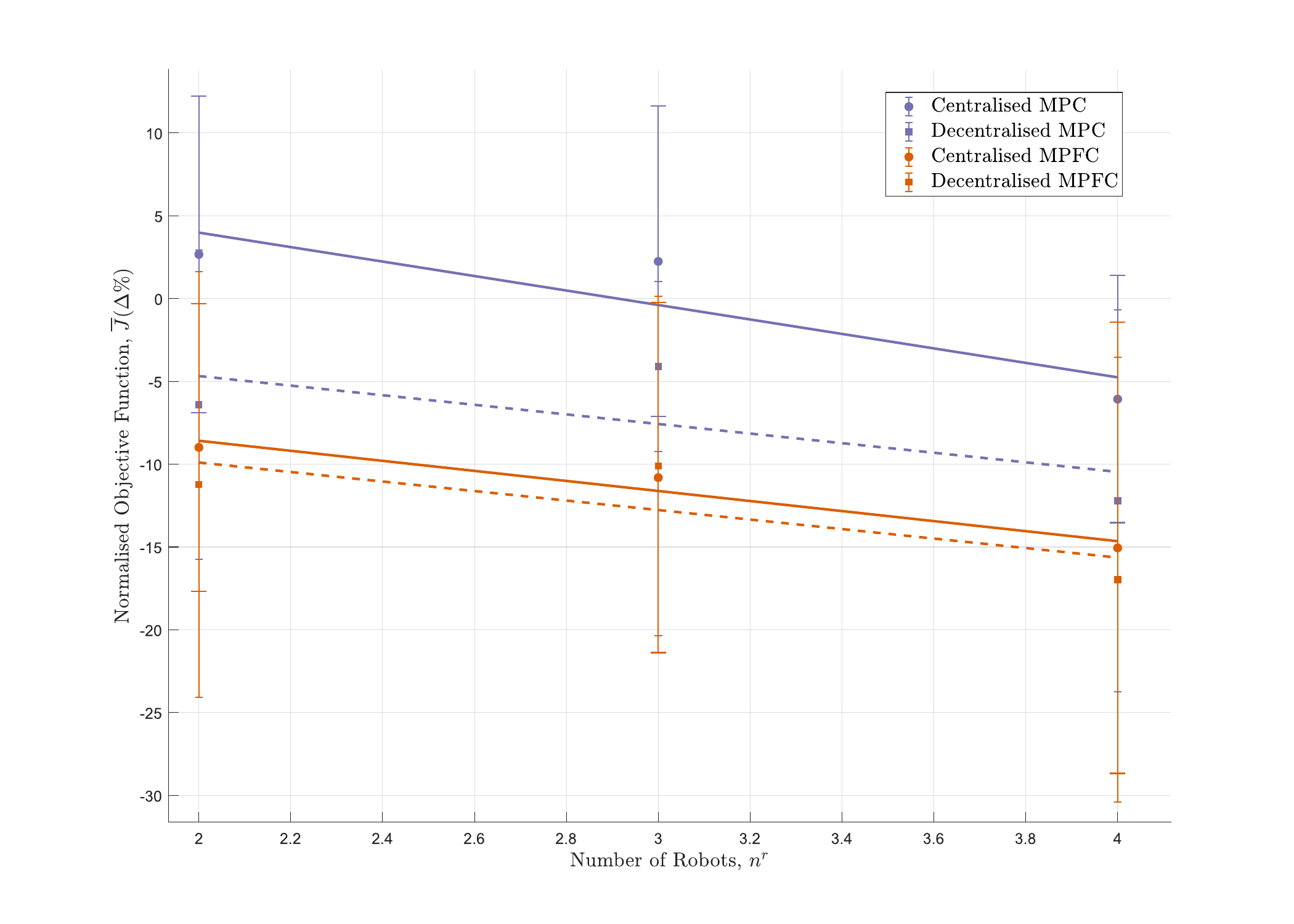}
\caption{Sensitivity with number of robots, $\nr$, 5 simulations each. Normalised Mean objective function, $\overline{J}(k)$.}
\label{fig:obj_n_a}
\end{figure}

\begin{figure}[t]
\centering
\includegraphics[width=.9\linewidth]{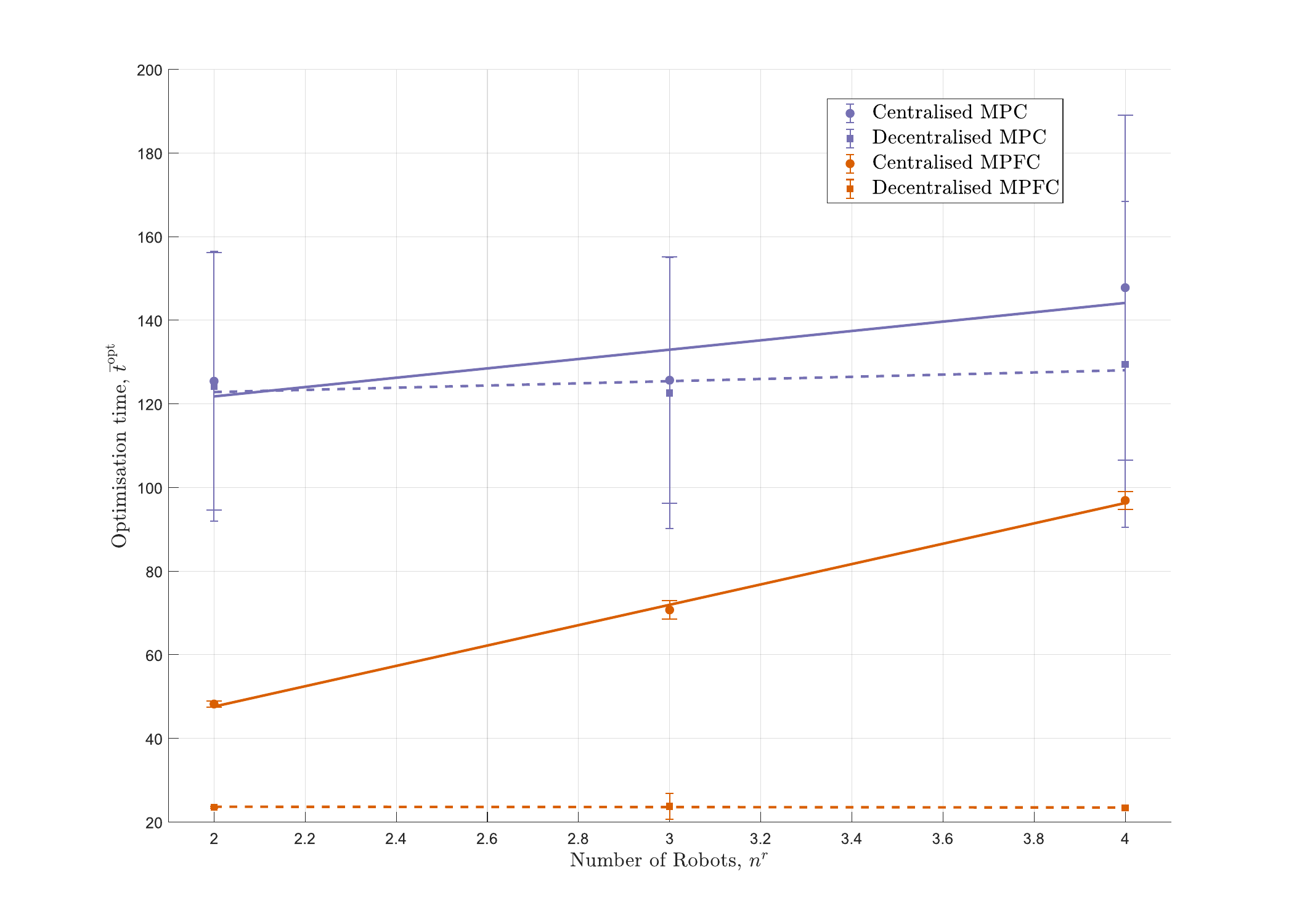}
\caption{Sensitivity with number of robots, $\nr$, 5 simulations each. Mean optimisation time, $\overline{t}^{\text{opt}}(k)$.}
\label{fig:t_opt_n_a}
\end{figure}

\begin{figure}[t]
\centering
\includegraphics[width=.9\linewidth]{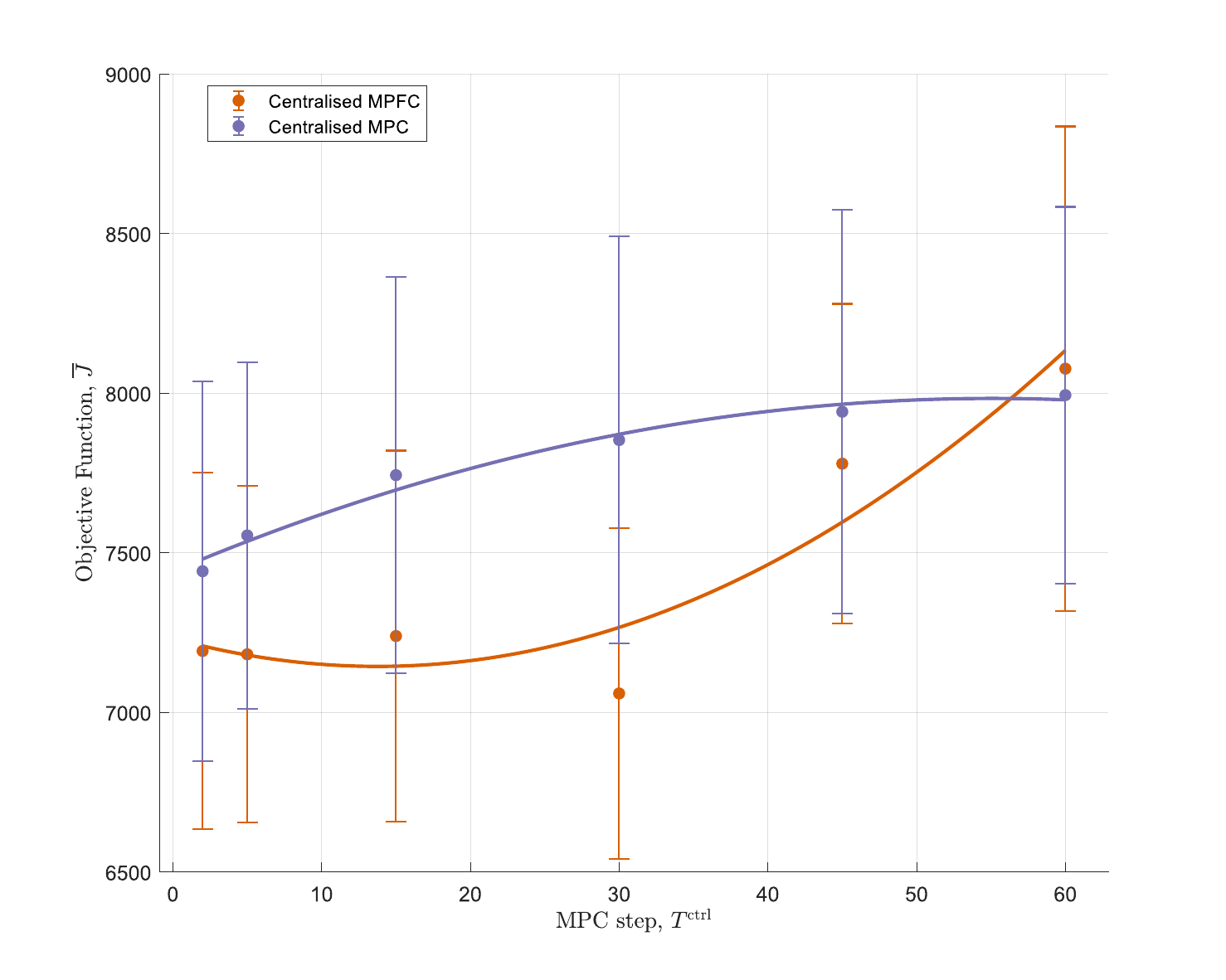}
\caption{Sensitivity of MPFC with MPC step, $\Tflc$, 5 simulations each. Mean objective function, $\overline{J}(k)$.}
\label{fig:obj_t_mpc}
\end{figure}

Figure \ref{fig:obj_n_a} presents the normalised instantaneous objective function for decentralised and centralised MPC and MPFC architectures with the number of robots in the range $\nr = [2, 3, 4]$.
The objective function values are normalised against the results for a Pre-tuned FLC controller and expressed as the percentage difference from the Pre-tuned FLC mean. 
The individual data points for each set of simulations are denoted by coloured dots, and the $95\%$ confidence intervals are represented by error bars.
To differentiate between controller architectures, centralised controller results are displayed using circular points and wider error bar tips, while decentralised controller results are displayed using square points and narrower error bar tips.
Additionally, a first-order polynomial trend line is fitted to the results of each architecture to visualise the correlation with the number of robots.
This is done in MATLAB using the function \texttt{polyfit}, that performs the least square error minimization for a polynomial of a chosen order (in this case of order one), where by increasing the order of the polynomial, that should be done, e.g., when there are more data points, we would expect to see an exponential trend (that is case for the following MPC step size, see \ref{fig:obj_t_mpc}).
A solid line is used for centralised architectures and a dashed line is used for decentralised architectures.
Figure \ref{fig:t_opt_n_a}, instead, shows the mean optimisation times for the simulated architectures. 

\paragraph{Disaster environment size} \label{sim_sens_n_env}

\begin{figure}[t]
\centering
\includegraphics[width=.9\linewidth]{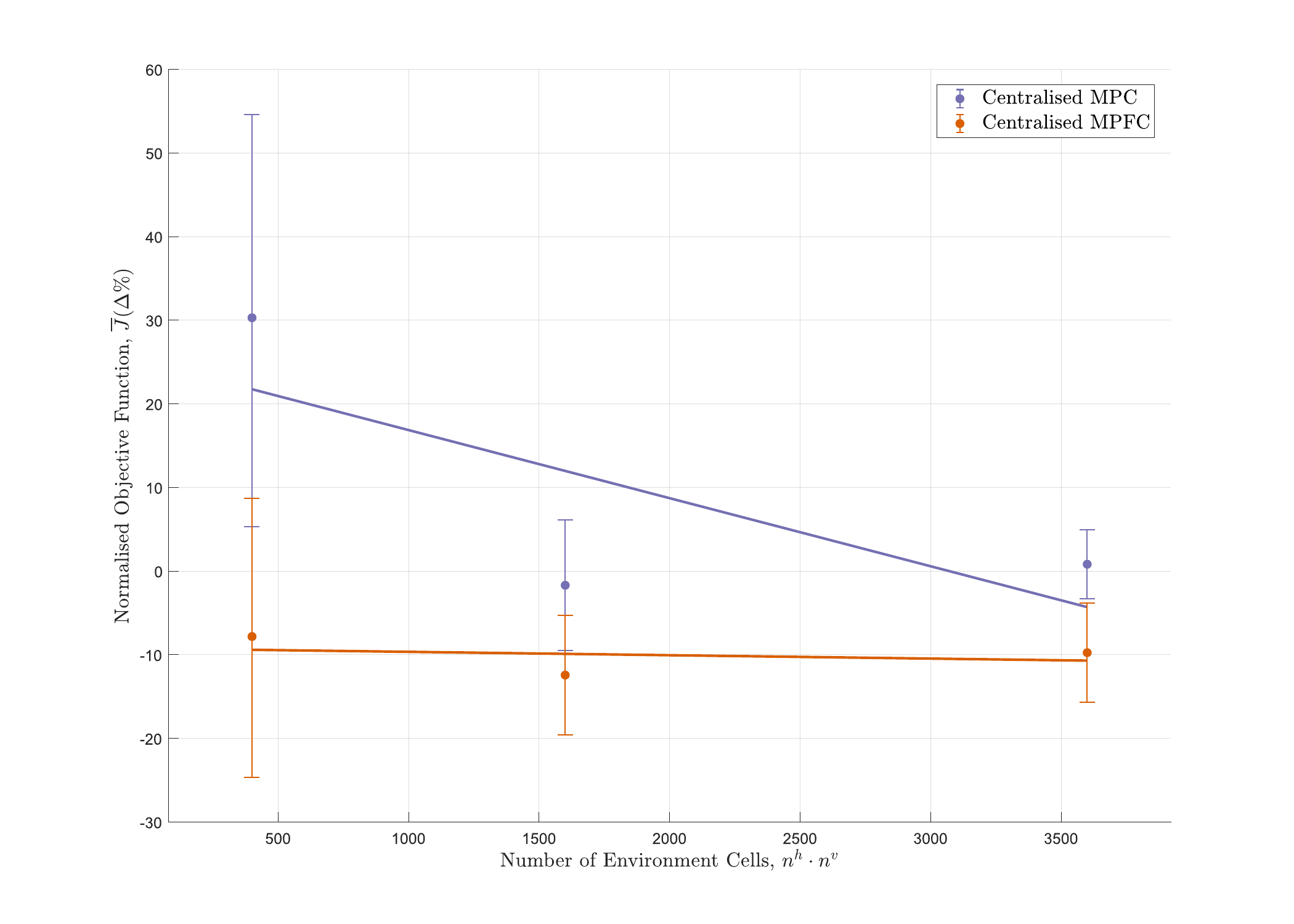}
\caption{Sensitivity with environment size, $\nxe \cdot \nye$, 5 simulations each. Normalised Mean objective function, $\overline{J}(k)$.}
\label{fig:obj_env_size}
\end{figure}

\begin{figure}[t]
\centering
\includegraphics[width=.9\linewidth]{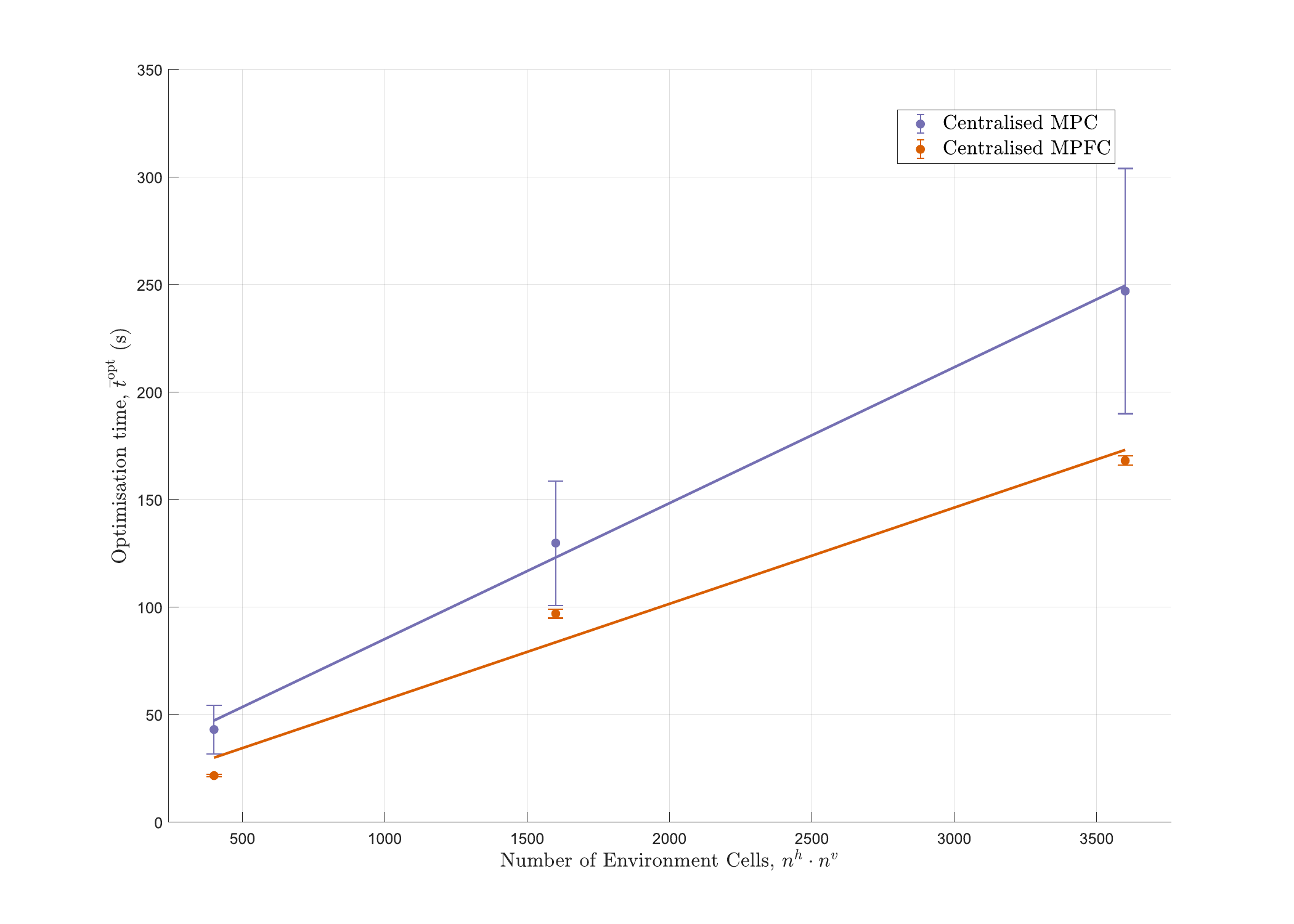}
\caption{Sensitivity with environment size, $\nxe \cdot \nye$, 5 simulations each. Mean optimisation time, $\overline{t}^{\text{opt}}(k)$.}
\label{fig:opt_time_env_size}
\end{figure}

Figure \ref{fig:obj_env_size} presents the normalised instantaneous objective function for the centralised MPC and MPFC controller architectures against the number of cells in the disaster environment.
In particular, the environment is not changing, whereas just the number of cells increases (in fact, in all cases there is still only one ignition point, while there are more victims that are distributed differently, because they are randomly located).
Figure \ref{fig:opt_time_env_size} shows the mean optimisation times for this sensitivity analysis.

\paragraph{MPC step size} \label{sim_sens_k_mpc}

\begin{figure}[t]
\centering
\includegraphics[width=.9\linewidth]{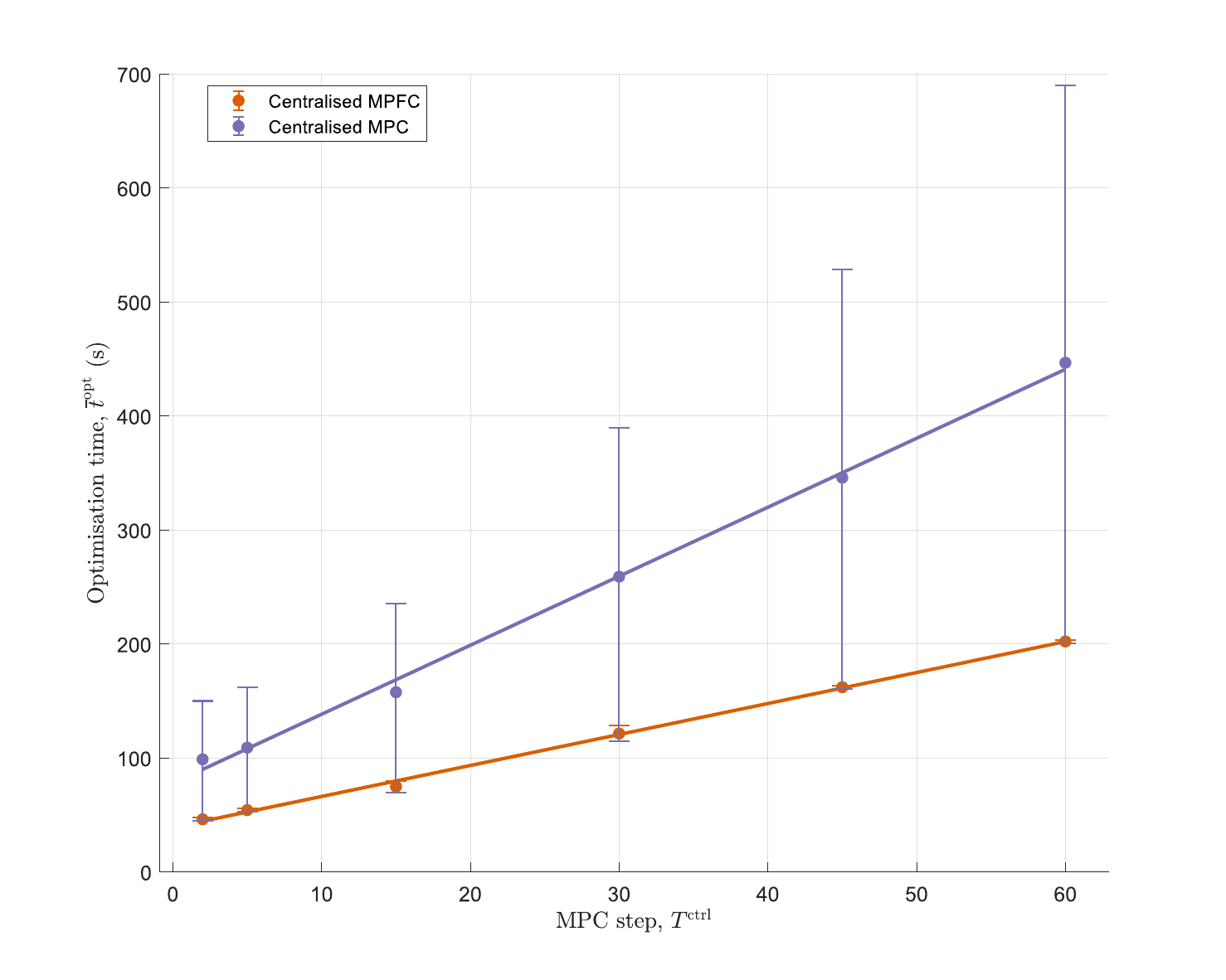}
\caption{Sensitivity of MPFC with MPC step, $\Tflc$, 5 simulations each. Mean optimisation time, $\overline{t}^{\text{opt}}(k)$.}
\label{fig:t_opt_t_mpc}
\end{figure}

In our simulation, the MPC step size dictates the interval at which control parameters are updated, as well as the interval at which the MPC is called.
In this sensitivity analysis, the MPC step size, $\Tflc$, is varied while the prediction horizon is maintained as $\Np = \Tflc + 15$.
Figure \ref{fig:obj_t_mpc} presents the instantaneous objective function, that, in contrast to the number of robots and the disaster environment size, is not normalised because the FLC has not this parameter (step size).
Instead, Figure \ref{fig:t_opt_t_mpc} shows the mean optimisation times for this sensitivity analysis.

\paragraph{Prediction Horizon} \label{sim_sens_k_pred}

\begin{figure}[t]
\centering
\includegraphics[width=.9\linewidth]{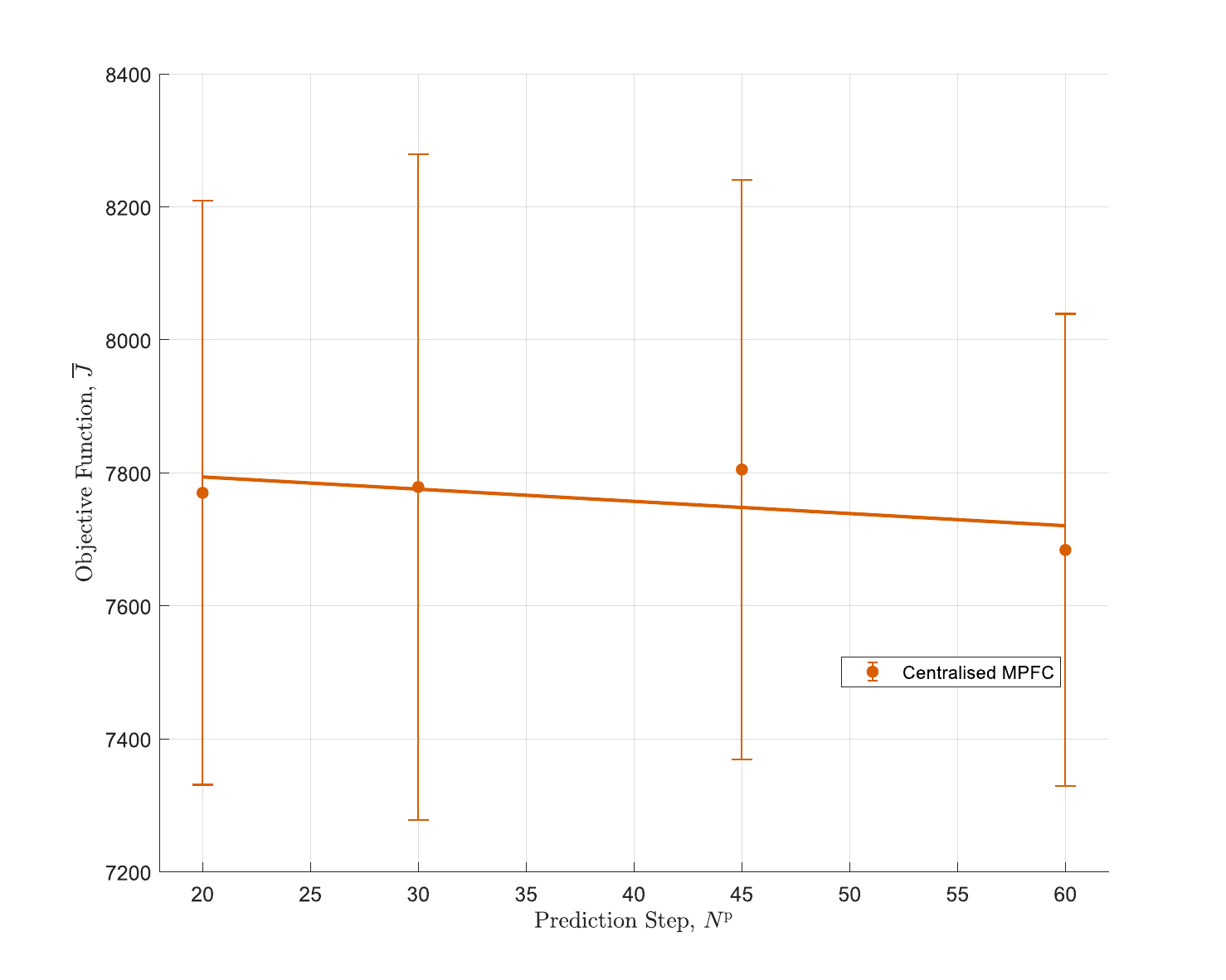}
\caption{Sensitivity of MPFC with prediction step, $\Np$, 5 simulations each. Mean objective function, $\overline{J}(k)$.}
\label{fig:obj_t_pred}
\end{figure}

\begin{figure}[t]
\centering
\includegraphics[width=.9\linewidth]{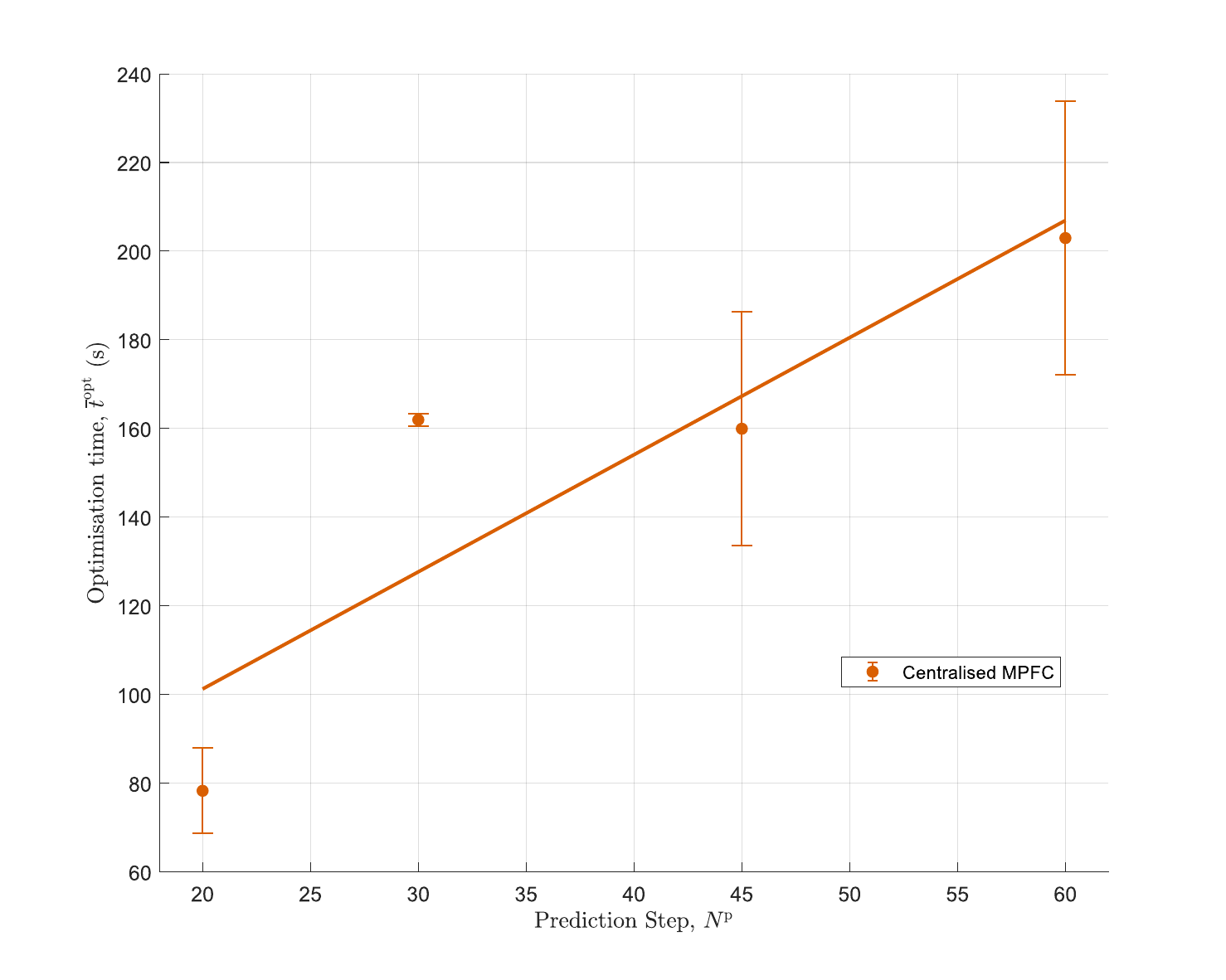}
\caption{Sensitivity of MPFC with prediction step, $\Np$, 5 simulations each. Mean optimisation time, $\overline{t}^{\text{opt}}(k)$.}
\label{fig:t_opt_t_pred}
\end{figure}

In this sensitivity analysis, the prediction horizon, $\Np$, is varied while the MPC step size is constrained to $\Tflc = 30$. 
This results in the prediction horizon defining how far beyond the control horizon the MPC or MPFC controller performs its prediction.
Figure \ref{fig:obj_t_pred} presents the instantaneous objective function (that is not normalised since the FLC has not the prediction horizon parameter) while Figure \ref{fig:t_opt_t_pred} shows the mean optimisation times for this sensitivity analysis.

\subsubsection{MPFC Design Exploration} \label{subsec:designExploration}
This section investigates potential design choices for MPFC, which may enhance its performance and computational efficiency or understand configuration options for the controller.

\paragraph{Prediction Modes} \label{sim_expl_prediction_mode}

\begin{figure}[t]
\centering
\includegraphics[width=.9\linewidth]{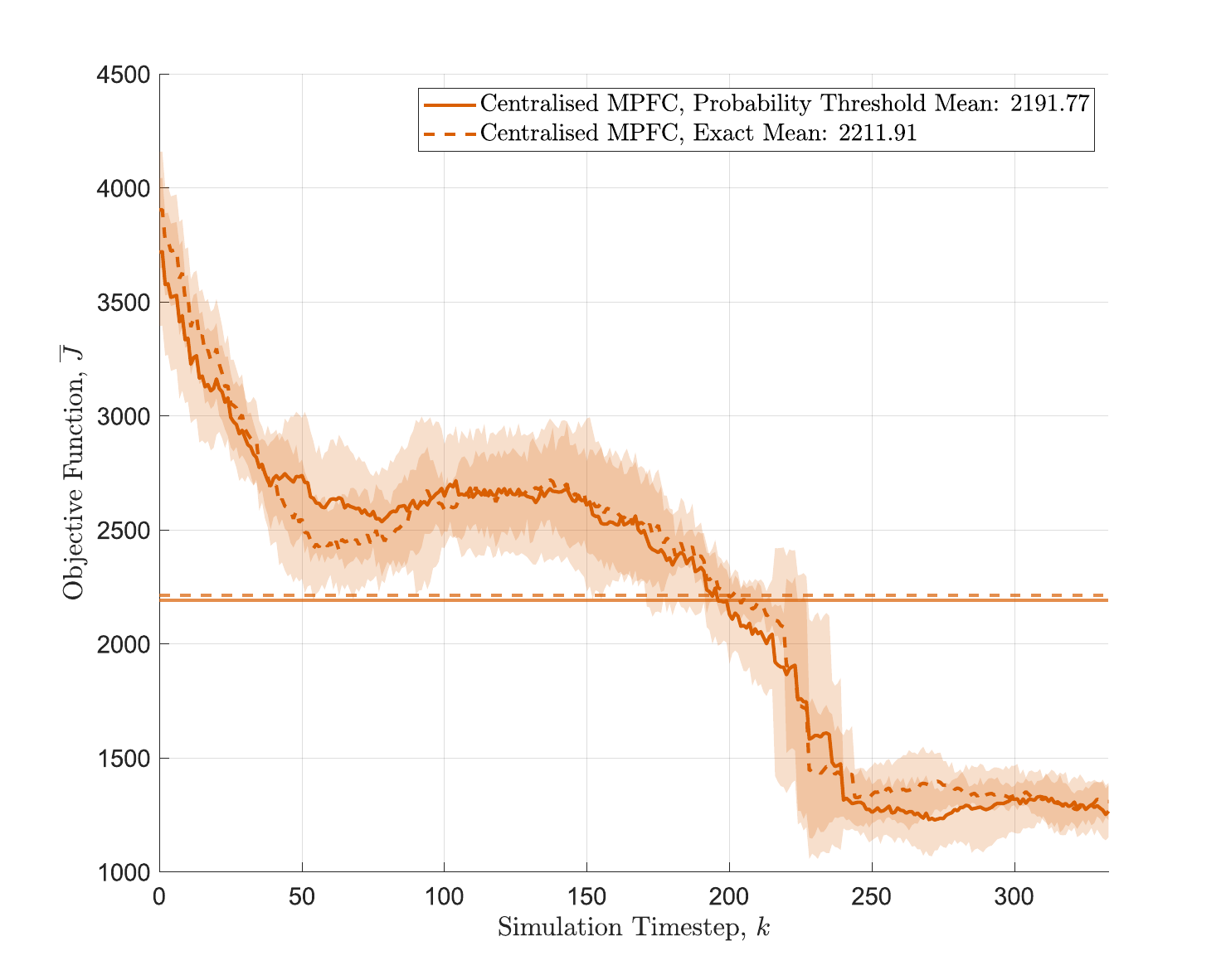}
\caption{$\overline{J}(k)$ for prediction modes, $\Tflc = 30$, 5 simulations.}
\label{fig:prediction_modes_obj}
\end{figure}

\begin{figure}[t]
\centering
\includegraphics[width=.9\linewidth]{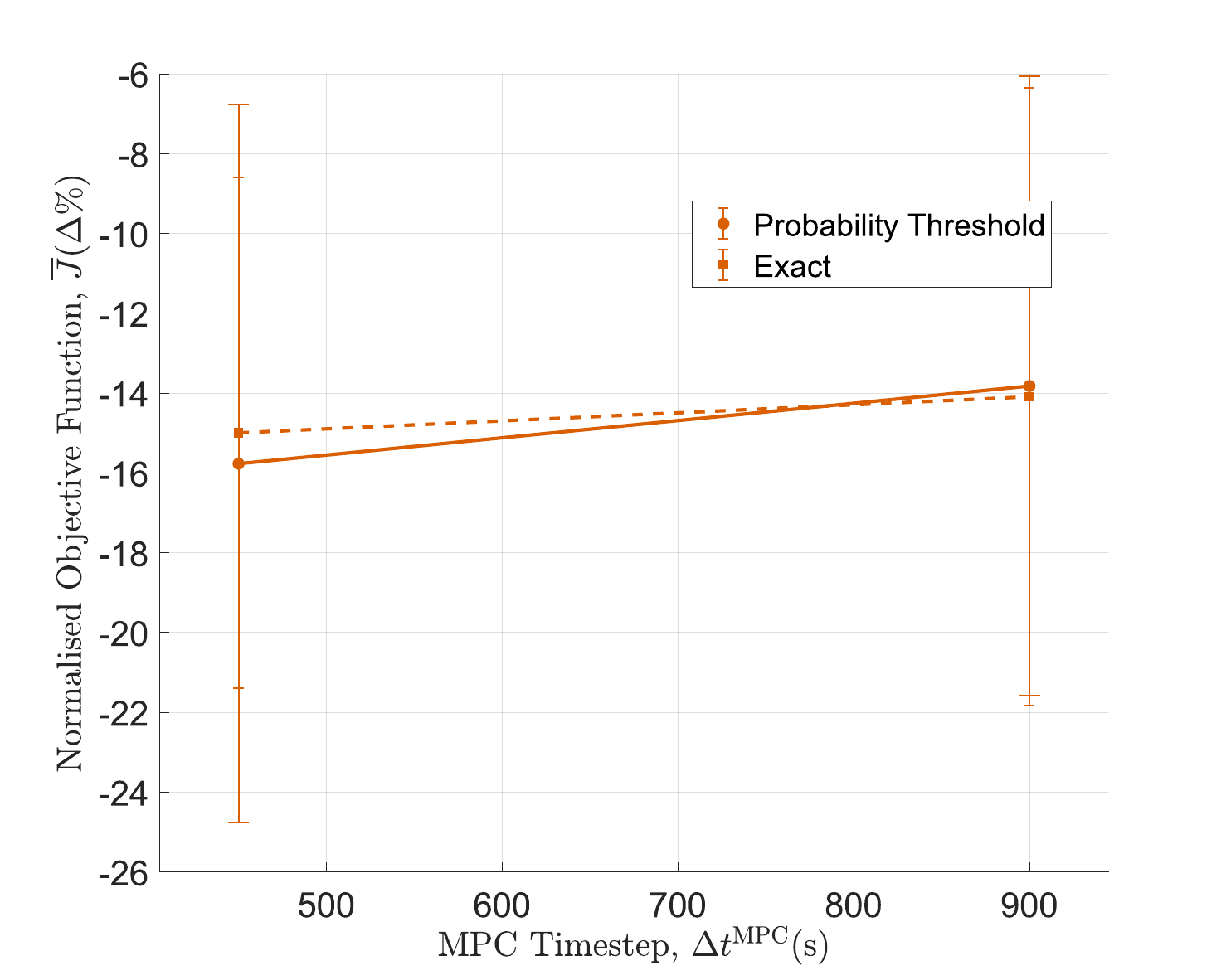}
\caption{Comparison of prediction modes: sensitivity of centralised MPFC with $t^{\text{MPC}}$, 5 simulations each.}
\label{fig:sens_prediction_mode}
\end{figure}

As mentioned in Section \ref{sec:comparison_approaches}, two prediction modes are implemented.
The \textit{probability threshold} prediction mode is used for all simulations in the analysis as we assume the controller cannot predict future probabilistic states perfectly.
In this simulation, we assess the performance of the controller with the probability threshold versus the \textit{exact} prediction mode for a two-robot system in a small disaster environment.
A large MPC step size is chosen to compare prediction modes as errors are more likely to accumulate over longer prediction horizons with the probability threshold prediction mode.
The performance for two MPFC controllers using each prediction mode is shown in Figure \ref{fig:prediction_modes_obj}.
To understand the relative difference in controller performance between these two prediction modes, a sensitivity analysis is run against $t^{\text{MPC}}$ for centralised MPFC in the range $[\SI{450}{\second}, \SI{900}{\second}]$.
Figure \ref{fig:sens_prediction_mode} shows the objective function evaluation for each prediction mode, normalised against the performance of a Pre-tuned FLC.

\paragraph{Type-1 vs Type-2 FLC} \label{sim_expl_flc}

\begin{figure}[t]
\centering
\includegraphics[width=.9\linewidth]{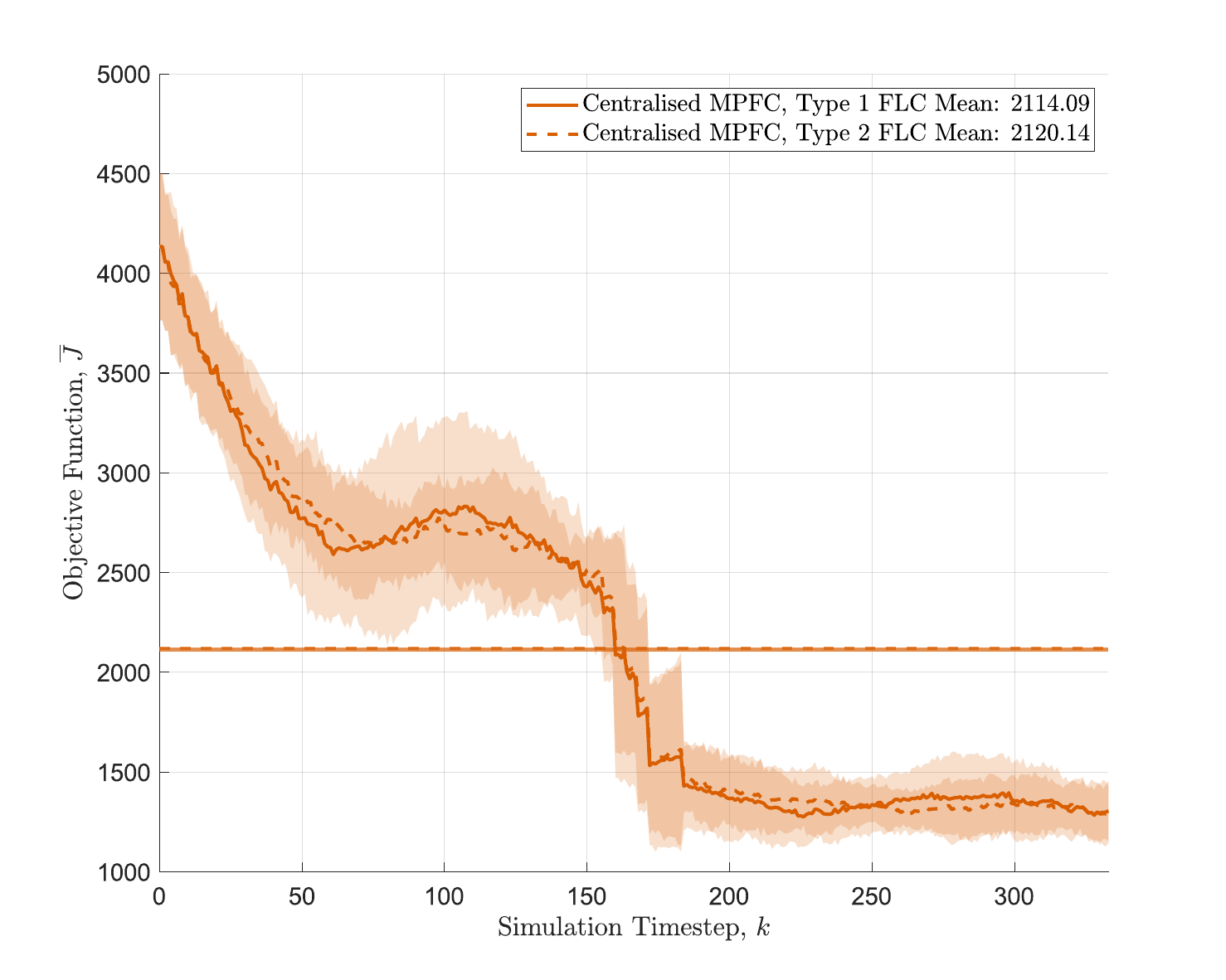}
\caption{$\overline{J}(k)$ for Type-1 vs Type-2 FLC in MPFC architecture, 5 simulations.}
\label{fig:flc_type_obj}
\end{figure}

\begin{figure}[t]
\centering
\includegraphics[width=.9\linewidth]{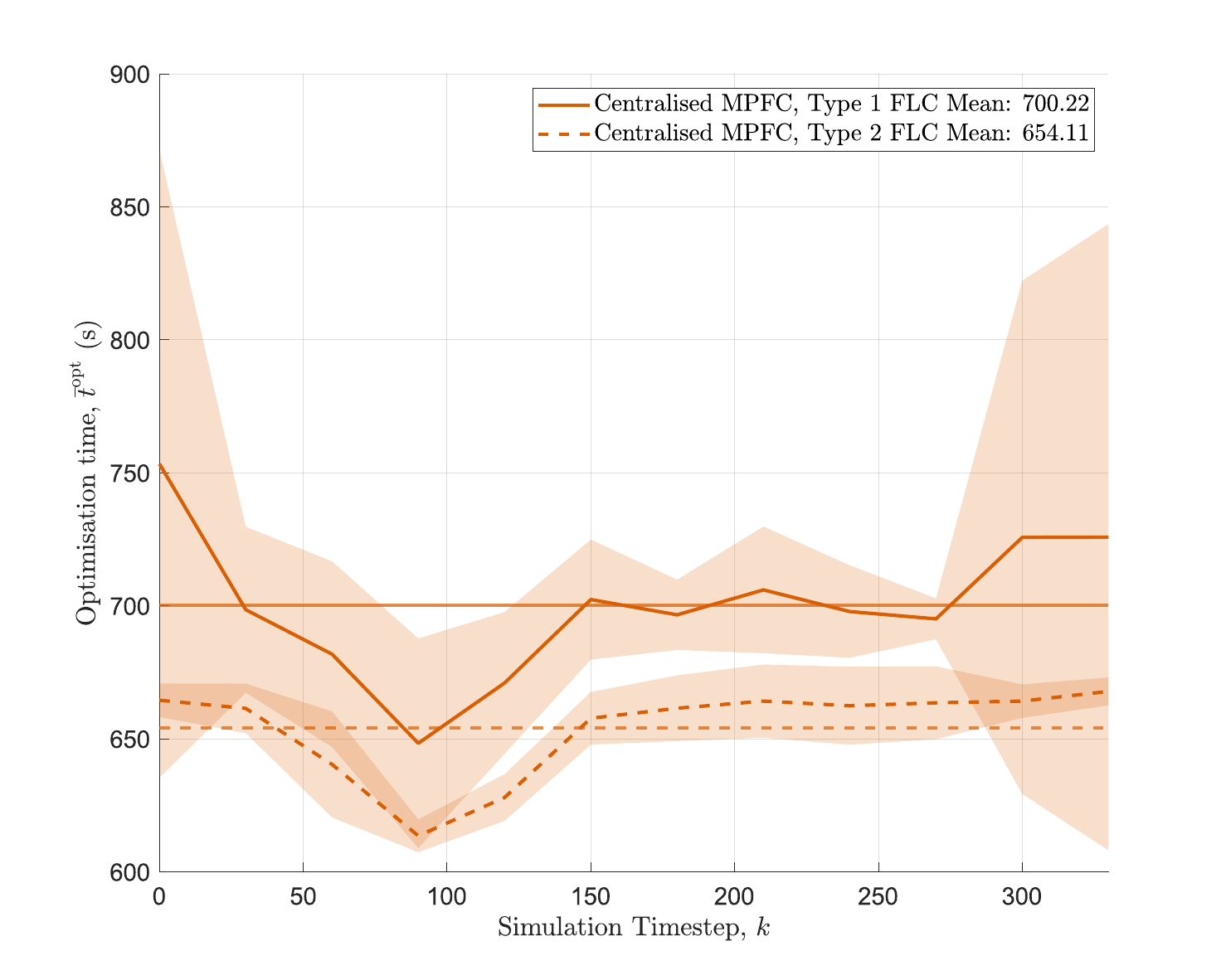}
\caption{Mean optimisation time for Type-1 vs Type-2 FLC in MPFC architecture, 5 simulations.}
\label{fig:flc_type_t_opt}
\end{figure}

In this simulation, we implement MPFC in a dynamic environment with a Type-2 TSK FLC and compare it against our previous formulation using a Type-1 FLC. 
Type-1 FLCs use crisp MFs where each input is associated with a single degree of membership. This simplicity allows for efficient computations and straightforward implementation. 
In contrast, Type-2 FLCs employ fuzzy MFs, which means that each input is associated with a range of degrees of membership, which can improve performance and robustness when faced with uncertain variables at the expense of more complex computations.
Figure \ref{fig:flc_type_obj} presents the normalised instantaneous objective function while Figure \ref{fig:flc_type_t_opt} shows the mean optimisation times.

\paragraph{Local Prediction Maps} \label{sim_expl_local_prediction_maps}

\begin{figure}[t]
\centering
\includegraphics[width=.9\linewidth]{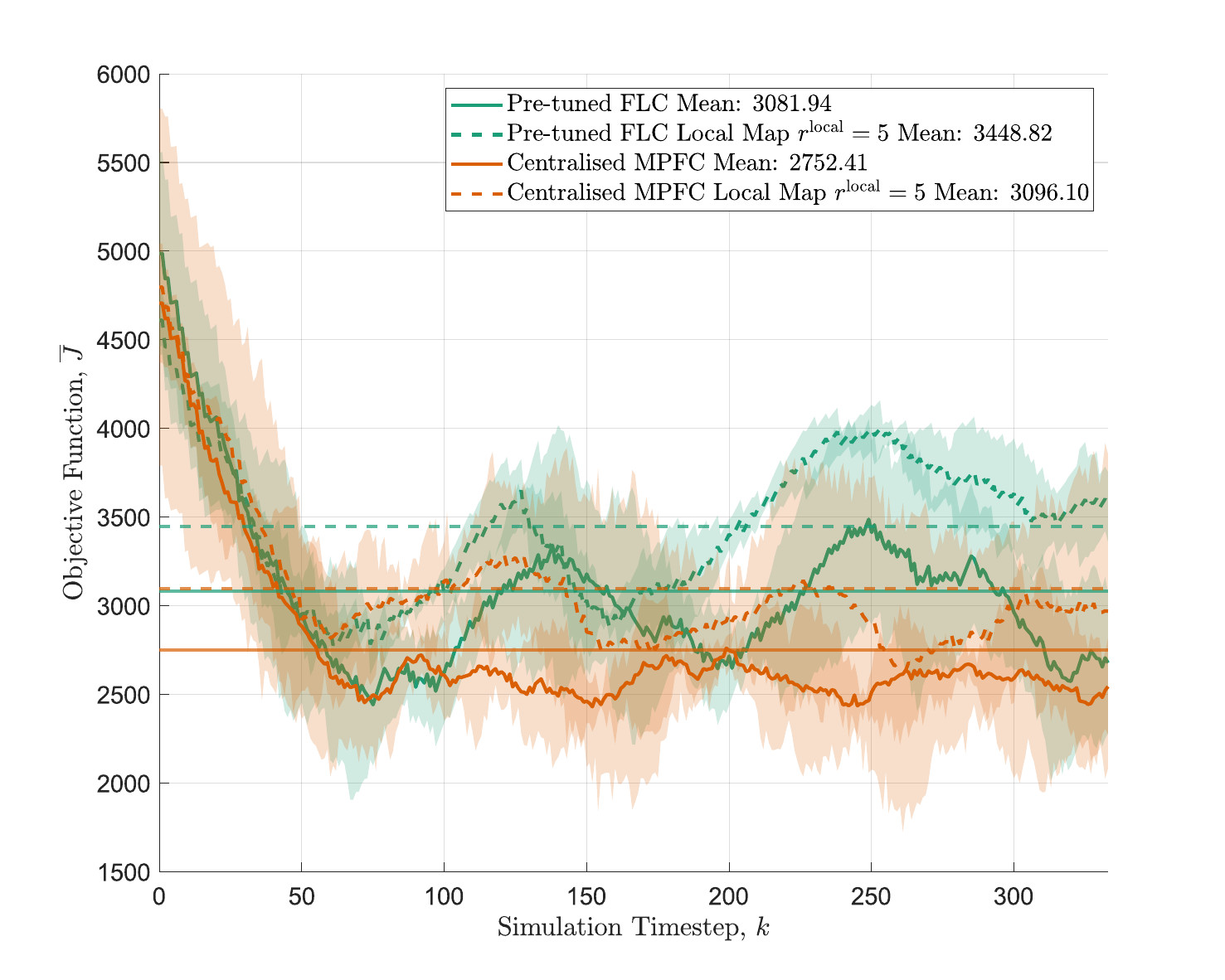}
\caption{$\overline{J}(k)$ with $r^{\text{local}}$, 3 simulations.}
\label{fig:obj_r_local_static_small}
\end{figure}

\begin{figure}[t]
\centering
\includegraphics[width=.9\linewidth]{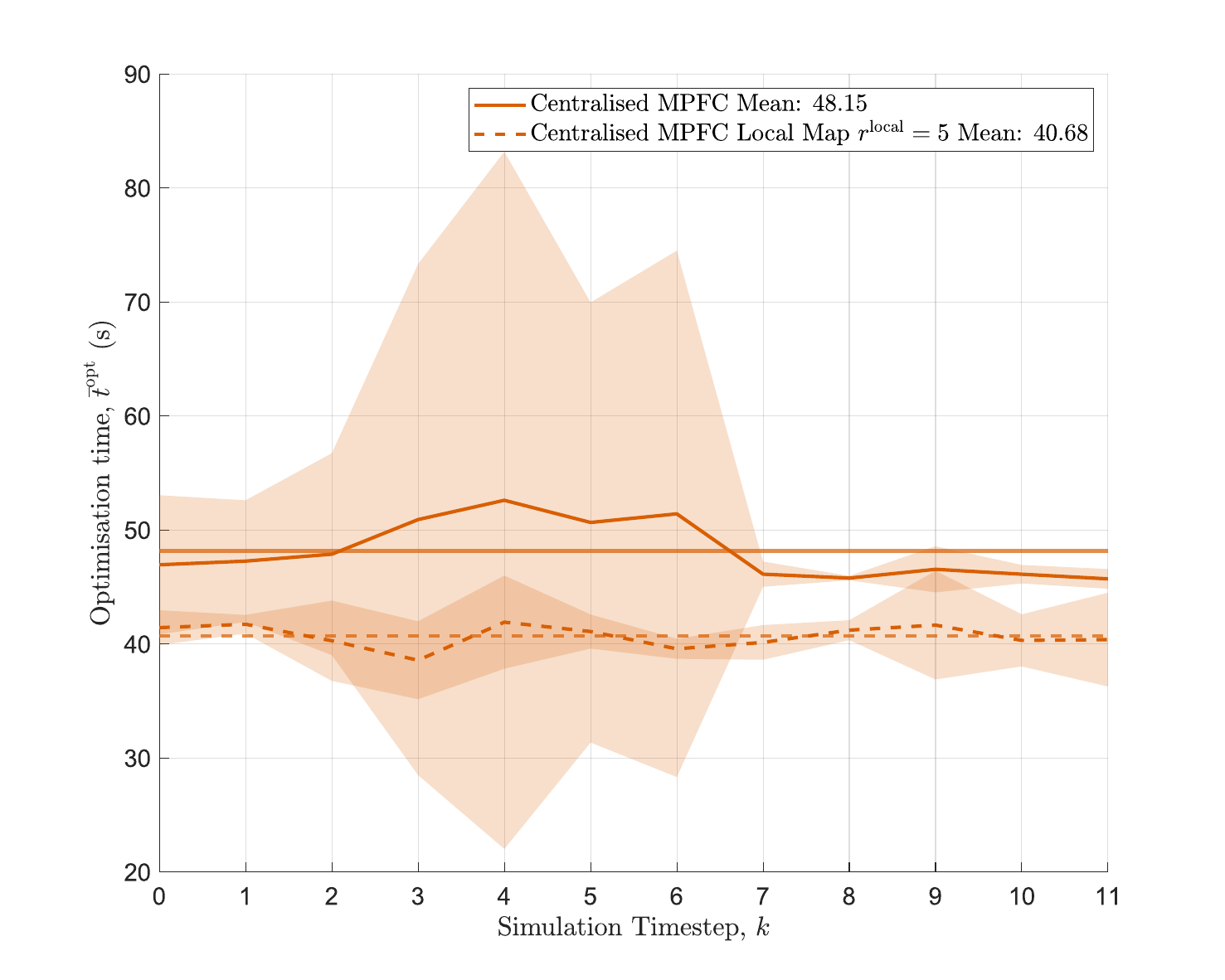}
\caption{Mean optimisation time with $r^{\text{local}}$.}
\label{fig:t_opt_r_local_static_small}
\end{figure}

\begin{figure}[t]
\centering
\includegraphics[width=.9\linewidth]{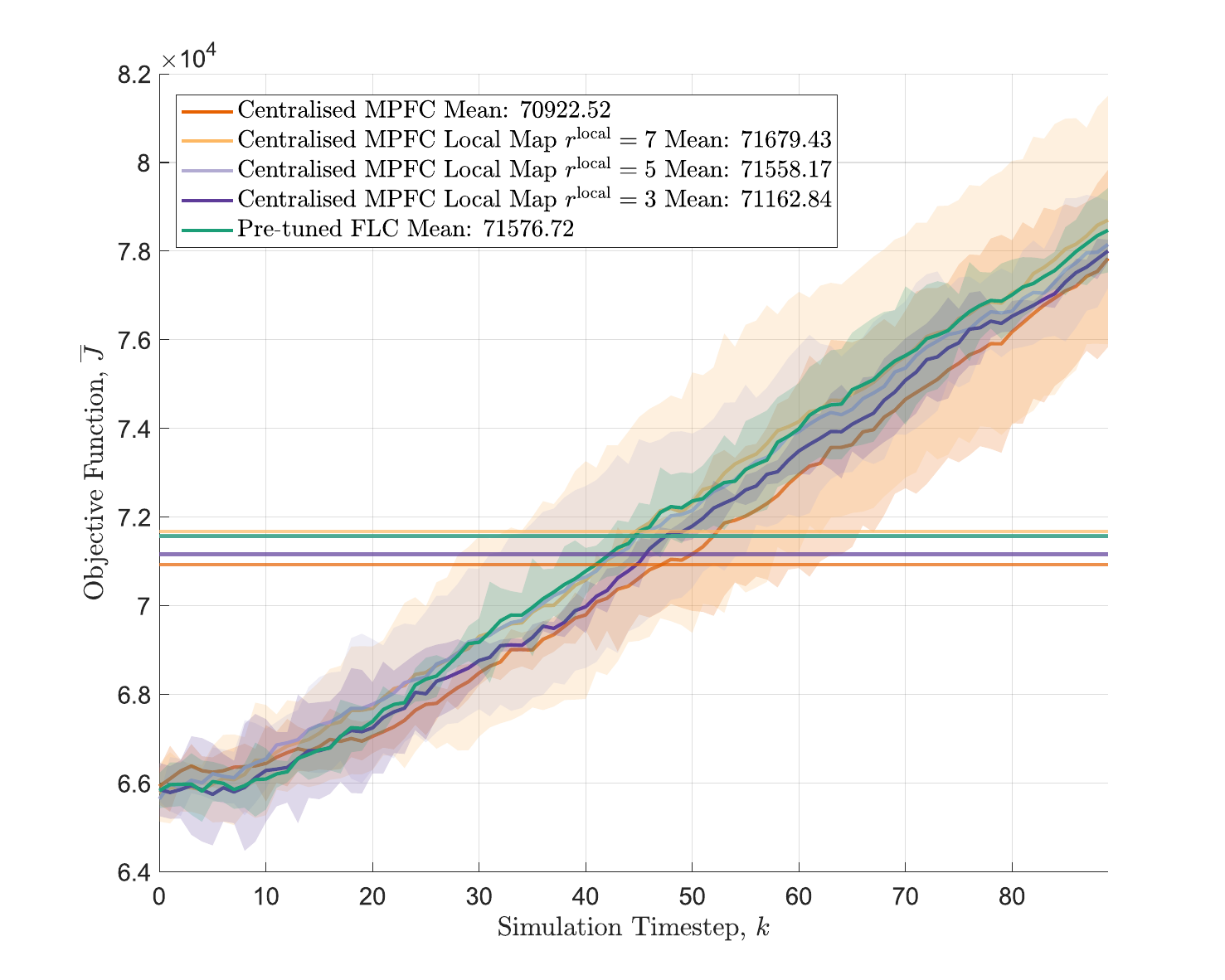}
\caption{$\overline{J}(k)$ with $r^{\text{local}}$, 3 simulations.}
\label{fig:obj_r_local}
\end{figure}

\begin{figure}[t]
\centering
\includegraphics[width=.9\linewidth]{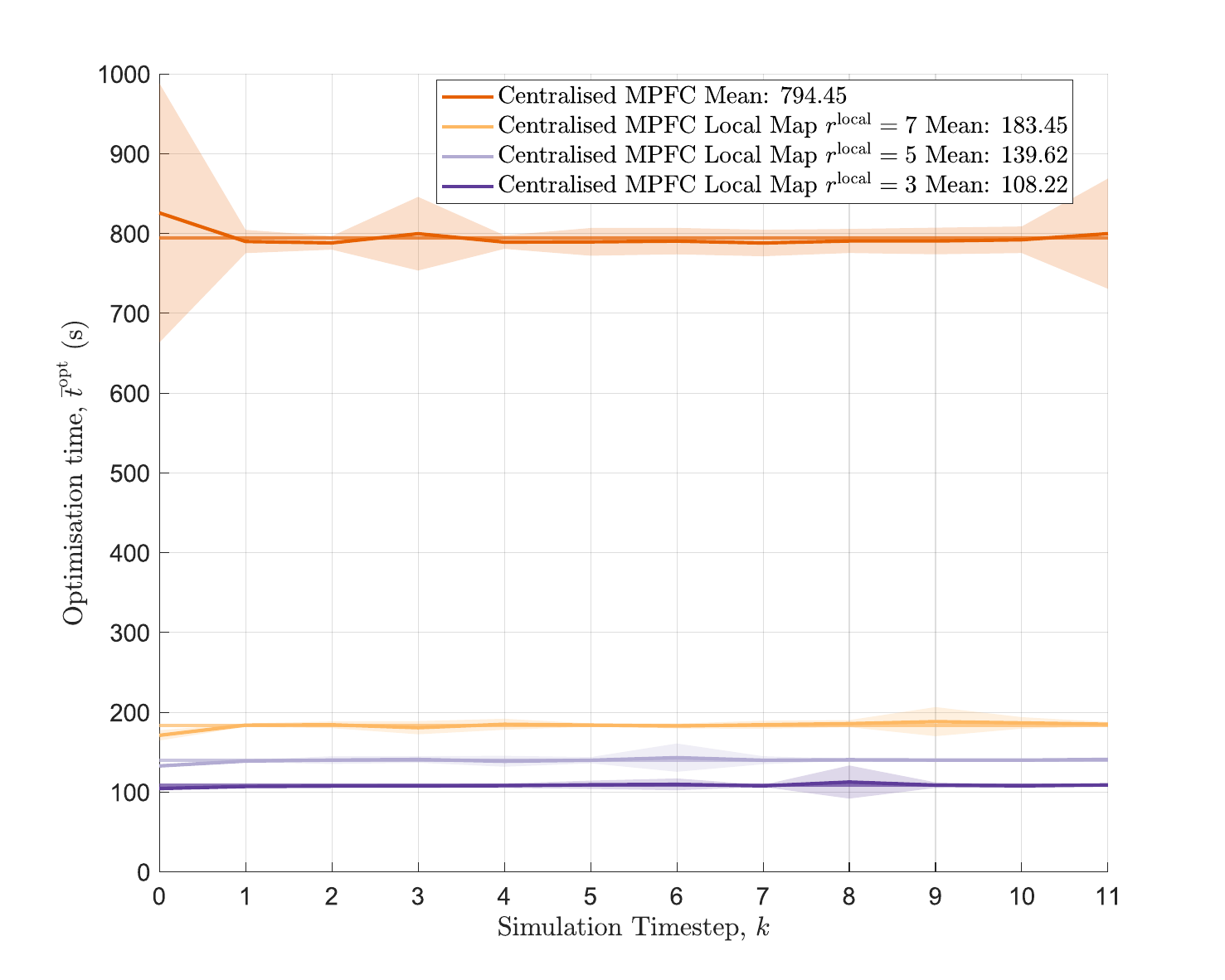}
\caption{Mean optimisation time with $r^{\text{local}}$, 3 simulations.}
\label{fig:t_opt_r_local}
\end{figure}

\begin{figure*}
\centering
\includegraphics[width=.8\linewidth]{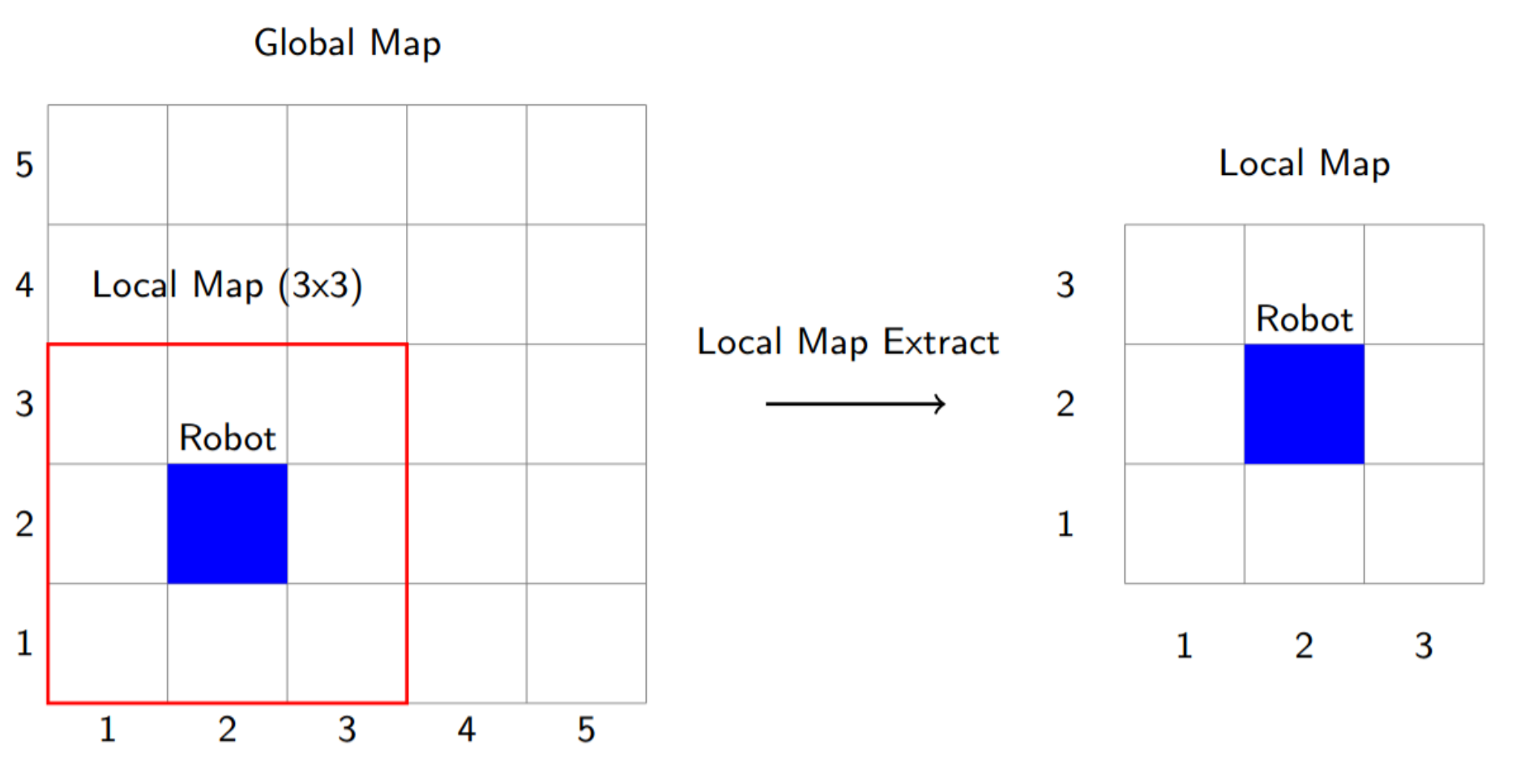}
\caption{Extraction of local map from global map, where $r^{\text{local}} = 1$.}
\label{fig:local_map_method}
\end{figure*}

As it can be seen from the sensitivity analysis of the disaster environment size, a significant weakness of MPFC is that the optimisation time is highly dependent on the environment and search map dimensions.
Due to the configuration of the control problem introduced in this case study, an robot will prioritise nearby cells over distant cells due to the lower response time to reach them.
Therefore, we can assume that the FLC does not need to consider cells outside a given radius, $r^{\text{local}}$, in the $x$ and $y$ axis around the robot.
To address the issue of long optimisation times for large environments, a \textit{local map} model is proposed to improve controller performance by limiting predictions to the local cells within a given radius of each robot, thereby reducing the computational complexity of the prediction step.
In this method, during the prediction step, a slice with dimensions [$2 \cdot r^{\text{local}} + 1$, $2 \cdot r^{\text{local}} + 1$] centred on the robot is taken for each coarsened matrix state.
A local attraction map is then calculated and restored to the global map before the robot target cell assignment.
Figure \ref{fig:local_map_method} illustrates the local map extraction.
With this method, the computational complexity of the MPFC prediction is effectively decoupled from the search map size.

First, the performance of the local map model is analysed for a two-robot system in the small static disaster environment of dimensions $\nxe = \nye = 20$.
We simulate MPFC and Pre-tuned FLC using both the global and local maps models with $r^{\text{local}} = 5$.
Figure \ref{fig:obj_r_local_static_small} presents the normalised instantaneous objective function while Figure \ref{fig:t_opt_r_local_static_small} shows the mean optimisation times.
After that, the next simulation focuses on exploring the performance using this method for a range of values of $r^{\text{local}}$ in a larger complex dynamic disaster environment, i.e., the third scenario.
MPFC is simulated using the local maps model for the range $r^{\text{local}} = [3, 5, 7]$.
Figure \ref{fig:obj_r_local} presents the normalised instantaneous objective function while Figure \ref{fig:t_opt_r_local} shows the mean optimisation times.

\subsection{Discussions} \label{sec:discussion}

\subsubsection{Two-robot system in small static disaster environment}

In the static scenario (see Figure \ref{fig:static_2_agents_obj}), the objective function starts around $5000$, then rapidly decreases as robots scan cells and stabilises once most cells have been observed.
This illustrates the effect of the sensor accuracy component of $\Mscan$ (see \eqref{eq:agentScanState}) in the objective function (see \eqref{eq:mpc_obj_func}), which has a high overall contribution to the objective function at the start of the simulation as all cells begin unscanned, and reaches a stable state when there is a balance between the rate that robots scan cells and the rate of degradation of $\Mscan$ due to the sensor accuracy component.
The Pre-tuned FLC demonstrates the poorest mean performance at $2925$, followed by the centralised MPFC with $2705$ (\SI{-7.5}{\percent}) and the centralised MPC with $2440$ (\SI{-16.6}{\percent}).
Note that the initial performance of the Pre-tuned FLC is very close to the centralised MPFC, suggesting that the initial tuning of the FLC is well-chosen.
Although centralised MPFC does not outperform centralised MPC, most of the difference in performance is during the first \SI{1000}{\second} of the simulation, where the centralised MPC is able to optimise the initial cells, while both stabilise around a similar objective function value over the remainder of the simulation.
Additionally, the confidence intervals for centralised MPFC are much tighter.
From Figure \ref{fig:static_2_agents_t_opt}, centralised MPFC has a mean of \SI{48}{\second}, almost half of the mean of \SI{93}{\second} of the centralised MPC.
The optimisation time remains stable for both control methods over the simulation.

\subsubsection{Two-robot system in small dynamic disaster environment}

From Figure \ref{fig:dynamic_2_agents_obj}, the effect of the fire propagation can be seen in the objective function, with active fires reaching a peak at around \SI{2000}{\second} and most fires burning out at around $\SI{3500}{\second}$.
In this case, the centralised MPC has a mean objective function of $2618$, followed by the Pre-tuned FLC with $2490$ (\SI{-4.9}{\percent}) and the centralised MPFC with $2375$ (\SI{-9.3}{\percent}). 
With the introduction of the uncertain fire spread component, the performance of the centralised MPC has degraded relative to the Pre-tuned FLC, while the centralised MPFC achieves the greatest performance.

From Figure \ref{fig:dynamic_2_agents_t_opt}, centralised MPFC has a mean optimisation time of \SI{56}{\second} while centralised MPC has as optimisation time of \SI{122}{\second} (+\SI{118}{\percent}).
The optimisation times for all control methods are higher due to the additional complexity of predicting fire spread. 
Unlike the static case, the optimisation time is highly variable for the centralised MPC, while the centralised MPFC maintains a stable optimisation time over the course of the simulation.

\subsubsection{Four-robot system in small dynamic disaster environment}

As shown in Figure \ref{fig:dynamic_4_agents_obj}, the Pre-tuned FLC demonstrates the poorest mean performance at $2176$, followed by the centralised MPC with $2044$ (\SI{-6.1}{\percent}) and the centralised MPFC with $1849$ (\SI{-15.0}{\percent}).
By introducing more robots in the same disaster environment, the objective function values are lower.
The robot actions become more closely coupled and the performance of the MPC and MPFC methods is improved relative to the Pre-tuned FLC, as they are able to optimise individual robot behaviours against the predicted future states.
Figure \ref{fig:dynamic_4_agents_t_opt} shows that the centralised MPFC has a mean optimisation time of \SI{97}{\second} while centralised MPC has as optimisation time of \SI{148}{\second} (+\SI{53}{\percent}).
Compared to the two-robot case, the centralised MPC optimisation time is closer to the centralised MPFC due to the greater increase in number of optimisation variables for the centralised MPFC versus the centralised MPC.

\subsubsection{Centralised vs decentralised MPFC for two-robot system}

From Figure \ref{fig:sim_perfo_architecture_n_a_2_obj}, the Pre-tuned FLC demonstrates the poorest mean performance at $2599$, followed by the decentralised MPFC with $2320$ (\SI{-10.7}{\percent}), and the centralised MPFC with $2275$ (\SI{-12.5}{\percent}).
In this case both architectures achieve a similar performance, but there is a trade-off between the number of optimisation variables due to the number of robots in the system.
Figure \ref{fig:sim_perfo_architecture_n_a_2_t_opt} shows that the centralised MPFC has a mean optimisation time of \SI{601}{\second} while decentralised MPFC has as optimisation time of \SI{604}{\second} (+\SI{0.5}{\percent}).

\subsubsection{Centralised vs decentralised MPFC for four-robot system}

As shown in Figure \ref{fig:sim_perfo_architecture_n_a_4}, the Pre-tuned FLC demonstrates the poorest mean performance at $2176$, followed by the centralised MPFC with $1849$ (\SI{-15.0}{\percent}) and the decentralised MPFC with $1807$ (\SI{-17.0}{\percent}).
Likewise, Figure \ref{fig:sim_perfo_architecture_n_a_4_t_opt} shows that the centralised MPFC has a mean optimisation time of \SI{97}{\second} while decentralised MPFC has as optimisation time of \SI{93}{\second} (\SI{-4}{\percent}).
In this case, the decentralised MPFC outperforms the centralised MPFC in both metrics despite the fact that robot behaviour is more coupled. 
This likely comes down to a trade-off between the degree of coupling between robot actions and the number of optimisation parameters which must be solved, which for centralised MPFC scales with the number of robots in the system.

\subsubsection{Two-robot system in complex dynamic disaster environment}

As shown in Figure \ref{fig:complex_2_agents_obj}, the Pre-tuned FLC demonstrates the poorest mean performance at $7181$, followed by the centralised MPFC with $6437$ (\SI{-10.4}{\percent}) and the decentralised MPFC with $6283$ (\SI{-12.5}{\percent}).
The centralised MPC achieves more stable performance over the entire simulation.
This may be due to the MPC having direct control over robot target cells, which allows it to always respond to active fires regardless of the positions of the fires or robots, while MPFC can only control robot actions indirectly via the output parameters of the FLC.
This highlights the importance of the design of the FLC.
For instance, by choosing alternative input parameters, it may be possible for MPFC to tune robots to prioritise certain geographical areas of the disaster environment.

\subsubsection{Sensitivity analysis: number of robots}

From Figure \ref{fig:obj_n_a}, all predictive controller configurations exhibit improved objective function performance relative to the Pre-tuned FLC with $\nr$.
As seen in the previous simulations, the decentralised MPFC architecture slightly outperforms the centralised MPFC architecture, and likewise for the decentralised and centralised MPC controller architectures.
All controller architectures demonstrate relatively similar trends with the number of robots.
For the full range of $\nr$ tested, centralised MPFC consistently outperforms centralised MPC by around \SI{10}{\percent} and decentralised MPFC consistently outperforms decentralised MPC by around \SI{5}{\percent}.
From Figure \ref{fig:t_opt_n_a}, the optimisation time for decentralised architectures remains stable with the number of robots,
however decentralised architectures also require $\nr$ separate optimisations to be performed instead of a single optimisation.
Therefore, if these optimisations can be performed in parallel they may be more efficient than if they must be performed on the same processor.
The confidence intervals also clearly show that the variability in optimisation times is far lower for the MPFC architectures than for the MPC architectures.

\subsubsection{Sensitivity analysis: disaster environment size}

From Figure \ref{fig:obj_env_size}, MPFC demonstrates around a fairly consistent \SI{10}{\percent} improvement over the Pre-tuned FLC, while the MPC performs better in large disaster environments than smaller ones (it should be noted that this happens because of the normalization, otherwise, the objective function increases when the number of cells increases, since the scenario is more complex).
The trend line for the MPC does not intersect the data point confidence intervals at $1600$ cells, so it may not be a strictly linear relationship.
The mean optimisation times for this sensitivity analysis (Figure \ref{fig:opt_time_env_size}) show that both have a linear correlation with the number of environment cells and do not demonstrate a clear difference in scalability between them.

\subsubsection{Sensitivity analysis: MPC step size}

From Figures \ref{fig:obj_t_mpc}-\ref{fig:t_opt_t_mpc}, the results appear to indicate roughly even performance in the range \SI{100}{\second} to \SI{300}{\second}, with the MPFC performance degrading when the MPC step size is increased further.
The figures demonstrate a strong linear correlation with optimisation time, although a decreased MPC time step will also require more frequent optimisations.
An optimal selection of $k^{\text{MPC}}$ may be to minimise it while selecting a value in the range that achieves the best overall objective function performance, possibly in the range \SI{100}{\second} to \SI{300}{\second} according to our results.

\subsubsection{Sensitivity analysis: prediction horizon}

Figure \ref{fig:obj_t_pred} shows no clear correlation between the global objective function with the prediction time step size for a strong linear correlation with the mean optimisation time (Figure \ref{fig:t_opt_t_pred}).
This suggests that extending the prediction horizon beyond the control horizon is of little value for the simulation implemented in this case study.

\subsubsection{Prediction modes}

From Figure \ref{fig:prediction_modes_obj}, the predicted objective function remains consistent across both prediction modes, with both results within the confidence interval of the other.
The probability threshold prediction mode performs slightly better, with a mean objective function of 2191.77 against the mean objective function of 2211.91 for the exact prediction mode.
The largest discrepancy between prediction modes is seen during the start of the simulation, when the most fire spread is occurring.
The exact prediction mode achieves a better overall performance during the first \SI{900}{\second} of the simulation, corresponding to the first two MPC steps.
From Figure \ref{fig:sens_prediction_mode}, both prediction modes achieve a relative decrease in objective function of around \SI{15}{\percent} when $t^{\text{MPC}} = \SI{450}{\second}$, degrading to around \SI{14}{\percent} when $t^{\text{MPC}} = \SI{900}{\second}$, however the confidence intervals remain wide.
The decrease in performance happens because when increasing the time, the error accumulates, therefore the performance decreases.
Overall, the performance of the probability threshold prediction mode correlates closely with the exact prediction mode, indicating it is a suitable estimation method for this simulation.

\subsubsection{Type-1 vs Type-2 FLC}

The results of the FLC type comparison show that there is no clear difference in performance between the Type-1 FLC and the Type-2 FLC for the selected simulation scenario.
A Type-2 FLC may be more suitable in scenarios where additional uncertain parameters are introduced or if the optimal system behaviour due to uncertain parameters were more complex.

\subsubsection{Local prediction maps}

The results in Figure \ref{fig:obj_r_local_static_small} show that in the small static disaster environment MPFC can maintain a similar performance using local maps to the Pre-tuned FLC using global maps, while the mean optimisation time is reduced from \SI{48.2}{\second} to \SI{40.7}{\second}.
The optimisation times are also far more consistent for the local maps model.
The results in this simulation validate the potential for local maps to reduce optimisation times without a significant decrease in the objective function.
The results for the large dynamic environment presented in Figure \ref{fig:obj_r_local} show
that the objective function increase continuously due to the size of the disaster environment compared to the number of robots and the propagation of the fire at the start of the simulation, resulting in an increasing objective function evaluated at each time step.
As expected, MPFC using global maps demonstrated the best mean performance of $70923$.
The remaining local map MPFC controllers achieve a performance of $71163$ for $r^{\text{local}} = 3$ (\SI{-0.3}{\percent}), and the other local map MPFC controllers performing worse than the Pre-tuned FLC.
These results, combined with the confidence interval ranges, mean there is no clear trend we can extract between the value of  $r^{\text{local}}$ and mean performance.
Despite the lack of correlation with $r^{\text{local}}$, all MPFCs using local maps demonstrated similar or improved performance than the Pre-tuned FLC with far lower optimisation times than MPFC using global maps, as shown in Figure \ref{fig:t_opt_r_local}.
These initial results indicate that this may be a feasible solution to implement MPFC which is scalable independent of the disaster environment size.

\section{Conclusion and recommendations for future work} \label{sec:conclusion}

In this paper we have introduced a novel hierarchical control formulation that includes a local FLC controller and a supervisory MPC controller for a multi-robot system in a disaster environment with fire spread.
The proposed approach has been tested in a case study compared to other controller types, MPC and Pre-tuned FLC, under different conditions of the environment and control parameters.
The results showcase that MPFC generally outperforms the other control approaches under various performance metrics.

As topics for future work, we may include uncertainties, due to the wind (velocity, direction or other), and therefore the need for an robust or stochastic MPC formulation.
Future research can focus on more accurate modelling and simulation of the disaster environment and robots and on the MPFC controller design.
The modelling can be improved by implementing accurate continuous 3-D models and by including models of other important states, disturbances, and uncertainties such as sensor models and measurement errors, environment geometry, and building damage.
Further research can also focus on the design of the MPFC controller, including the choice of inputs, outputs, membership functions, and rule base for the FLC controllers and the objective function, the optimisation solver, and using probability-based prediction models for the MPC controller.

\section{Acknoweledgement}

This research has been supported partially by the NWO Talent Programme Veni project ``Autonomous drones flocking for search-and-rescue'' (18120), 
which has been financed by the  Netherlands Organisation for Scientific Research (NWO), and partially by 
the TU Delft AI Labs \& Talent programme.

\section*{Data Availability}

The code used to perform the simulations of the case study is publicly available in the 4TU.ResearchData repository \cite{maxwell_dataset_2024}.

\bibliographystyle{cas-model2-names}

\bibliography{./bibliography}

\newpage

\appendix
\section{Appendix: example simulations} \label{app:examples}

\begin{figure*}
\centering
\includegraphics[width=0.5\linewidth]{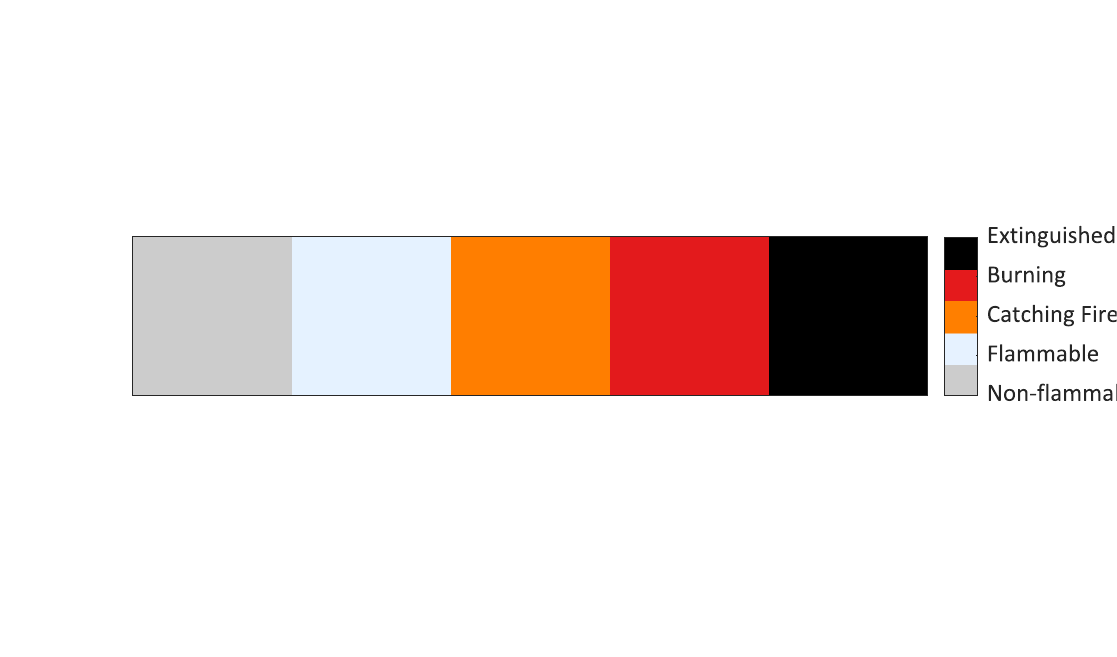}
\caption{Colour scheme for the states of the fire matrix $\Mf$}
\label{fig:fireMapLegend}
\end{figure*}

This appendix presents some example simulations for the fire spread and the motion of the robots based on the windspeed.
For visualisation of the fire states, the colour scheme in Figure~\ref{fig:fireMapLegend} is established.

\begin{figure*}
\centering
\includegraphics[width=.9\linewidth]{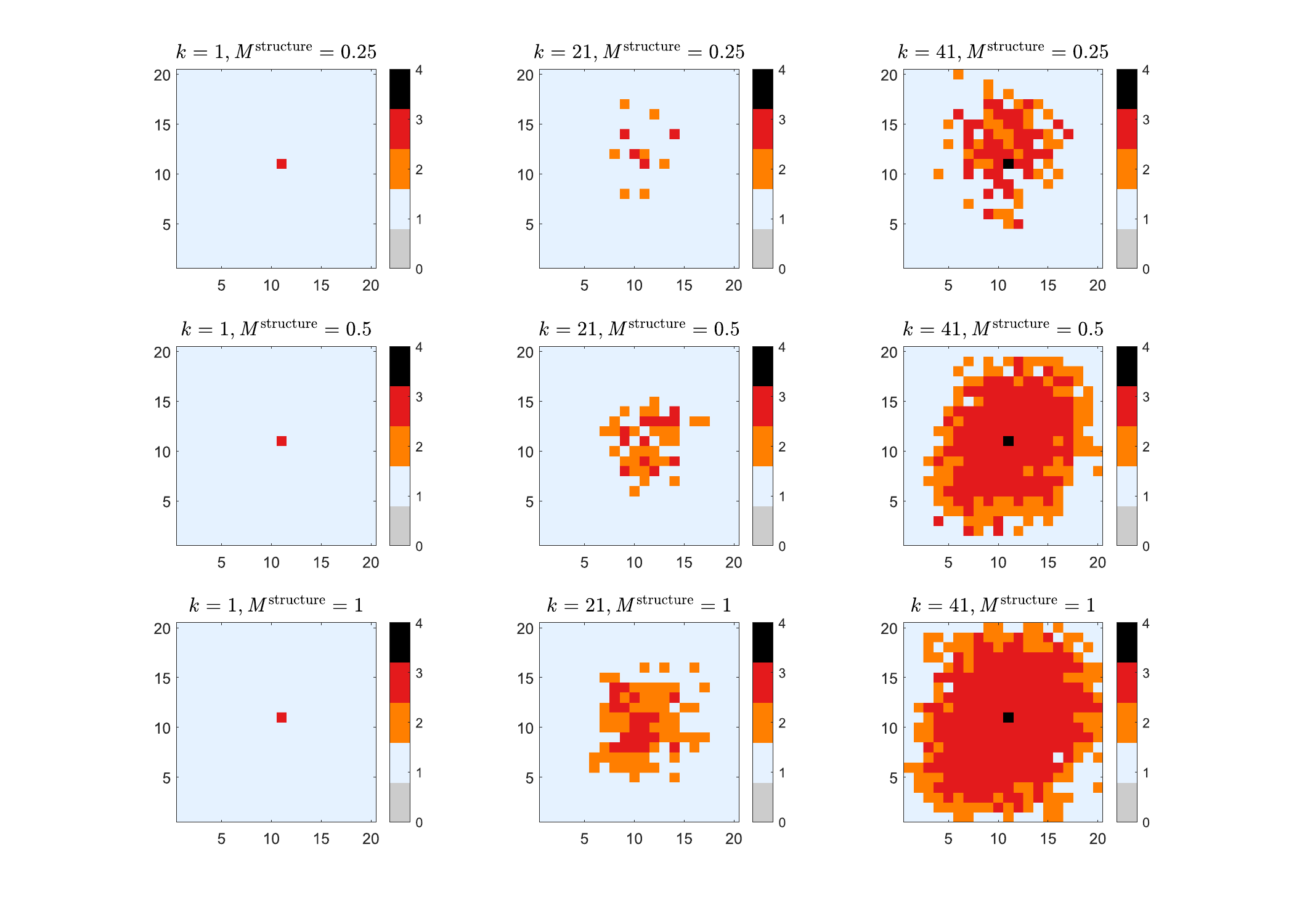}
\caption{Fire spread over time with $\Mstr$}
\label{fig:fire_spread_structures}
\end{figure*}

Figure \ref{fig:fire_spread_structures} displays the correlation between fire spread and $\Mstr$ for an example environment where $\mvel = \SI{0}{\meter \per \second}$, where cells with active fires are given as red and cells with have burnt out are given as black.
Higher values of $\Mstr$ greatly increase the speed at which the fire propagates.

\begin{figure*}
\centering
\includegraphics[width=.9\linewidth]{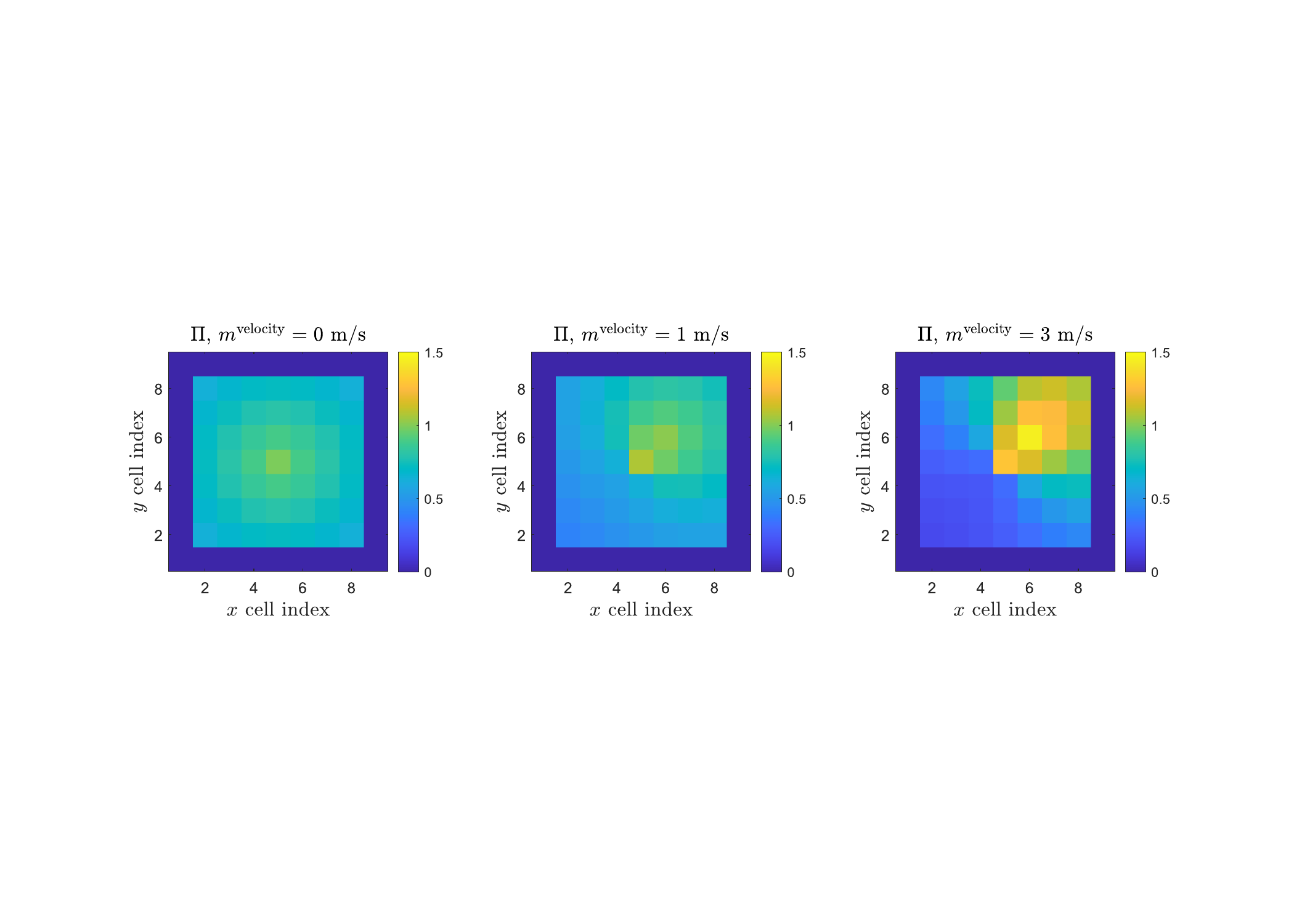}
\caption{Variation of $\pf$ with $\mvel$}
\label{fig:F_wind}
\end{figure*}

In Figure \ref{fig:F_wind}, an example of $\pf$ for a range of values of $\mvel$ is shown for a disaster environment of size $\nxe = \nye = 9$, where an active fire is initialised in the centre, $M^{\textrm{fire}}_{5, 5}(k) = 3$.
In this example, we set $r^{\text{wind}} = 3$ and $\mdir = \frac{\pi}{4}$.
For this example, we can see that there is zero possibility of the fire spreading further than the wind range from the active fire.

\begin{figure*}
\centering
\includegraphics[width=.9\linewidth]{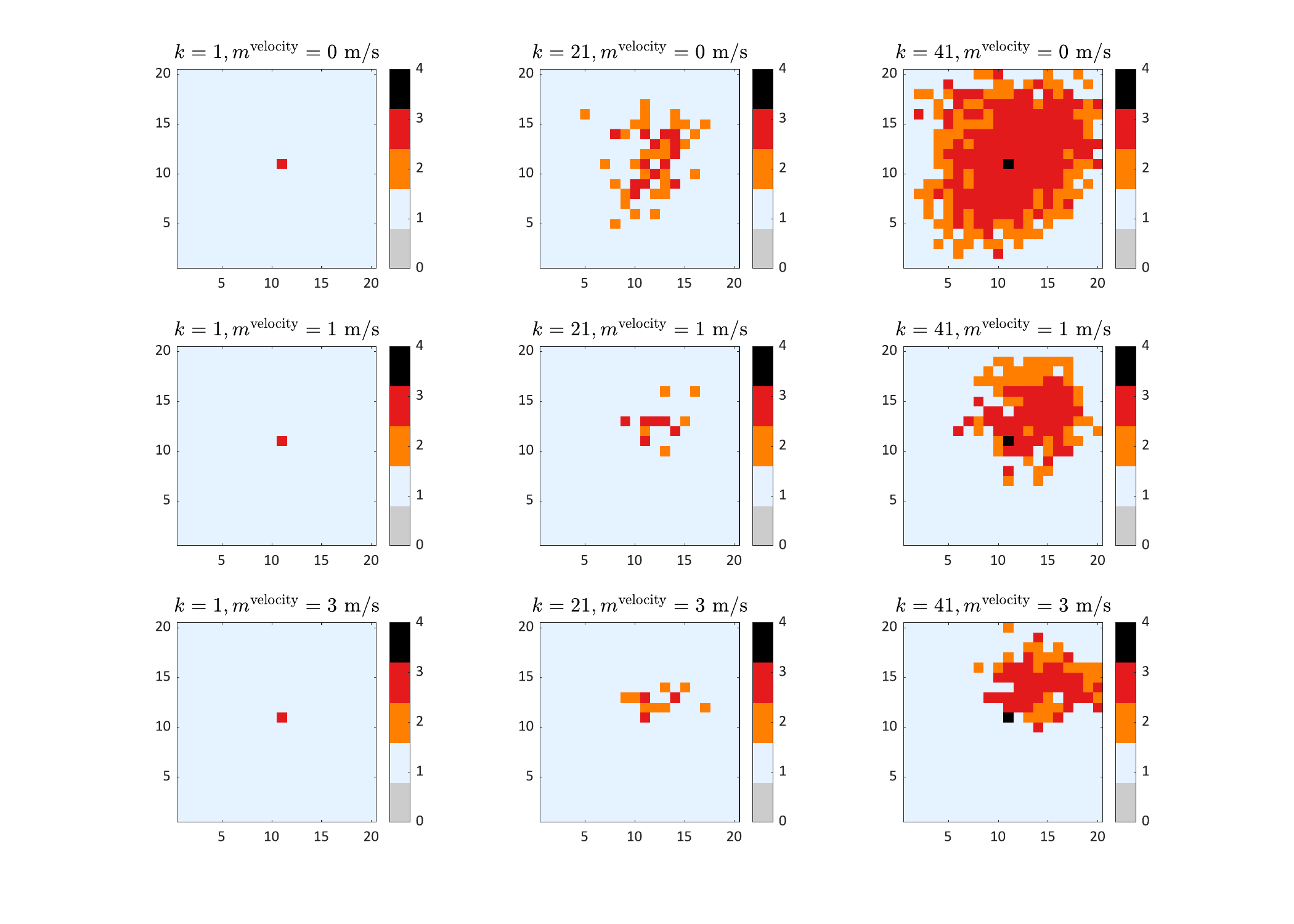}
\caption{Fire spread over time with $\mvel$}
\label{fig:fire_spread_wind}
\end{figure*}

Figure \ref{fig:fire_spread_wind} demonstrates the relationship between fire spread and wind velocity.
In this simulation, a fire is initiated in the centre of a $(20 \times 20)$ cell disaster environment with the wind direction set as $\mdir = \frac{\pi}{4} ~\textrm{rad}$ and $\mstr_{ij} = 0.2$ for $i, j = 1, \ldots, 20$.
It can be seen that the correlation between fire spread and wind direction is greater at higher wind velocities, however, the behaviour of this model may be tuned using the constants defined in ~\eqref{eq:fireSpreadProbability}.
In this example we set $c^{\text{fs1}} = 0.1$, $c^{\text{fs2}} = 1.2$, $c^{\text{wm1}} = 0.15$, $c^{\text{wm2}} = 1$, and $c^{\text{wmd}} = 1$.

\begin{figure*}
\centering
\includegraphics[width=.9\linewidth]{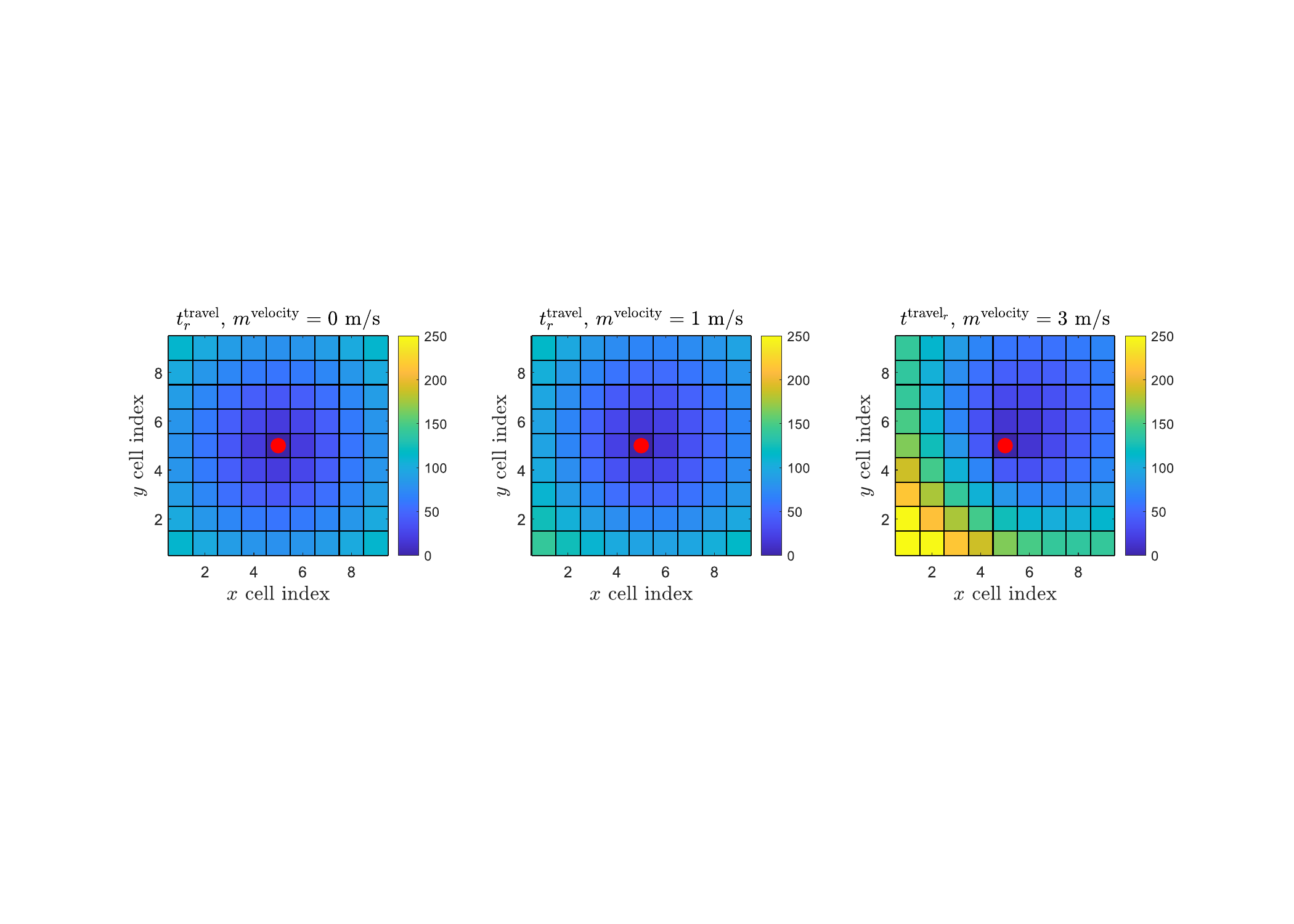}
\caption{Robot travel time, $t^{\textrm{travel}}_r$, with $\mvel$}
\label{fig:agent_travel_times}
\end{figure*}

Moreover, Figure \ref{fig:agent_travel_times} displays an example of the travel times of an robot to each cell in the disaster environment at differing wind velocities.
In this example $\va = \SI{5}{\meter \per \second}$, $ \mdir = \frac{\pi}{4} $, $\nxe = \nye = \SI{10}{\meter}$.
The robot is positioned in cell $(5,5)$, represented by the red circle.

\section{Appendix: algorithms} \label{app:algorithms}

This appendix presents the algorithms used for the computation of some variables in fire model, robot model, FLC and MPC.

Algorithm \ref{alg:fire_risk_time} shows how the fire time risk is computed.
The cell fire risk time, $t^{\text{risk}}_{ij}(k)$, depends on how close is the nearest cell with a growing fire ($\mf = 2$), that becomes a fully developed fire in $2~\textrm{min}$, considering the worst case, i.e., making the assumption that the fire will spread immediately in state $\mf = 3$ (which is in fact the case when $\pf$ is above the threshold), and based on the wind velocity.
The cell fire risk time matrix $T^{\text{risk}}$ is given by Algorithm \ref{alg:fire_risk_time}, where the neighbourhoods with radius 1, 2 and 3 are as in Figure~\ref{fig:ohgai_cells}, respectively. For farther cells, we compute the neighbourhood for radius 1, 2 or 3 and propagate it.
The fire risk time $t^{\text{risk}}_{ij}(k)$ is put to $100$ (a very large number) if the cell is unburnable ($\mf = 0$), so that it does not influence the attraction of those cells.

\begin{algorithm*} 
\caption{Fire time risk calculation}
\label{alg:fire_risk_time}
\begin{algorithmic}[1]
\State \textbf{Input:} Fire matrix $M^{\textrm{fire}}$, wind velocity $\mvel$
\State \textbf{Output:} Fire time risk matrix $T^{\text{risk}}$

\State Initialise $T^{\text{risk}} \gets 100$ for all cells

\For{each cell $(i,j)$}
    \For{$r \in \{1, 2, 3\}$}
        \State Get neighbours in radius $r$ around $(i,j)$
        \State Extract valid neighbours within grid bounds
        \State Apply fire rules:
        \If{$\mvel < 1 \frac{\textrm{m}}{\textrm{s}}$}
            \If{any neighbour in radius $1$ has $M^{\textrm{fire}}_{ij} = 3$} 
                \State $t^{\text{risk}}_{ij} \gets 0$
            \ElsIf{any neighbour in radius $1$ has $M^{\textrm{fire}}_{ij} = 2$} 
                \State $t^{\text{risk}}_{ij} \gets \min(t^{\text{risk}}_{ij}, 2~\text{min})$
            \ElsIf{any neighbour in radius $2$ has $M^{\textrm{fire}}_{ij} = 3$}
                \State $t^{\text{risk}}_{ij} \gets \min(t^{\text{risk}}_{ij}, 2~\text{min})$
            \ElsIf{any neighbour in radius $2$ has $M^{\textrm{fire}}_{ij} = 2$}
                \State $t^{\text{risk}}_{ij} \gets \min(t^{\text{risk}}_{ij}, 4~\text{min})$
            \ElsIf{any neighbour in radius $3$ has $M^{\textrm{fire}}_{ij} = 3$}
                \State $t^{\text{risk}}_{ij} \gets \min(t^{\text{risk}}_{ij}, 4~\text{min})$
            \ElsIf{any neighbour in radius $3$ has $M^{\textrm{fire}}_{ij} = 2$}
                \State $t^{\text{risk}}_{ij} \gets \min(t^{\text{risk}}_{ij}, 6~\text{min})$
            \EndIf
        \ElsIf{$1 \frac{\textrm{m}}{\textrm{s}} \leq \mvel < 5 \frac{\textrm{m}}{\textrm{s}}$}
            \If{any neighbour in radius $\leq 2$ has $M^{\textrm{fire}}_{ij} = 3$}
                \State $t^{\text{risk}}_{ij} \gets 0$
            \ElsIf{any neighbour in radius $\leq 2$ has $M^{\textrm{fire}}_{ij} = 2$}
                \State $t^{\text{risk}}_{ij} \gets \min(t^{\text{risk}}_{ij}, 2~\text{min})$
            \EndIf
        \ElsIf{$\mvel > 5 \frac{\textrm{m}}{\textrm{s}}$}
            \If{any neighbour in radius $\leq 3$ has $M^{\textrm{fire}}_{ij} = 3$}
                \State $t^{\text{risk}}_{ij} \gets 0$
            \ElsIf{any neighbour in radius $1$ has $M^{\textrm{fire}}_{ij} = 2$}
                \State $t^{\text{risk}}_{ij} \gets \min(t^{\text{risk}}_{ij}, 2~\text{min})$
            \EndIf
        \EndIf
    \EndFor
\EndFor
\end{algorithmic}
\end{algorithm*}

Algorithm \ref{alg:fire_spread_probability} is used to calculate the fire spread probability $\pf$ for all active fires at each time step $k$ and consolidate them into a single fire spread probability map parameter, $\Pi$, where the matrices $O$, $F$, $\Delta$, $V$, $\Psi$ include, respectively, all the elements $o_{ij}$, $f_{ij}$, $\delta \big( (l, q), (i, j) \big)$, $\vij$ and $\psiij$ of the environment.
\begin{algorithm*}
\caption{Update fire spread probability}
\label{alg:fire_spread_probability}
\begin{algorithmic}[1]
\State \textbf{Input:} $\Pi$, $\alpha_1$, $\alpha_2$, $\alpha_3$, $\alpha_4$, $\Mstr$, $O$, $F$, $\Delta$, $V$, $\Psi$, $\nxe$, $\nye$, $r^{\text{wind}}$
\State \textbf{Output:} $\Pi$
\For {each active fire cell $(i, j)$ in the environment}
    \State Compute row range: $R^{\text{row}} \leftarrow [ \max(1, i - r^{\text{wind}}), \min(\nxe, i + r^{\text{wind}}) ] $
    \State Compute column range: $R^{\text{col}} \leftarrow [ \max(1, j - r^{\text{wind}}) : \min(\nye, j + r^{\text{wind}}) ]$
    \State Update fire spread probability: $\Pi(R^{\text{row}}, R^{\text{col}}) \leftarrow \Pi(R^{\text{row}}, R^{\text{col}}) + \alpha_1 \cdot \Mstr(R^{\text{row}}, R^{\text{col}}) \cdot O(R^{\text{row}}, R^{\text{col}}) \cdot f_{ij} \cdot$ 
    \State \qquad $\exp\Big(\alpha_2 \Delta(R^{\text{row}}, R^{\text{col}}) + \alpha_3 V(R^{\text{row}}, R^{\text{col}}) + \alpha_4 V(R^{\text{row}}, R^{\text{col}}) \cos \big( \Psi(R^{\text{row}}, R^{\text{col}}) \big) \Big)$
\EndFor
\end{algorithmic}
\end{algorithm*}

The action algorithm for robot $r$, i.e., the algorithm which executes the robot tasks, operates as described in Algorithm~\ref{alg:agent_action_flc}.
There, \( \arg \max A_r \) returns the index of the cell with the maximum value in the attraction matrix \( A_r \), which contains all the attraction values $a_{(\tau_x,\tau_y)_r}$; a robot task $\sigma_r = 0$ represents ``travel'' while a robot task $\sigma_r = 1$ represents ``scan''; $T$ represents the simulation time step; $t^{\text{scan:}(\tau_x,\tau_y)_{r,k}}$ is the scan time for cell $(\tau_x,\tau_y)_{r,k}$.
Instead, for MPC, the algorithm operates as described in Algorithm~\ref{alg:agent_action_mpc}.

\begin{algorithm}
\caption{Robot action algorithm for MPFC/FLC}
\label{alg:agent_action_flc}
\begin{algorithmic}[1]
\Require $A_r(k)$, $(\tau_x,\tau_y)_{r,k}$, $\sigma_r(k)$, $t_{r}^{\text{scan}}(k)$, $t_{r}^{\text{travel}}(k)$, $k$
\If{$\sigma_r(k) = 0$ and $t_{r}^{\text{travel}}(k) > 0$}
    \State $t_{r}^{\text{travel}}(k+1) \gets t_{r}^{\text{travel}}(k) - T$
\ElsIf{$\sigma_r(k) = 0$ and $t_{r}^{\text{travel}}(k) \leq 0$}
    \State $\sigma_r(k) \gets 1$
    \State $(i, j)_{r,k} \gets (\tau_x,\tau_y)_{r,k}$
    \State $t_{r}^{\text{travel}}(k) \gets t^{\text{travel}} \big( (i, j)_r, (\tau_x,\tau_y)_{r,k,q} \big)$
\ElsIf{$\sigma_r(k) = 1$ and $t_{r}^{\text{scan}}(k) > 0$}
    \State $t_{r}^{\text{scan}}(k+1) \gets t_{r}^{\text{scan}}(k) - T$
\ElsIf{$\sigma_r(k) = 1$ and $t_{r}^{\text{scan}}(k) \leq 0$}
    \State $\sigma_r(k) \gets 0$
    \State $(\tau_x,\tau_y)_{r,k} \gets \arg \max A_r(k)$
    \State $t_{r}^{\text{scan}}(k) \gets t^{\text{scan:}(\tau_x,\tau_y)_{r,k}}$
\EndIf
\end{algorithmic}
\end{algorithm}

\begin{algorithm}
\caption{Robot action algorithm for MPC}
\label{alg:agent_action_mpc}
\begin{algorithmic}[1]
\Require $(\tau_x,\tau_y)_{r,k}$, $\sigma_r(k)$, $t_{r}^{\text{scan}}(k)$, $t_{r}^{\text{travel}}(k)$, $q=1$, $k$
\If{$\sigma_r(k) = 0$ and $t_{r}^{\text{travel}}(k) > 0$}
    \State $t_{r}^{\text{travel}}(k+1) \gets t_{r}^{\text{travel}}(k) - T$
\ElsIf{$\sigma_r(k) = 0$ and $t_{r}^{\text{travel}}(k) \leq 0$}
    \State $\sigma_r(k) \gets 1$
    \State $(i, j)_{r,k} \gets (\tau_x,\tau_y)_{r,k}(q)$
    \State $q \gets q + 1$
    \State $t_{r}^{\text{travel}} \gets t^{\text{travel}} \big( (i, j)_r, (\tau_x,\tau_y)_{r,k,q} \big)$
\ElsIf{$\sigma_r(k) = 1$ and $t_{r}^{\text{scan}}(k) > 0$}
    \State $t_{r}^{\text{scan}}(k+1) \gets t_{r}^{\text{scan}}(k) - T$
\ElsIf{$\sigma_r(k) = 1$ and $t_{r}^{\text{scan}}(k) \leq 0$}
    \State $q = q + 1$
    \State $\sigma_r(k) \gets 0$
    \State $t_{r}^{\text{scan}}(k) \gets t^{\text{scan-}(\tau_x,\tau_y)_{r,k}}$
\EndIf
\end{algorithmic}
\end{algorithm}

Lastly, Algorithm \ref{alg:downwind_map} shows how the downwind matrix is computed.

\begin{algorithm*}
\caption{Calculate downwind matrix}
\label{alg:downwind_map}
\begin{algorithmic}[1]
\Require $\bm{M}^{\text{fire}}$, $\nxe$, $\nye$
\Ensure $\bm{M}^{\text{downwind}}$
\State Initialise $\bm{M}^{\text{downwind}}$ as a zero matrix of size $\nxe \times \nye$
\For{$i = 1$ to $\nxe$}
    \For{$j = 1$ to $\nye$}
        \If{$\bm{m}_{ij}^{\text{fire}} = 2$ or $\bm{m}_{ij}^{\text{fire}} = 3$} \Comment{Active or burning fire}
            \State Calculate $\bm{m}^{\text{downwind direction}}$ and $\bm{m}^{\text{downwind distance}}$ using Equation~\ref{eq:m_downwind_dir} and Equation~\ref{eq:m_downwind_dist}
            \State $\bm{m}^{\text{downwind}} = \max\left(\bm{m}^{\text{downwind}}, \bm{m}^{\text{downwind direction}} \cdot \bm{m}^{\text{downwind distance}} \right)$
        \EndIf
    \EndFor
\EndFor
\State $\bm{M}^{\text{downwind}} = 1 - \bm{M}^{\text{downwind}}$ \Comment{Invert the downwind effect}
\State \Return $\bm{M}^{\text{downwind}}$
\end{algorithmic}
\end{algorithm*}

\section{Appendix: scenarios} \label{app:scenarios}

This appendix presents the three scenarios used in the case study.

The first scenario consists of a basic small disaster environment, where $\nxe = \nye = 40$. 
This initialisation is scaled appropriately for the following larger disaster environments unless otherwise noted.

\begin{figure*}
\centering
\includegraphics[width=0.8\linewidth]{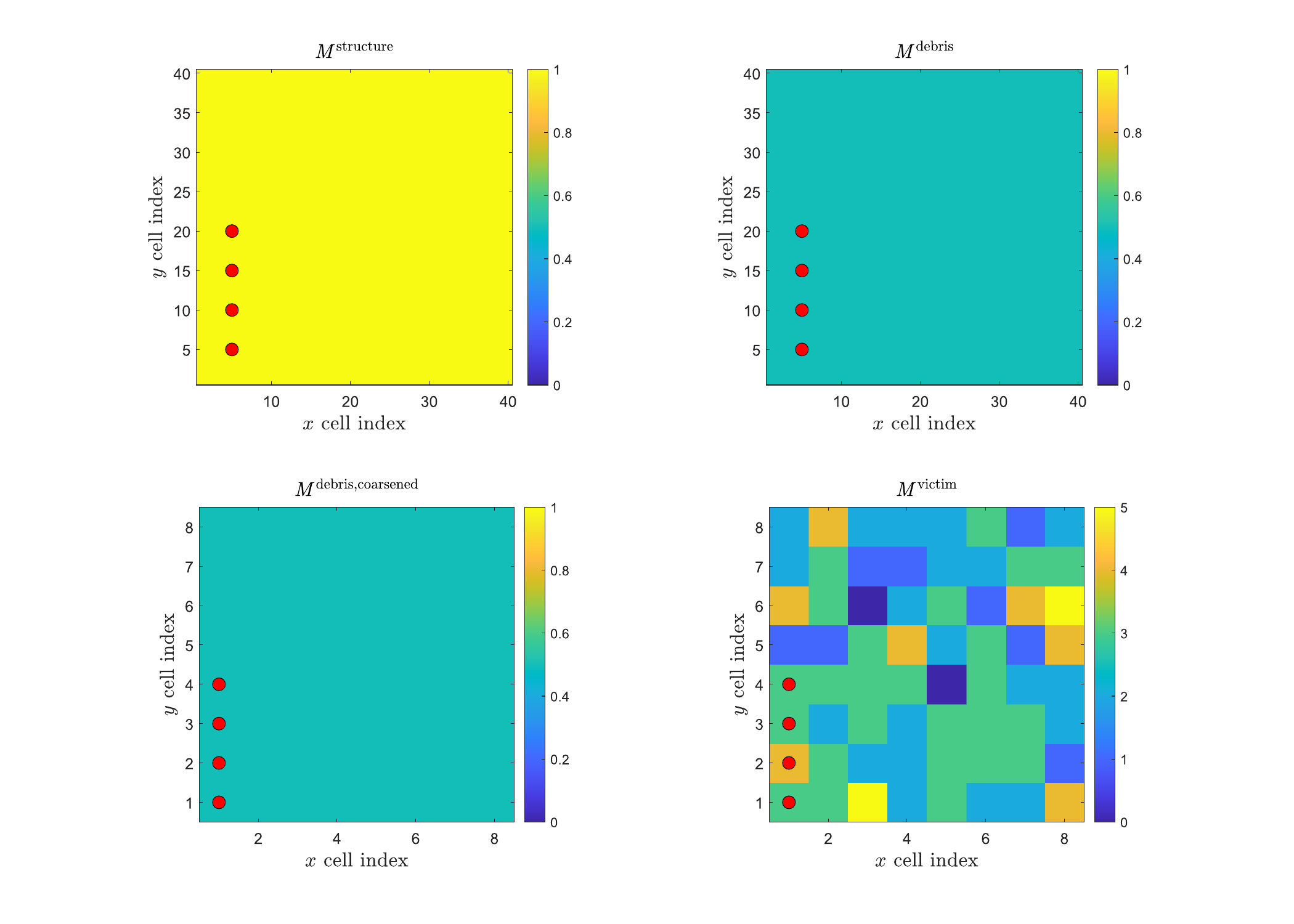}
\caption{Single instance of environment setup for basic small disaster environment, red dots indicate robot positions. 
Each subplot represents the value of the labelled environmental state variable for the corresponding cell, from the lower limit (dark blue) to upper limit (yellow) of the parameter range.}
\label{fig:env_states_small_static}
\end{figure*}

Figure \ref{fig:env_states_small_static} shows the initialisation of the environment map parameters for the first scenario, i.e., the basic small environment with two robots.
As shown, uniform distributions for $\Mstr$ and $\Mdebris$ are defined.
The figure shows one instance of the coarsened victim matrix, $\Mvictim$, where it can be seen that the distribution of victims is random over the disaster environment due to the uniform debris occupancy matrix, however, this parameter varies with each simulation seed.
Robots are initialised in a row of adjacent coarsened maps cells in the bottom left hand corner of the disaster environment.

\begin{figure*}
\centering
\includegraphics[width=\linewidth]{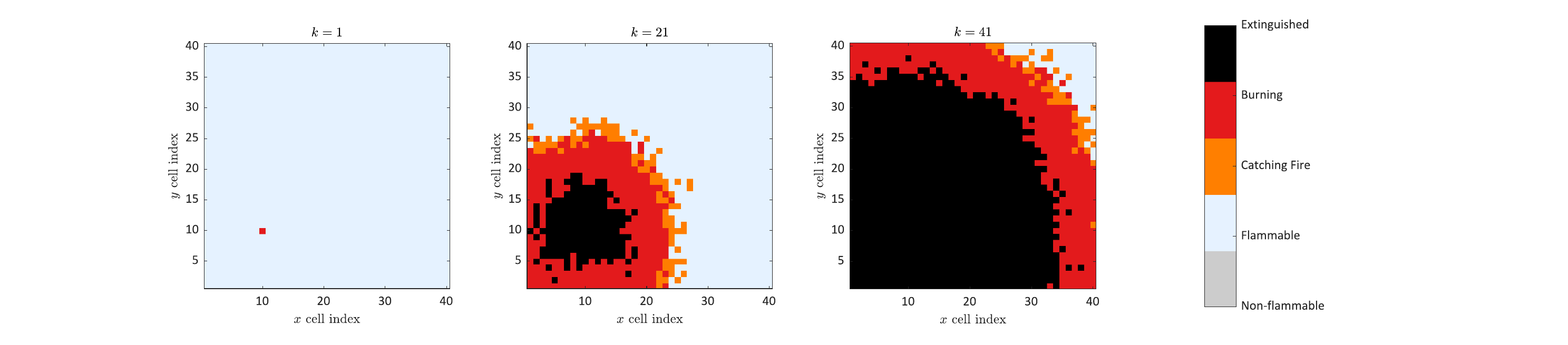}
\caption{Example simulation of the fire spread progression for a basic small dynamic environment.}
\label{fig:fire_spread_small_dynamic}
\end{figure*}

Static environments are initialised with the fire matrix $\Mf$ with elements $\mf_{ij}(k) = 1$, so that there are no active fires during the simulation and there is no fire spread.

Likewise, Figure \ref{fig:fire_spread_small_dynamic} shows the propagation of the fire model for one simulation.
The fire map is initialised with a uniform distribution of flammable cells, and an active fire is initialised for four cells, where $\mf_{20, 21}(k) = 3$.
This setup ensures the fire remains active in each simulation, rather than extinguishing prematurely.

\begin{figure*}
\centering
\includegraphics[width=0.8\linewidth]{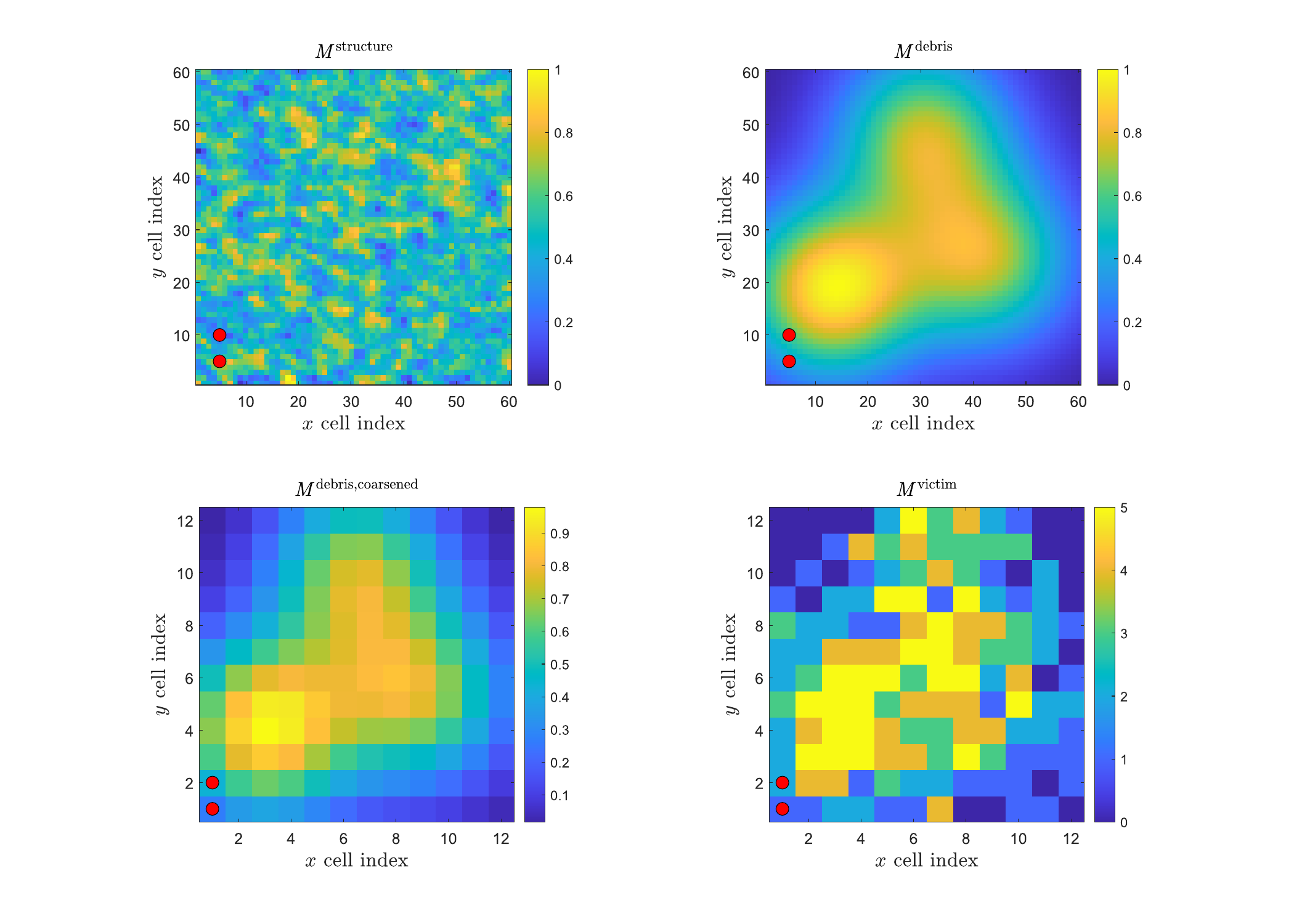}
\caption{Single instance of environment setup for complex disaster environment, red dots indicate robot positions. 
Each subplot represents the value of the labelled environmental state variable for the corresponding cell, from the lower limit (dark blue) to upper limit (yellow) of the parameter range.}
\label{fig:complex_2_agents_params}
\end{figure*}

The second scenario consists of a complex environment, with more complex parameters.
Figure \ref{fig:complex_2_agents_params} shows the environment set up for a single simulation seed.
We initialise $\Mstr$ as a Perlin noise matrix to emulate clusters of buildings with different flammabilities, and $\Mdebris$ is defined by three probability density functions representing several population centres across the disaster environment.

The wind velocity is set as $\mvel_{ij}(k) = \SI{1}{\meter \per \second}$ and the wind direction is set as $\mdir_{ij}(k) = \frac{\pi}{4} ~\textrm{rad}$.
The fire model is configured to increase the influence of wind direction and velocity on fire spread and to slow down fire spread throughout the simulation.
This is achieved by setting the constants of the fire and wind models as $c^{\text{fs1}} = 0.8$, $c^{\text{fs2}} = 2.5$, $c^{\text{wm1}} = 0.1$, $c^{\text{wm2}} = 1.5$, and $c^{\text{wmd}} = 0.9$.
The fire map, $\Mf$, is initialised with two active fires in random cells in the disaster environment.

\begin{figure*}
\centering
\includegraphics[width=\linewidth]{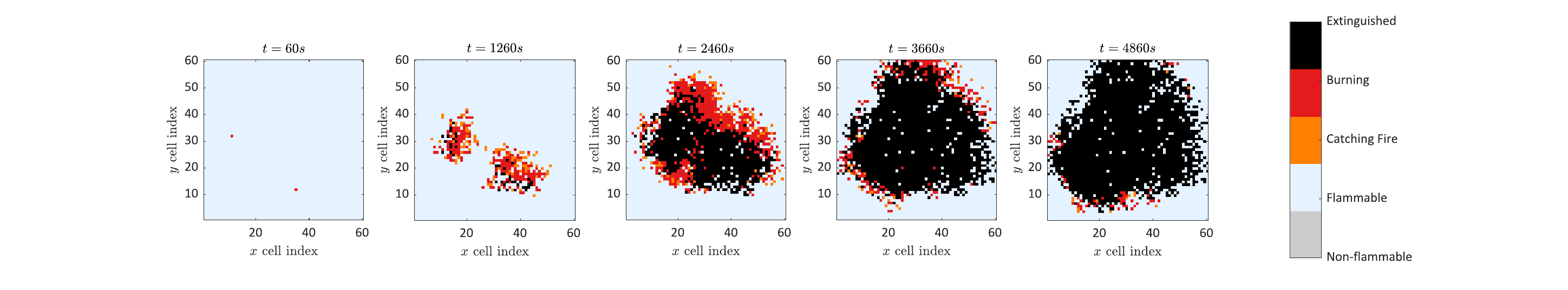}
\caption{Single instance of fire spread for complex disaster environment.}
\label{fig:complex_2_agents_f_hist}
\end{figure*}

Figure \ref{fig:complex_2_agents_f_hist} shows the fire spread over time for a single simulation seed.
In this case, we can see that several fire fronts form and spread more strongly with the wind direction to the Northeast, while some cells do not catch fire due to their low flammability.
In this simulation case, fire spread occurs over the entire duration of the simulation, and does not have a clear peak early on as in previous simulations.

\begin{figure*}
\centering
\includegraphics[width=0.8\linewidth]{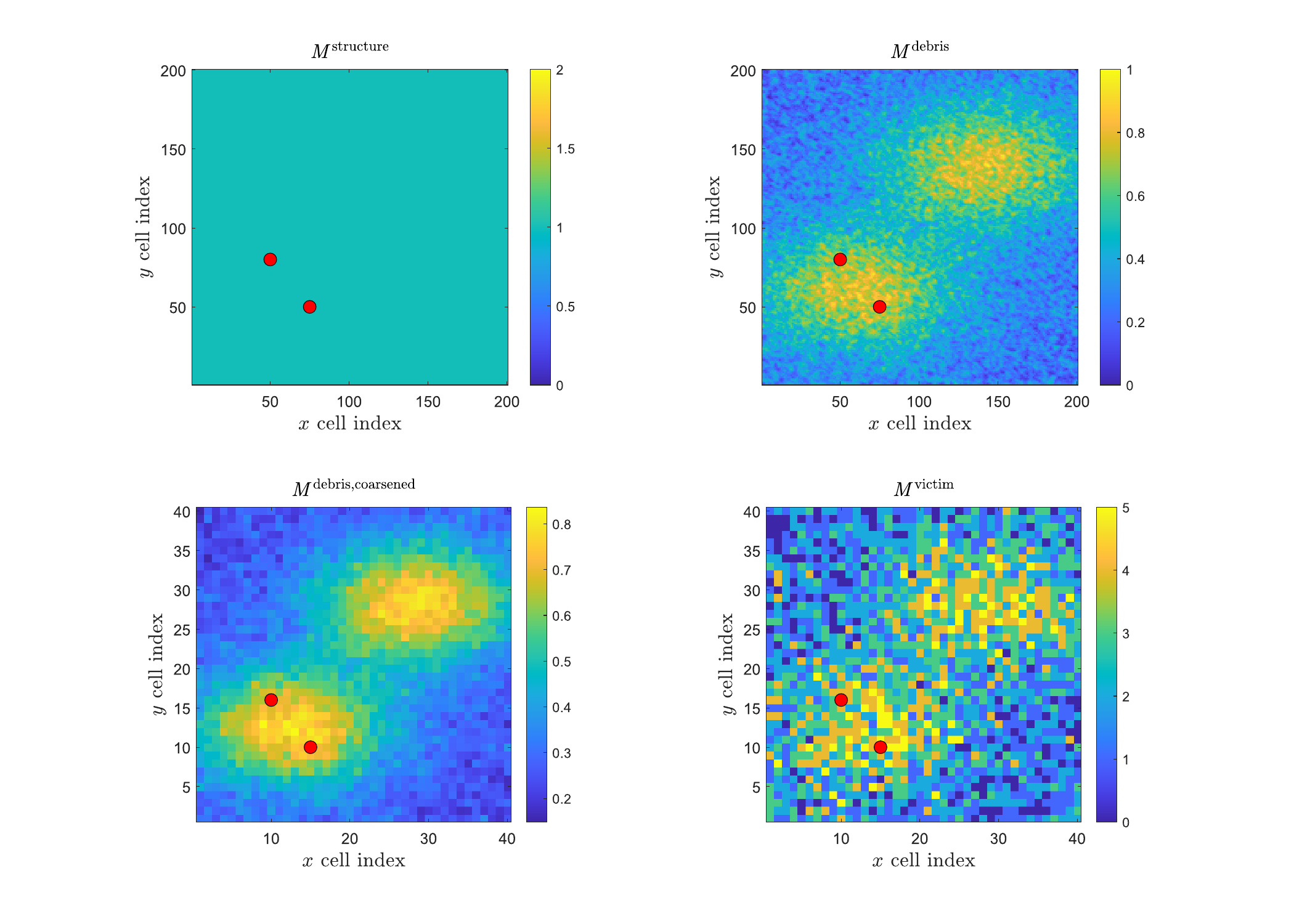}
\caption{Disaster environment setup for local map simulation, red dots indicate robot positions.}
\label{fig:env_states_r_local}
\end{figure*}

The third scenario is also with a two-robot system in a large complex dynamic disaster environment, of dimensions $\nxe = \nye = 200$.
The initialisation of the environment and robot states is shown in Figure \ref{fig:env_states_r_local}, where red circles show the initial robot locations.
The building map, $\Mdebris$, is initialised with two population centres with Gaussian distributions; the structure map, $\Mstr$ is fixed to ones; and the victim map $\Mvictim$ is initialised according to the building distribution.

\begin{figure*}
\centering
\includegraphics[width=\linewidth]{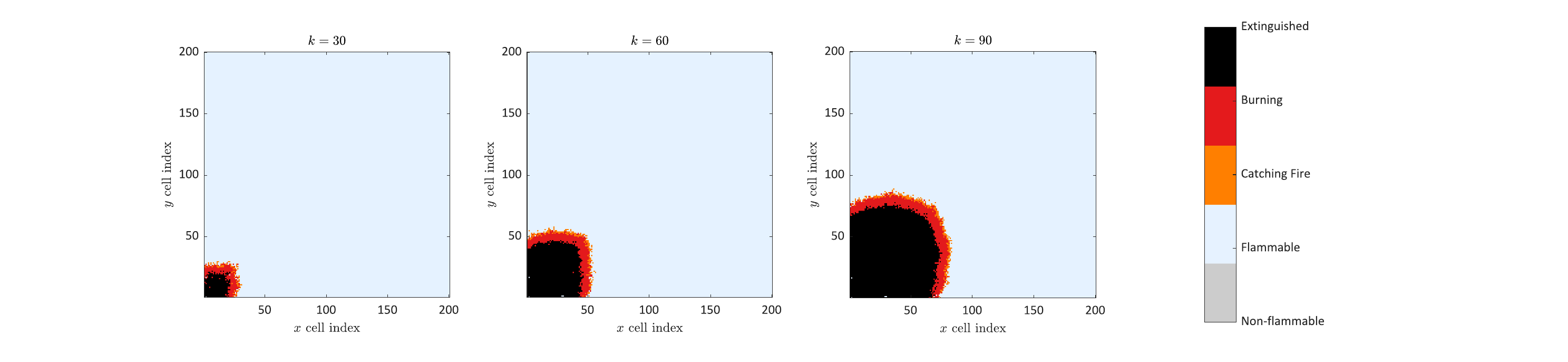}
\caption{Fire spread progression for local map simulation, example simulation.}
\label{fig:fire_spread_r_local}
\end{figure*}

The fire map is initialised with active fires in the bottom left corner.
The progression of the fire spread over one of the simulations is shown in Figure \ref{fig:fire_spread_r_local}.
Due to the size of the disaster environment and the building map, the fire spread rapidly propagates out from the initial active fire point with a wide radial active fire front.

\end{document}